\documentclass{elsart}

\usepackage[sort&compress]{natbib}

\usepackage{eucal}
\usepackage{mathrsfs}
\usepackage{amsmath}
\usepackage{amssymb}
\usepackage{amsfonts}
\usepackage{latexsym}

\usepackage{graphics,graphicx}
\usepackage{epsfig,fancybox}

\usepackage{longtable}

\newcommand{\be}{\begin{eqnarray}}
\newcommand{\ee}{\end{eqnarray}}
\newcommand{\bse}{\begin{subequations}}
\newcommand{\ese}{\end{subequations}}

\newcommand{\bit}{\begin{itemize}}
\newcommand{\eit}{\end{itemize}}
\newcommand{\ben}{\begin{enumerate}}
\newcommand{\een}{\end{enumerate}}

\newcommand{\bpm}{\begin{pmatrix}}
\newcommand{\epm}{\end{pmatrix}}

\newcommand{\mbb}{\mathbb}
\newcommand{\mcal}{\mathcal}
\newcommand{\mfr}{\mathfrak}
\newcommand{\mrm}{\mathrm}

\newcommand{\bs}{\boldsymbol}

\newcommand{\p}{\partial}
\newcommand{\f}{\frac}
\newcommand{\diff}{\mrm{d}}
\newcommand{\lan}{\left\langle}
\newcommand{\ran}{\right\rangle}

\def\dbar{\mathchar'26\mkern-11mu \diff}

\newcommand{\R}{\mbb{R}}

\newcommand{\Z}{\mcal{Z}}

\newcommand{\M}{\mrm{M}}
\newcommand{\J}{\mrm{J}}
\newcommand{\MJ}{\mrm{MJ}}

\newcommand{\ga}{\alpha}
\newcommand{\gb}{\beta}
\newcommand{\gc}{\gamma}
\newcommand{\Gc}{\Gamma}
\newcommand{\gd}{\delta}
\newcommand{\Gd}{\Delta}
\newcommand{\gl}{\lambda}
\newcommand{\Gl}{\Lambda}

\newcommand{\gk}{\kappa}
\newcommand{\go}{\omega}
\newcommand{\Go}{\Omega}
\newcommand{\veps}{\varepsilon}
\newcommand{\eps}{\epsilon}
\newcommand{\vphi}{\varphi}
\newcommand{\gr}{\varrho}

\newcommand{\Gs}{\Sigma}
\newcommand{\gs}{\sigma}

\newcommand{\Gt}{\Theta}
\newcommand{\gt}{\theta}

\newcommand{\csp}{\;,\qquad\qquad}
\newcommand{\fa}{\forall\;}

\newcommand{\Obs}{\mcal{O}}
\newcommand{\eve}{\mathscr{E}} 
\newcommand{\Obsm}{\Obs'}

\newcommand{\Gsm}{\Sigma'}
\newcommand{\Gsc}{\Sigma_*}

\newcommand{\hyp}{\mfr{H}}  
\newcommand{\cone}{\mfr{C}} 

\newcommand{\iso}{\mfr{T}}  

\newcommand{\kB}{k_\mrm{B}}
\newcommand{\vB}{v_\mrm{B}}
\newcommand{\pB}{p_\mrm{B}}

\newcommand{\MC}{\mrm{MC}}

\newcommand{\Energy}{\mcal{U}}
\newcommand{\HamEn}{\mcal{E}}

\newcommand{\energy}{\lan \eps \ran}
\newcommand{\vir}{\lan \bs p\cdot \bs v \ran}

\newcommand{\Ent}{\mcal{S}}
\newcommand{\Press}{\mcal{P}}
\newcommand{\Temp}{\mcal{T}}
\newcommand{\Vol}{V}
\newcommand{\Volume}{\mbb{V}}
\newcommand{\ent}{\mfr{S}}
\newcommand{\Force}{\mcal{F}}

\newcommand{\Mom}{\mcal{G}}
\newcommand{\Work}{\mcal{A}}
\newcommand{\Heat}{\mcal{Q}}

\newcommand{\Prob}{\mathbb{P}} 

\newcommand{\Ind}{\mcal{I}}

\newcommand{\diag}{\mrm{diag}}

\newcommand{\drift}{\mcal{K}} 
\newcommand{\D}{\mcal{D}} 
\newcommand{\Wie}{{\bs B}} 

\newcommand{\bath}{\mrm{b}}

\newcommand{\dP}{\diff P}
\newcommand{\dV}{\diff V}

\newcommand{\dX}{\diff X}
\newcommand{\dx}{\diff x}
\newcommand{\dt}{\diff t}
\newcommand{\ds}{\diff s}
\newcommand{\dB}{\diff B}

\newcommand{\dtau}{\diff \tau}

\newcommand{\dbx}{\diff \bs x}

\newcommand{\parder}[2]{\frac{\p {#1}}{\p {#2}}}

\newcommand{\EW}[1]{\lan{#1} \ran}

\begin{document}

\begin{frontmatter}


\title{Relativistic Brownian Motion}
\author[label1]{J\"orn Dunkel}, 
\ead{jorn.dunkel@physics.ox.ac.uk}
\ead[url]{http://www-thphys.physics.ox.ac.uk/people/JornDunkel/}
\author[label2]{Peter H\"anggi}
\ead{peter.hanggi@physik.uni-augsburg.de}
\ead[url]{http://www.physik.uni-augsburg.de/theo1/hanggi/}

\address[label1]{Rudolf Peierls Centre for Theoretical Physics,
University of Oxford, 1 Keble Road, Oxford OX1 3NP, United Kingdom}
\address[label2]{Institut f\"ur Physik, Universit\"at Augsburg,
Universit\"atsstra{\ss}e 1, D-86135 Augsburg, Germany}

\begin{abstract}
Over the past one hundred years Brownian motion theory has contributed substantially to our understanding of various microscopic phenomena. Originally proposed as a phenomenological  paradigm for atomistic matter interactions, the theory has since evolved into a broad and vivid research area, with an ever increasing number of applications in biology, chemistry, finance, and physics. The mathematical description of stochastic processes has led to new approaches in other fields, culminating in the path integral formulation of modern quantum theory. Stimulated by experimental progress in high energy physics and astrophysics,  the  unification of relativistic and stochastic concepts has re-attracted considerable interest during the past decade. Focusing on the framework of special relativity, we review here recent progress in the phenomenological description of relativistic diffusion processes. After a brief historical overview, we will summarize basic concepts from the Langevin theory of nonrelativistic Brownian motions and discuss relevant aspects of relativistic equilibrium thermostatistics. The introductory parts are followed by a detailed discussion of  relativistic Langevin equations in phase space.  We address the choice 
of time parameters, discretization rules, relativistic fluctuation-dissipation theorems, and Lorentz transformations of stochastic differential equations. The general theory is illustrated through analytical and numerical results for the diffusion of free relativistic Brownian particles. Subsequently, we discuss how Langevin-type equations can be obtained as approximations to  microscopic models.  The final part of the article is dedicated to relativistic diffusion processes in Minkowski spacetime. Since the velocities of relativistic particles are bounded by the speed of light,  nontrivial relativistic Markov processes in spacetime do not exist; i.e., relativistic generalizations of the nonrelativistic diffusion equation and its Gaussian solutions must necessarily be non-Markovian. We compare different proposals that were made in the literature and discuss their respective benefits and drawbacks. The review concludes with a summary of open questions, which may serve as a starting point for future investigations and extensions of the theory.
\end{abstract}
\begin{keyword}
Brownian motion \sep 
special relativity\sep
stochastic processes \sep 
relativistic Langevin equations\sep 
Fokker-Planck equations\sep
diffusion processes\sep
relativistic thermodynamics
\PACS 
02.50.Ey \sep 
05.40.-a \sep 
05.40.Jc \sep 
47.75.+f      
\end{keyword}
\end{frontmatter}


\newpage
\tableofcontents

\section{Introduction}

In his \textit{annus mirabilis} 1905 Albert Einstein published four manuscripts~\cite{1905Ei_Photo,1905Ei_SRT1,1905Ei_SRT2,1905Ei_BM} that would forever change the world of physics. Two of those papers~\cite{1905Ei_SRT1,1905Ei_SRT2} laid the foundations for the special theory of relativity, while another one~\cite{1905Ei_BM}  solved the longstanding problem of classical (nonrelativistic) Brownian motion.\footnote{Einstein's first paper~\cite{1905Ei_Photo} provided the  explanation for the photoelectric effect.} Barring gravitational effects~\cite{MiThWe00,Weinberg}, special relativity has proven to be the correct framework for describing physical processes on all terrestrial scales~\cite{2005La_R,SexlUrbantke}. Accordingly, during the past century extensive efforts have been made to adapt established nonrelativistic theories such as, e.g., thermodynamics, quantum mechanics or field theories~\cite{WeinbergQFT1} to the requirements of special relativity. Following this tradition, the present review focuses on recent progress in the theory of special relativistic Brownian motion and diffusion processes~\cite{1996MaWe,1997DeMaRi,1998DeRi,2001BaDeRi,2001BaDeRi_2,2007AnFr,2007ChKr,2005DuHa,2005DuHa_2,2006DuHiHa,2005Zy,2006DuHa,2007DuHa,2007DuTaHa,2006Fa,2007Li,2007ChDe_1,2007ChDe_2, 2008ChDe,2005Ko,1998BoPoMa,2007Ko,2008DuHaWe,2008Ba_1,2008Ba_2}.

\subsection{Historical background}

Historically, the term \lq Brownian motion\rq~refers to the irregular dynamics exhibited by a test particle (e.g., dust or pollen) in a liquid environment. This phenomenon, already mentioned in 1784 by the Dutch physician Ingen-Housz~\cite{1784IH,1789IH}, was first analyzed in detail by the Scottish botanist Robert Brown~\cite{1827Br} in 1827. About eighty years later, Sutherland~\cite{1905Su}, Einstein~\cite{1905Ei_BM} and von Smoluchowski~\cite{EiSm} were able to theoretically explain these observations. They proposed that Brownian motion is caused by quasi-random, microscopic interactions with molecules forming the liquid. In 1909 their theory was confirmed experimentally by Perrin~\cite{1909Pe}, providing additional evidence for the atomistic structure of matter. During the first half of the 20th century the probabilistic description of Brownian motion processes was further elaborated in seminal papers by Langevin~\cite{1908La,1997LeGy}, Fokker~\cite{1914Fo},  Planck~\cite{1917Pl}, Klein~\cite{1921Kl},  Uhlenbeck and Ornstein~\cite{1930UhOr} and Kramers~\cite{1940Kr}.\footnote{Excellent early reviews are given by Chandrasekhar~\cite{1943Ch}, and Wang and Uhlenbeck~\cite{1945WaUh}.}
\par
In parallel with these early theoretical studies in the field of physics, outstanding mathematicians like Bachelier~\cite{1900Ba}, Wiener~\cite{1923Wi,1930Wi,1933PaWiZy}, Kolmogoroff~\cite{1931Ko,1933Ko,1937Ko}, Feller~\cite{1936Fe}, and L\'evy~\cite{1939Le,1948Le} developed a rigorous basis for the theory of Brownian motions and stochastic processes. Between 1944 and 1968 their groundbreaking work was complemented by Ito~\cite{1944Ito,1951Ito}, Gihman~\cite{1947Gi,1950Gi_1,1950Gi_2}, Fisk~\cite{1963Fisk,1965Fisk} and Stratonovich~\cite{1964St,1966St,1968St}, who introduced and characterized different types of stochastic integrals or, equivalently, stochastic differential equations (SDEs). SDEs present a very efficient tool for modeling random processes and their analysis has attracted an ever-growing interest over the past decades~\cite{MacKean,1967Nelson,1972CoxMiller,1996Risken,KaSh91,VanKampen,Grigoriu}.\footnote{The history of the mathematical literature on Brownian motions and stochastic processes is discussed extensively in Section 2.11 of Ref.~\cite{KaSh91}; see also Sections 2-4 in Nelson~\cite{1967Nelson}.} Nowadays, the modern theory of stochastic processes goes far beyond the original problem considered by Einstein and his contemporaries, and the applications cover a wide range of different areas including physics~\cite{1982HaTh,1990HaTaBo,2002Re,2002AsHa,2005HaMa_R,2005FrKr,2005SoKl,2005EbSo}, biology~\cite{2002Allen,2006Wilkinson}, economy and finance~\cite{ElKo99,BoPo01,2004Glasserman}. 
\par
A central topic of this review concerns the question how SDE-based Brownian motion models can be generalized within the framework of special relativity. In the physics literature~\cite{VanKampen}, SDEs are often referred to as Langevin equations~\cite{1908La,1997LeGy}, and we shall use both terms synonymously here. From a mathematical perspective, SDEs~\cite{KaSh91} determine well-defined models of stochastic processes; from the physicist's point of view, their usefulness for the description of a real system is \textit{a priori} an open issue. Therefore, the derivation of nonrelativistic Langevin equations from microscopic models has attracted considerable interest over the past sixty  years~\cite{1945Bo,1959Ma,1965FoKaMa,1985CoWeLi,1986Po,1988LeShPo,1997Ha,2006DuHa}. Efforts in this direction not only helped to clarify the applicability of SDEs to physical problems but led, among others, also to the concept of quantum Brownian motion~\cite{1965FoKaMa,1960Se,1961Se,1983CaLe,1983GrTa,1984GrWeTa,1985CaLe,1985RiHaWe,1987FoKa,1988FoLeCo,1991Pe,2000TsPe,2005KoLeHa,2005HaIn,2006HaIn}.\footnote{The vast literature on classical Brownian motion processes and their various applications in nonrelativistic physics is discussed in several survey articles~\cite{1990HaTaBo,2002Re,2002AsHa,2005HaMa_R,2005FrKr,2005SoKl,1978Fo,1994Kl,1995Klimontovich,1995HaJu,1996AbEtAl}. Nonrelativistic generalizations of the standard theory as, e.g., anomalous diffusion processes have been summarized in~\cite{2005SoKl,1990BoGe,2000MeKl}, while review articles on nonrelativistic quantum Brownian motion can be found in~\cite{1988GrScIn,1998GrHa,2005KoLeHa,2005HaIn}.}
\par 
If one aims at generalizing the classical Brownian motion concepts to special relativity, then several elements from relativistic equilibrium thermodynamics and relativistic statistical mechanics play an important role. More precisely, thermostatistic principles govern the stationary behavior of Brownian particles and, thus, impose constraints on the structure of relativistic Langevin equations. The first papers on relativistic thermodynamics were published in 1907 by Planck~\cite{1907Pl,1908Pl} and Einstein~\cite{1907Ei}. A main objective of theirs was to identify the Lorentz transformation laws of thermodynamic variables (temperature, pressure, etc.).\footnote{See also Pauli~\cite{1921Pa}, Eddington~\cite{1923Ed}, Tolman~\cite{1934Tolman} and van Dantzig~\cite{1939Da} for early discussions of this problem.} In 1963 the results of Einstein and Planck became questioned by Ott~\cite{1963Ott}, whose work initiated an intense debate about the correct relativistic transformation behavior of thermodynamic quantities~\cite{1965Bo,1965Ga,1965Su,1965Ar,1965Ar_1,1966Ar,1966Ro,1966La_1,1966Ki,1966Pa,1966Pa_2,1966La,1966Fr,1967La,1967No,1967Wi,1967Wi_2,1967Re,1967Mo,1968Li,1968Ge,1968LaJo,1968VK,1969VK,1969VK_2,1970VK,1969Mu,1969LeVKMaJa,1969Ha,1971Ho_2,1971CaHo,1970Ge,1970LaJo,1971Ja,1973Gr,1975Ei,1976Is,1977Ag,1978Kr,1979IsSt,1980La,1981La,1987Is,1986ShMuRu,1995Ko,1996LaMa,1997KrMu,2001ReSmDu,1992ScTr,2008Re}.\footnote{The pre-1970 literature on this disputed issue has been reviewed by Yuen~\cite{1970Yu} and Ter Haar and Wegland~\cite{1971TeWe}, see also Israel~\cite{1987Is}; more recent surveys can be found in~\cite{Neugebauer,1992Liu,1994Liu,1999-lrr-1}.} However, as clarified by van Kampen~\cite{1968VK} and Yuen~\cite{1970Yu}, the controversy surrounding relativistic thermodynamics can be resolved by realizing  that thermodynamic quantities can be \emph{defined} in different, equally consistent~ways. 
\par
While some authors considered relativistic thermodynamics as a purely macroscopic theory, others tried to adopt a more fundamental approach by focussing on relativistic equilibrium statistical mechanics. Early  pioneering work in the latter direction was provided by Planck's students von Mosengeil~\cite{1907Mo} and von Laue~\cite{1911vL}, and his collaborator J\"uttner~\cite{1911Ju}, who derived in 1911 the relativistic generalization of Maxwell's velocity distribution~\cite{1867Ma}.\footnote{A general introduction to relativistic gases is given in   Synge's textbook classic~\cite{1957Sy}.  Relativistic generalizations of equipartition and virial theorems~\cite{1918To} are discussed by Pauli~\cite{1921Pa} and Einbinder~\cite{1948Ei}; more recent investigations of these two specific topics can be found in, e.g., Refs.~\cite{1985Kr,1990LuSc,1995Ko_2}.  Equilibrium distributions for ideal relativistic quantum gases were also derived by J\"uttner~\cite{1928Ju} in 1928. A few recent papers~\cite{2005Sc,2006Le,2006Ka,1981HoScPi,1989HoShSc,1995BuHo,2007DuHa} have raised doubts about the correctness of J\"uttner's results~\cite{1911Ju,1928Ju}, but relativistic molecular dynamics simulations confirm J\"uttner's  prediction~\cite{2006AlRoMo,2007CuEtAl}, cf. discussion in Sections~\ref{s:juettner_gas} and \ref{s:propertime} below.} Research on relativistic equilibrium thermostatistics experienced its most intense phase between 1960 and 1970 \cite{1966Pa,1966Pa_2,1967Mo,1968VK,1951Be,1953ScKr,1962KrHa,1964Ch,1968Mo,1967EbKu,1965Ba,1967Ba,1967BaKo,1967BaKoPi,1967Ha,1968Ba,1969Ba,1971BaBr,1968Na,1970Ge,1970JoLa_1,1970JoLa_2}. An excellent exposition on the conceptual foundations and difficulties of relativistic statistical mechanics was given by Hakim~\cite{1967Ha_2,1967Ha_3,1967Ha_4} in 1967. Over the past years the field has continued to attract interest~\cite{1977Re,1978Hak,1983Ce,1985Kr,1995BY,1996MaKrSz,1998Tr,2001DuNaTr,2001DeRiLe,2001MoRo,2002Ka,2004Be,2005ArLoAn,2005Sc,2005SiLi,2005Ka,2006Ka,2006Si,2006Le,2007BeFe,2007LiJiZh,2006AlRoMo,2007CuEtAl,1981HoScPi,1989HoShSc,1995BuHo,2006Na,2007DuHa,2008ClEtAl,2008De}. 
\par
The recurring debate on relativistic thermostatistics can be traced back to the difficulty of treating many-particle interactions in a relativistically consistent manner.\footnote{
 Seminal contributions to the theory of relativistic many-particle interactions were provided by Fokker~\cite{1929Fo}, Wheeler and Feynman~\cite{1945WhFe,1949WhFe}, Pryce~\cite{1948Pr},  Havas and Goldberg~\cite{1964Ha,1962HaGo}, and  Van Dam and Wigner~\cite{1965DaWi,1966DaWi}.  Over the past decades several no-interaction theorems were proven~\cite{1966DaWi,1963CuJoSu,1965Le,1984MaMuSu}. These forbid,  within their respective qualifications, certain types of interaction models within the framework of special relativity. The mathematical structure of relativistic many-particle interactions was analyzed in detail by Arens and Babbitt~\cite{1969ArBa}; various semi-relativistic approximations have been discussed, e.g., in Refs.~\cite{1965Chou,1969HaSt,1972WoHa}. Kerner~\cite{1972Kerner} has edited a reprint collection covering large parts of the pre-1972 literature on relativistic action-at-a-distance models; more recent contributions include~\cite{1998Tr,2001DuNaTr,1970DV,1973GiLi,1973Ra,1974Ma,1980GaYuTr,1980GaYuTr_2,1983GaYuTr,1986SoTr,1989SoStGr,1994GaTrYa}. An alternative, intensely studied method for describing relativistic interactions is based on the so-called constraint formalism~\cite{2001DuNaTr,2006Le,1949Di,1951AnBe,1976To,1978Ko_a,1978Ko_b,1981GoSuMu,1981KiMaMu,1981Ro,1982Ro,1982Sa,1981HoRo,1985HoRo,1985MaEtAl,1986LoLu,1999MoRoTh,2001WoCr,2006GeKhPa}. The foundations of this approach were worked out by Dirac~\cite{1949Di}, who aimed at constructing a  relativistic quantum theory for interacting many-particle systems~\cite{1966DaWi,1963CuJoSu,1965Le,1984MaMuSu}. For a detailed discussion of relativistic many-particle theory, we refer to the insightful considerations in the original papers of Van Dam and Wigner~\cite{1965DaWi,1966DaWi} and Hakim~\cite{1967Ha_2,1967Ha_3,1967Ha_4,1978Hak} as well as to the recent review by Hakim and Sivak~\cite{2006HaSi}.}  
In nonrelativistic physics interactions may propagate at infinite speed, i.e., they can be modeled via instantaneous interaction potentials which enter additively in the Hamilton function; from that point on, nonrelativistic statistical mechanics emerges without much difficulty~\cite{Becker,Huang}. Unfortunately, the situation becomes significantly more complicated in the relativistic case:  Due to their finite propagation speed, relativistic interactions should be modeled by means of fields that can exchange energy with the particles~\cite{Weinberg}. These fields add an infinite number of degrees of freedom to the particle system. Eliminating the field variables from the dynamical equations may be possible in some cases but this procedure typically leads to retardation effects, i.e., the particles' equations of motions become non-local in time~\cite{1967Ha_2,1967Ha_3,1945WhFe,1949WhFe,1965DaWi,1966DaWi}. Thus, in special relativity it is usually very difficult or even impossible to develop a consistent field-free Hamilton formalism of interacting particles.
\par
In spite of the difficulties impeding a rigorous treatment of classical  relativistic many-particle systems, considerable progress was made during the second half of the past century in constructing an approximate relativistic kinetic theory~\cite{1981Er,1940Ec,1940LiMa,1953KlDGMa,1956BeBu,1963Is,1979IsSt,1983Ce,1965Chou,1967Kl,1964Ma,1965GoPr,1966Li,1970Ak,1979DG,1981Go,1982PaJoCa,1999CeKr,2000CeKr,2000Le,2003Ho,2004StGu,2006HsHu,2008BrEtAl} based on relativistic Boltzmann equations for the \emph{one-particle} phase space probability density functions (PDFs).\footnote{Comprehensive introductions to relativistic Boltzmann equations can be found in the textbooks by Stewart~\cite{Stewart}, de Groot et al.~\cite{1980DG}, and Cercignani and Kremer~\cite{CercignaniKremer}, or also in the reviews by Ehlers~\cite{1971Eh} and Andr\'easson~\cite{2005-lrr-2}.}  From such a kinetic theory it is only a relatively small step to formulating a theory of relativistic Brownian motion processes in terms of Fokker-Planck equations and Langevin equations. While the relativistic Boltzmann equation~\cite{1980DG,CercignaniKremer} is a \emph{nonlinear} partial integro-differential equation for the PDF, Fokker-Planck equations are linear partial differential equations and, therefore, can be more easily solved or analyzed~\cite{1996Risken}. 
\par
The present article focuses primarily on relativistic stochastic processes that are characterized by \emph{linear} evolution equations for their respective one-particle (transition) PDFs. The corresponding phenomenological theory of relativistic Brownian motion and diffusion processes has experienced considerable progress during the past decade, with applications in various areas of high-energy physics~\cite{1988Sv,2005HeRa,2006WoEtAl,2006HeGrRa,2006RaGrHe,2004AbGa,2005AbGa,2004Wo} and astrophysics \cite{1984De,2006Be,2006WoMe,2006DiDrSh,2008AlBeGa}. From a general perspective, relativistic stochastic processes provide a useful approach whenever one has to model the quasi-random behavior of relativistic particles in a complex environment. Therefore, it may be expected that relativistic Brownian motion and diffusion concepts will play an increasingly important role in future investigations of,  e.g., thermalization and relaxation processes in astrophysics~\cite{1984De,2006Be,2006WoMe,2006DiDrSh} or high-energy collision experiments~\cite{1988Sv,1997RoEtaAl,2005HeRa,2006HeGrRa,2006HeGrRa_2,2006RaGrHe}.

\subsection{Relativistic diffusion processes: problems and general strategies}

According to our knowledge, the first detailed mathematical studies on relativistic diffusion processes were performed independently by {\L}opusza{\`n}ski~\cite{1953Lo}, Rudberg~\cite{1957Ru}, and Schay~\cite{1961Sc} between 1953 and 1961. In the 1960s and 70s  their pioneering work was further elaborated by Dudley who published a series of papers~\cite{1965Du,1967Du,1973Du,1974Du} that aimed at providing an axiomatic approach to Lorentz invariant Markov processes~\cite{KaSh91} in \emph{phase space}. Independently, a similar program was  pursued by Hakim \cite{1967Ha_2,1967Ha_3,1967Ha_4,1965Ha,1968Ha}, whose insightful analysis helped to elucidate the conceptual subtleties of relativistic stochastic processes~\cite{1968Ha}. Dudley (Theorem~11.3 in~\cite{1965Du}) and Hakim (Proposition~2 in~\cite{1968Ha}) proved the non-existence of nontrivial\footnote{A diffusion process is considered as \lq nontrivial\rq~if a typical path has a non-constant, non-vanishing velocity.} Lorentz invariant Markov processes in Minkowski spacetime, as already suggested by {\L}opusza{\`n}ski~\cite{1953Lo}. This fundamental  result implies that it is difficult to find acceptable relativistic generalizations of the well-known nonrelativistic diffusion equation~\cite{1855Fi,Becker}
\be\label{e:intro_diffusion_equation}
\f{\p}{\p t}\gr=\D\,\nabla^2\gr,
\ee
where $\D>0$ is the diffusion constant and $\gr(t,\bs x)\ge 0$ the PDF for the particle positions $\bs x\in \R^d$ at time $t$. In order to circumvent this \lq no-go\rq~theorem for relativistic Markov processes in spacetime, 
one usually adopts either of the following two strategies:\footnote{The mathematical interest in relativistic diffusion processes increased in the 1980s and 1990s, when several authors~\cite{1976Ya,1975Ca_1,1975Ca_2,1976Ca,1976Caubet,1978Ne,1977Ca,1981Su,1978GuRu,1978GuRu_2,1980Na,1980Roy,1981Roy,1984Ca,1985Ca,1992Ne,1992Ga,1995MoVi,1995GaKlOl,1996Na,1997Po,2001Pa,1993Fa} considered the possibility of extending Nelson's stochastic quantization approach~\cite{1966Ne} to the framework of special relativity.These studies, although interesting from a mathematical point of view, appear to have relatively little 
physical relevance because Nelson's stochastic dynamics~\cite{1966Ne} fails to reproduce the correct quantum correlation functions even in the nonrelativistic case~\cite{1979GrHaTa}.} 
\begin{itemize}
\item
One considers non-Markovian diffusion processes $\bs X(t)$ in Minkowski  spacetime~\cite{1854Kelvin,1950Go,1974Ka,1996MaWe,2007DuTaHa}.
\item
One constructs relativistically acceptable Markov processes in phase space by considering not only the position coordinate $\bs X(t)$ of the diffusing particle but also its momentum coordinate $\bs P(t)$~\cite{1961Sc,1965Du,1967Du,1973Du,1974Du,1967Ha_2,1967Ha_3,1967Ha_4,1965Ha,1968Ha,1997DeMaRi,1998DeRi,2001BaDeRi,2001BaDeRi_2,2007AnFr,2007ChKr,2005DuHa,2005DuHa_2,2006DuHiHa,2005Zy,2006DuHa,2007DuHa,2006Fa,2008ChDe,2007ChDe_1,2007ChDe_2}.
\end{itemize}

\subsubsection{Non-Markovian diffusion models in Minkowski spacetime}

A commonly considered \lq relativistic\rq\space generalization of Eq.~\eqref{e:intro_diffusion_equation} is the telegraph equation~\cite{1854Kelvin,1950Go,1974Ka,1996MaWe}\footnote{Masoliver and Weiss~\cite{1996MaWe} discuss several possibilities of deriving Eq.~\eqref{e:intro_telegraph_equation} from different underlying models.}
\be\label{e:intro_telegraph_equation}
\tau_v\f{\p^2}{\p t^2}\gr+\f{\p}{\p t}\gr=\D\,\nabla^2\gr,
\ee  
where $\tau_v>0$ denotes a finite relaxation time scale. Unlike the classical diffusion equation~\eqref{e:intro_diffusion_equation}, which is recovered for $\tau_v=0$, the telegraph equation~\eqref{e:intro_telegraph_equation} contains a second order time-derivative and, therefore, describes a non-Markovian process. While the classical diffusion equation~\eqref{e:intro_diffusion_equation} permits superluminal propagation speeds, the diffusion fronts described by Eq.~\eqref{e:intro_telegraph_equation} travel at finite absolute velocity \mbox{$v=(\D/\tau_v)^{1/2}$}; cf. the discussion in Section~\ref{s:rel_diff} below. 
\par
Historically, Eq.~\eqref{e:intro_telegraph_equation} was first obtained by Thomson (Lord Kelvin) in 1854~\cite{1854Kelvin}, while studying the signal transduction for the first transatlantic cable. The first probabilistic derivation of Eq.~\eqref{e:intro_telegraph_equation} for the one-dimensional case was given by Goldstein~\cite{1950Go} in 1950. His approach was based on a so-called persistent random walk model originally introduced by F\"urth \cite{1917Fu,1922Fu} in 1917 as a paradigm for diffusive motion in biological systems and later also considered by Taylor~\cite{1922Ta} in an attempt to treat turbulent diffusion.\footnote{See also Kac~\cite{1974Ka} and Bogu{\~{n}}\'a et al.~\cite{1998BoPoMa,1999BoPoMa}.} In contrast to standard non-directed random walk models, which lead to the classical diffusion equation~\eqref{e:intro_diffusion_equation} when performing an appropriate continuum limit~\cite{KaSh91}, the random jumps of a persistent walk take into account the history of a path by assigning a larger probability to those jumps that point in the direction of the motion before the jump~\cite{1950Go,1974Ka}. Persistent random walk models can be used to describe the transmission of light in multiple scattering media~\cite{1989Is} such as foams~\cite{2003MiSt,2005MiMaSt,2006MiSaFa} and thin slabs~\cite{1999BoPoMa,2008K2}. Similarly, the telegraph equation~\eqref{e:intro_telegraph_equation} has been applied in various areas of physics over the past decades, e.g., to model the propagation of electric signals and heat waves.\footnote{A detailed review of the pre-1990 research on heat waves was provided by Joseph and Preziosi~\cite{1989JoPr,1990JoPr}, while more recent discussions and applications of  Eq.~\eqref{e:intro_telegraph_equation} can be found in~\cite{1996MaWe,2000KoLi,2001KoLi,2001HePa,2004AbGa,2005AbGa,2005Ko,2007Ko}.} 
\par
Moreover, an interesting connection between the free particle Dirac equation~\cite{Peskin} and the telegraph equation~\eqref{e:intro_telegraph_equation} was pointed out by Gaveau et al.~\cite{1984GaEtAl} in 1984: The solutions of both equations may be linked by means of an analytic continuation quite similar to the relation between the classical diffusion equation~\eqref{e:intro_diffusion_equation} and the free-particle Schr\"odinger equation in the nonrelativistic case.\footnote{For further reading about path integral representations of the Dirac propagator  we refer to the papers of Ichinose~\cite{1982Ic,1984Ic}, Jacobson and Schulman~\cite{1984JaSc}, Barut and Duru~~\cite{1984BaDu}, and Gaveau and Schulman~\cite{1987GaSc}; see also footnote 7 in Gaveau et al.~\cite{1984GaEtAl} and pp. 34-36 in Feynman and Hibbs~\cite{FeynmanHibbs}.} On the other hand, the telegraph equation~\eqref{e:intro_telegraph_equation} is not the only possible generalization of Eq.~\eqref{e:intro_diffusion_equation} and a rather critical discussion of Eq.~\eqref{e:intro_telegraph_equation} in the context of relativistic heat transport was given by van Kampen~\cite{1970VK} in 1970. Section~\ref{s:rel_diff} below takes a closer look at the properties of Eq.~\eqref{e:intro_telegraph_equation} and also addresses potential alternatives~\cite{1984Ka,2007DuTaHa}.

\subsubsection{Relativistic Markov processes in phase space}

As an alternative to non-Markovian diffusion models in spacetime, one can consider relativistic Markov processes in phase space~\cite{1961Sc,1965Du,1967Du,1973Du,1974Du,1967Ha_2,1967Ha_3,1967Ha_4,1965Ha,1968Ha,1997DeMaRi,1998DeRi,2007ChDe_1,2007ChDe_2,2001BaDeRi,2001BaDeRi_2,2007AnFr,2007ChKr,2005DuHa,2005DuHa_2,2006DuHiHa,2005Zy,2006DuHa,2007DuHa,2006Fa,2008ChDe,2008Ba_1,2008Ba_2}. Typical examples are processes described by Fokker-Planck equations (FPEs) or Langevin equations~\cite{2006DuHa,1997DeMaRi,1998DeRi,2001BaDeRi,2001BaDeRi_2,2005OrHo,2003CaEg,2005DuHa,2005DuHa_2,2006Fa,2006CaCa,2004Zy,2005Zy,2008ChDe,2007Li,2007AnFr,2007Fi,2007ChDe_1,2007ChDe_2,2008Ba_1,2008Ba_2}. Similar to the relativistic Boltzmann equation, relativistic FPEs in phase space can be used to describe non-equilibrium and relaxation phenomena in relativistic many-particle systems.  FPEs can be derived from  Langevin equations, as approximations to more general linear master equations~\cite{VanKampen,1961Sc} or by approximating the collision integrals in nonlinear Boltzmann equation by differential expressions that contains effective friction and diffusion coefficients~\cite{1970Ak,2007ChKr}. In particular, the latter method was successfully applied in different areas of physics over the past decades, including plasma physics~\cite{1970Ak,1980Da,1981VE,1981Fi,1985KaFi,1989GoGa,1990GoGa,1992GoGa,1995KlTe,1997DaBeHaGu,2000BaMo,2002BaMo,2006YoEtAl,2007KaTeKr}, high-energy physics~\cite{1988Sv,1997RoEtaAl,2005HeRa,2006WoEtAl,2006HeGrRa,2006HeGrRa_2,2006RaGrHe}, and astrophysics~\cite{1997MaGo_1,1997MaGo_2,1998ItKoNo,2006WoMe,2006BeLeDe,2006Be}.\footnote{Relativistic Fokker-Planck-type equations also played a role in the debate about whether or not the black body radiation spectrum is compatible with J\"uttner's relativistic equilibrium distribution~\cite{1979Bo,1979Bo_2,1981BY,1983BlPeSa,1985Kr}.} In the 1980s and 90s this approach was further elaborated~\cite{1980Da,1981VE,1981Fi,1985KaFi,1989GoGa,1990GoGa,1992GoGa,1995KlTe,1997DaBeHaGu} and several numerical methods for solving FPEs were developed~\cite{1985KaFi,1986Ka,1987ShFuBe,1998PeSh}.\footnote{A main reason for the lively interest in relativistic FPEs at that time was the prospect of building plasma fusion reactors.} Recent applications include the modeling of diffusion and thermalization processes in  quark-gluon plasmas, as produced in relativistic heavy ion collision experiments~\cite{1988Sv,1997RoEtaAl,2005HeRa,2006HeGrRa,2006HeGrRa_2,2006RaGrHe}, as well as the description of complex high-energy processes in astrophysics~\cite{1984De,1997MaGo_1,1997MaGo_2,1998ItKoNo,2006WoMe,2006BeLeDe,2006Be}.
\par
A complementary approach towards relativistic sto\-chastic processes in phase space starts from Langevin equations~\cite{2006DuHa,1997DeMaRi,1998DeRi,2001BaDeRi,2001BaDeRi_2,2005OrHo,2003CaEg,2005DuHa,2005DuHa_2,2006Fa,2006CaCa,2004Zy,2005Zy,2008ChDe,2007Li,2007AnFr,2007Fi,2007ChDe_1,2007ChDe_2}. Stochastic differential equations of the Langevin type yield explicit sample trajectories for the stochastic motion of a relativistic Brownian particle. Relativistic Langevin equations may either be postulated as phenomenological model equations~\cite{1997DeMaRi,2005DuHa} or obtained from more precise microscopic models by imposing a sequence of approximations~\cite{2006DuHa}. Compared with the nonrelativistic case, the latter task becomes considerably more complicated due to the aforementioned conceptual and technical difficulties in formulating  consistent relativistic many-particle theories. The phenomenological Langevin approach to relativistic Brownian motion was initiated by Debbasch et al.~\cite{1997DeMaRi}, who in 1997 proposed a simple relativistic generalization of the classical Ornstein-Uhlenbeck process~\cite{1930UhOr}, representing a special limit cases of a larger class of relativistic Langevin processes~\cite{2006DuHiHa,2007AnFr,2008ChDe}. From a practical point of view, relativistic Langevin equations provide a useful tool for modeling the dynamics of relativistic particles in a random environment, because these SDEs may be simulated by using well-established Monte-Carlo techniques that are numerically robust and efficient~\cite{Grigoriu,2004Glasserman,1999KloedenPlaten}. Recent applications include the analysis of thermalization effects in quark-gluon plasmas~\cite{2006HeGrRa,2006HeGrRa_2,2006RaGrHe,2008RaHe} and  ultrarelativistic plasma beam collisions~\cite{2006DiDrSh}.

\subsection{Structure of the review}

The present article intends to provide a comprehensive introduction to the theory of relativistic Brownian motions with a particular emphasis on relativistic Langevin equations. For this purpose, the subsequent parts are organized as follows. Section~\ref{s:nonrelativistic} summarizes the Langevin theory of nonrelativistic Brownian motions in phase space.  Section~\ref{s:Rel_Thermo} discusses aspects of relativistic equilibrium thermostatistics as far as relevant for the subsequent discussion. Relativistic Langevin equations in phase space and their associated FPEs are considered in Section~\ref{s:RBM_lab_time}.  Section~\ref{s:rel_diff} is dedicated to relativistic diffusion processes in Minkowski spacetime; as outlined above, such processes must necessarily be non-Markovian. Our review concludes with a summary of open questions in Section~\ref{s:summary}, which may serve as a starting point for future investigations and extensions of the theory. Since the topic of this article resides in the intersection between statistical and high-energy physics, we hope that the presentation is accessible for members of both communities. To keep the discussion as self-contained as possible, the article includes three appendices which summarize a few essentials about stochastic integrals, surface integrals in Minkowski spacetime and relativistic thermodynamics. 

\section{Nonrelativistic Brownian motion}
\label{s:nonrelativistic}

This section summarizes basic definitions and mathematical tools  as well as relevant results from the nonrelativistic theory of nonrelativistic Brownian motions. In particular, the Langevin and Fokker-Planck equations considered in this part define the nonrelativistic limit case of the relativistic theory described Section~\ref{s:RBM_lab_time}. In Section~\ref{s:LEs_and_FPEs} we shall briefly recall the general mathematical structure of Langevin and Fokker-Planck equations, the relevance of discretization rules and the choice of fluctuation-dissipation relations. Section~\ref{s:nonrelativistic_microscopic_models} focusses on the question how stochastic differential equations (SDEs) can be motivated and/or derived  from microscopic models. As typical examples, the well-known harmonic oscillator  model~\cite{1945Bo,1959Ma,1965FoKaMa,1985CoWeLi,1986Po,1988LeShPo,1997Ha} and a recently proposed binary collision model~\cite{2006DuHa} will be considered. In contrast to the oscillator model, the collision model can be generalized to the framework of special relativity, and its relativistic version will be discussed in Section~\ref{s:relativistic_binary_collision_model}.

\subsection{Langevin and Fokker-Planck equations} 
\label{s:LEs_and_FPEs}

The condensed discussion in this part is primarily based on the papers of Uhlenbeck and Ornstein~\cite{1930UhOr},  Wang and Uhlenbeck~\cite{1945WaUh}, and  Klimontovich~\cite{1994Kl}. For further reading about nonrelativistic stochastic processes and their numerous applications in physics and mathematics, we refer to the review articles of Chandrasekhar~\cite{1943Ch}, Fox~\cite{1978Fo}, H\"anggi and Thomas~\cite{1982HaTh}, Bouchaud and Georges~\cite{1990BoGe}, Metzler and Klafter~\cite{2000MeKl}, H\"anggi and Marchesoni~\cite{2005HaMa_R}, Frey and Kroy~\cite{2005FrKr}, or the textbooks references~\cite{VanKampen,Gardiner,KaSh91,Grigoriu}. For conceptual clarity, we restrict ourselves to the simplest case where motions are confined to one space dimension $(d=1)$. The generalization to higher space dimensions is obvious.

\subsubsection{Langevin equations and discretization rules}
\label{s:nonlinear_langevin_equations}
As a standard paradigm for Brownian motion, we consider the one-dimensional motion of a point-like Brownian particle (mass $M$), which is surrounded by a stationary homogeneous heat bath consisting, e.g., of smaller liquid particles (mass $m\ll M$) at constant temperature $\Temp$.  The inertial rest frame\footnote{By definition, the mean velocity of the heat bath particles vanishes in $\Gs$.} $\Gs$ of the heat bath will be referred to as \emph{lab frame} hereafter. The position of the Brownian particle in $\Gs$ at time $t$ is denoted by $X(t)$ and its velocity is given by $V(t):=\diff X(t)/\diff t$. The associated nonrelativistic momentum of the Brownian particle is defined by~$P(t):=MV(t)$.
\par
According to the Langevin picture of Brownian motion, the stochastic  dynamics of the Brownian particle due to the interaction with the bath  and in the presence of a conservative external force field $\Force(t,x)$  can be described by the differential equations~\cite{1908La,1997LeGy,1930UhOr,1943Ch,1945WaUh,1982HaTh,1994Kl,2004Eb,2007Li}
\bse\label{e:free_langevin}
\be
\label{e:free_langevin_a}
\f{\dX}{\dt}&=&\f{P}{M},\\
\label{e:OUP_free_langevin_b}
\f{\dP}{\dt}&=&\Force(t,X)-\ga(P) P+\mcal{L}(P,t),
\ee
complemented by the initial conditions\footnote{Without loss of generality we fix the initial time $t_0=0$.}  $X(0)=x_0$ and $P(0)=p_0$. The second term on the rhs. of Eq.~\eqref{e:OUP_free_langevin_b} is a  friction force with the shape of the friction coefficient function $\ga(p)>0$ depending on the microscopic details of the particle-bath interaction. The stochastic Langevin force 
\be
\mcal{L}(P(t),t)=[2D(P)]^{1/2}\odot\zeta(t)
\ee
\ese
reflects fluctuations in the surrounding heat bath. The symbol $\odot$ signals the choice of a specific discretization rule to be discussed in more detail below. The amplitude of the fluctuating force $\mcal{L}$ is tuned by function $D(p)>0$. For a spatially inhomogeneous heat bath the functions $\ga$ and $D$ would also depend on the position coordinate $x$. The stochastic \lq driving\rq\space function $\zeta(t)$ is often taken to be a Gaussian white noise, i.e., $\zeta(t)$ is characterized by:
\bse\label{e:white_noise}
\be
\lan \zeta(t)\ran &=&0,\\
\lan  \zeta(t)\, \zeta(s)\ran &=& \gd(t-s),\label{e:white_noise_b}
\ee
\ese
with all higher cumulants being zero. In Eqs.~\eqref{e:white_noise}, the bracket $\lan\, \cdot\,\ran$ symbolizes an average over all possible realizations of the noise process $\zeta(t)$.
\par
In the mathematical literature~\cite{KaSh91,Grigoriu}, SDEs like the Langevin Eq.~\eqref{e:free_langevin} are usually written in the differential notation
\bse\label{e:langevin_m} 
\be
\diff X(t)&=&(P/M)\,\diff t,\\
\label{e:nonlinear_langevin}
\diff P(t)&=&\Force(t,X)\,\dt-\ga(P)\, P\, \dt+[2D(P)]^{1/2}\odot\diff B(t); 
\ee 
Here, $\dX(t):=X(t+\diff t)-X(t)$ denotes the position increment and  $\dP(t):=P(t+\diff t)-P(t)$ the momentum change. The random function $B(t)$ is a standardized one-dimensional Brownian motion or, equivalently, a standard Wiener process~\cite{1923Wi,1982HaTh,KaSh91,Grigoriu}, whose increments
\be
\diff B(t):=B(t+\dt)-B(t)
\ee
are defined to be stochastically independent\footnote{This means that the joint probability density of an arbitrary collection of subsequent increments $\dB(t_i)$ is a product of the Gaussian distributions see, e.g.,~\cite{KaSh91,Grigoriu} for a precise mathematical definition.} and characterized by the Gaussian probability distribution
\be\label{e:OUP_langevin_math_density}
\Prob\{\dB(t)\in[y,y+\diff y]\}=\left(2\pi\,\diff t\right)^{-1/2}
\exp\bigl[-{y^2}/(2\,\diff t)\bigr]\;\diff y;
\ee 
\ese
i.e., the increments $\diff B(t)$ are independent random numbers drawn from a normal distribution with variance $\dt$. From Eq.~\eqref{e:OUP_langevin_math_density} and the independence of the increments at different times $s\ne t$, it follows that
\be\label{e:white_noise_math} 
\lan \dB(t)\ran =
0,\qquad\qquad \lan \dB(t)\,\dB(s) \ran=
\begin{cases}
0, & t\ne s\\
\diff t, &t=s,
\end{cases}
\ee
where now the expectation $\lan\,\cdot\, \ran$ is taken with respect to the probability measure $\Prob$ of the Wiener process $B(t)$. The two different representations \eqref{e:free_langevin} and \eqref{e:langevin_m} may be connected by formally identifying
$$
\dB(t)= \zeta(t)\,\diff t.
$$
In the remainder, we will write SDEs primarily in the differential notation of Eq.~\eqref{e:langevin_m}, which may also be viewed as a simple numerical integration scheme, cf., e.g., Ref.~\cite{1999KloedenPlaten,2004Glasserman} and Appendix~\ref{as:stochastic_calculus}. 
\par
It is worthwhile to summarize the physical assumptions, implicitly underlying Eqs.~\eqref{e:free_langevin} and~\eqref{e:langevin_m}:
\begin{itemize}
\item The heat bath is spatially homogeneous and stationary; i.e.,  relaxation processes within the heat bath occur on time scales much shorter than the relevant dynamical time scales associated with the motion of the heavy Brownian particle.\footnote{Interaction with a spatially inhomogeneous non-stationary heat bath can be modeled, e.g., by considering friction and noise amplitude functions of the form  $\ga(t,x,p)$ and  $D(t,x,p)$.}
\item Stochastic impacts between the Brownian particle and the constituents of the heat bath occur virtually uncorrelated.
\item On a macroscopic level, the interaction between Brownian particle and heat bath is sufficiently well described by the friction coefficient $\ga$ and the stochastic Langevin force $\mcal{L}$.
\item Equations~\eqref{e:free_langevin} and~\eqref{e:langevin_m} hold in the lab frame $\Gs$, corresponding to the specific inertial system, where the average velocity of the heat bath particles vanishes for all times $t$.
\end{itemize}
In Section~\ref{s:nonrelativistic_microscopic_models} we shall review how stochastic dynamical equations similar to Eqs.~\eqref{e:free_langevin} and~\eqref{e:langevin_m} can be derived or, at least, motivated by means of specific microscopic models.

\paragraph*{Discretization rules} 
A stochastic force with a momentum dependent noise amplitude function $D(p)$ as in Eq.~\eqref{e:nonlinear_langevin} is usually referred to as \lq multiplicative\rq\space noise. When considering SDEs that contain multiplicative noise terms, the specification of the discretization rule $\odot$ is necessary since, for fixed functions $\ga(p)$ and $D(p)$, different discretization schemes in general lead to non-equivalent stochastic processes~\cite{KaSh91,Grigoriu,1982HaTh,Gardiner}; put differently, the values of the stochastic integral $P(t)$ defined by Eq.~\eqref{e:nonlinear_langevin} depend on the choice of discretization rule. This is the most essential difference compared with ordinary differential equations, whose integral curves (i.e.,  solutions) are independent of the underlying discretization scheme when taking the continuum limit $\dt\to 0$.\footnote{For example, the discretization rule is irrelevant when the driving process $B(t)$ is a regular function, e.g., if $B(t)=\sin(\go t)$ such that $\dB(t)=\cos(\go t)\;\dt$. If, however,  the driving process is a strongly fluctuating function (of unbounded variation) as, e.g., a  Wiener (white noise) process, then different integration rules may yield non-equivalent trajectories; cf. Problem 2.29 in~\cite{KaSh91}.}  The three most commonly considered discretization rules are the following ones~\cite{1978Ha}:\footnote{A brief summary of the different discretization rules and their consequences with regard to stochastic differential calculus is given in Appendix~\ref{as:stochastic_calculus}.}
\begin{itemize}
 \item 
The pre-point discretization of Ito~\cite{1944Ito,1951Ito}, denoted by \lq\lq$\odot=*$\rq\rq, is defined by computing the function $C(P)=[D(p)]^{1/2}$ at $P(t)$, i.e.,
\bse
\be
C(P)*\dB(t):=C(P(t))\dB(t).
\ee
\item 
The post-point rule \lq\lq$\odot=\bullet$\rq\rq, sometimes also referred to as kinetic~\cite{1994Kl} or  backward Ito rule \cite{1978Ha}, is defined by  evaluating the function $C(P)=[D(p)]^{1/2}$ at $P(t+\dt)$, i.e.,
\be
C(P)\bullet\dB(t):=C(P(t+\dt))\dB(t).
\ee
\item 
The mid-point rule \lq\lq$\odot=\circ$\rq\rq\space of Stratonovich~\cite{1964St,1966St,1968St} and Fisk~\cite{1963Fisk,1965Fisk} is defined by taking the mean value of the Ito and the backward Ito  stochastic integral, i.e., 
\be
C(P)\circ \dB(t):=\f{1}{2}[C(P)*\dB(t)+C(P)\bullet \dB(t)].
\ee
\ese
\end{itemize}
From the mathematical point of view, the choice of the discretization rule reduces more or less to a matter of convenience~\cite{KaSh91,Grigoriu}. To briefly illustrate this,  consider Eqs.~\eqref{e:langevin_m} with post-point rule \lq\lq$\odot=\bullet$\rq\rq, reading 
\bse\label{e:langevin_m_post} 
\be
\diff X(t)&=&(P/M)\,\diff t,\\
\label{e:nonlinear_langevin_post}
\diff P(t)&=&[\Force(t,X)-\ga(P)\, P]\, \dt+[2D(P)]^{1/2}\bullet\diff B(t). 
\ee 
\ese
Then,  for each pair of sufficiently smooth functions $(\ga(p),D(p))$, one can determine new friction coefficients  $\ga_{\circ|*}(p)$ such that the pairs $(\ga_{\circ|*}(p),D(p))$ describe exactly the same stochastic dynamics as Eq.~\eqref{e:langevin_m_post} when combined with the corresponding discretization rule $\circ$ and $*$, respectively. Specifically, one can replace Eq.~\eqref{e:nonlinear_langevin_post} by the equivalent Stratonovich-Fisk SDE
\bse\label{e:nonlinear_langevin_sf} 
\be
\diff P(t)&=&[\Force(t,X)-\ga_\circ(P)\, P]\, \dt+[2D(P)]^{1/2} \circ \diff B(t),\\ 
\ga_\circ(p)&:=&\ga(p)-D'(p)/(2p),
\label{e:nonlinear_langevin_sf-b} 
\ee
\ese 
or, alternatively, by the equivalent Ito SDE 
\bse\label{e:nonlinear_langevin_ito} 
\be
\diff P(t)&=&[\Force(t,X)-\ga_*(P)\, P]\, \dt+[2D(P)]^{1/2} *\diff B(t),\\ 
\ga_*(p)&:=&\ga(p)-D'(p)/p,
\label{e:nonlinear_langevin_ito-b} 
\ee
\ese
where $D'(p):=\diff D(p)/\diff p$. The modified friction coefficients in Eqs.~\eqref{e:nonlinear_langevin_sf-b} and \eqref{e:nonlinear_langevin_ito-b}  account for the fact that the three discretization rules are characterized by different conditional expectations, respectively: 
\bse\label{e:conditional_mean_values}
\be
\lan [2D(P)]^{1/2}\bullet\dB(t)\; |\; P(t)=p\ran
&\;=\;&\label{e:conditional_mean_values-hk}
D'(p)\;\diff t,\\
\lan [2D(P)]^{1/2}\circ\dB(t)\; |\; P(t)=p\ran
&\;=\;&\label{e:conditional_mean_values-sf}
D'(p)\;\diff t/2,\\
\lan [2D(P)]^{1/2}*\dB(t)\; |\; P(t)=p\ran
&\;=\;&\label{e:conditional_mean_values-ito}
0.
\ee
\ese
From a practical point of view, each of the three above mentioned discretization methods has its own merits and drawbacks: Ito's pre-point rule~($*$) is particularly convenient for numerical simulations, but care is required when considering nonlinear transformations $G(P)$ of the momentum coordinate due to modifications of the differential calculus, cf. Appendix \ref{as:stochastic_calculus}. By contrast, if one adopts the Stratonovich-Fisk mid-point rule ($\circ$), then the transformation rules from ordinary differential calculus carry over, but it becomes more difficult to implement this mid-point rule in numerical simulations. The latter disadvantage also applies to the post-point rule employed in Eq.~\eqref{e:nonlinear_langevin}. However, as we shall see next, the post-point rule ($\bullet$) leads to a particularly simple form of the fluctuation-dissipation relation.\footnote{The general conversion formulae for the different discretization rules are summarized in Appendix~\ref{as:stochastic_calculus}.}

\paragraph*{Fokker-Planck equation} 
When studying SDEs of the type~\eqref{e:langevin_m}, one is typically interested in the probability $$f(t,x,p)\,\diff x\, \diff p\,$$ of finding the Brownian particle at time $t$ in the infinitesimal phase space interval $[x,x+\diff x]\times [p,p+\diff p]$. The non-negative phase space probability density function (PDF) $f(t,x,p)\ge 0$  of the Brownian particle is normalized at all times, i.e.
\be\label{e:OUP_normalization}
1=\int\diff x \diff p\, f(t,x,p)
\csp \fa t>0;
\ee
where, here and below, unspecified integrals range over the full phase space, position space, or momentum space, respectively. Given the phase space PDF $f(t,x,p)$, the marginal momentum PDF $\phi(t,p)$ and the marginal position PDF $\gr(t,x)$ are defined by
\bse
\be
\phi(t,p)&=&\int\diff x\, f(t,x,p),\\
\gr(t,x)&=&\int \diff p\, f(t,x,p).
\ee
\ese
Deterministic initial data $X(0)=x_0$ and $P(0)=p_0$ translate into the initial conditions 
\bse\label{e:OUP_initial_conditions}
\be
\label{e:OUP_initial_conditions_a}
f(0,x,p)&=&\gd(x-x_0)\,\gd(p-p_0),\\
\label{e:OUP_initial_conditions_b}
\phi(0,p)&=&\gd(p-p_0),\\
\label{e:OUP_initial_conditions_c}
\gr(0,x)&=&\gd(x-x_0).
\ee
\ese
Adopting the post-point rule ($\bullet$), the FPE describing the phase space density $f(t,x,p)$ of the stochastic process~\eqref{e:langevin_m} is given by\footnote{If we had considered Eq.~\eqref{e:nonlinear_langevin} with another stochastic integral interpretation (e.g., pre-point or mid-point discretization), then the corresponding FPE would be different from Eq.~\eqref{e:nonrelativistic_FPE_nonlinear}.} 
\be\label{e:external_FPE}
\f{\p f}{\p t} +\f{p}{M} \f{\p f}{\p x} + \Force(t,x)\f{\p f}{\p p} 
=
\f{\p}{\p p} \left[\ga(p)\, p\, f + D(p) \f{\p f}{\p p} \right].
\ee
Equation~\eqref{e:external_FPE} is a linear partial differential equation in $f$, i.e., more general general solutions can be obtained by integrating the special solution with deterministic initial condition~\eqref{e:OUP_initial_conditions} over some arbitrary initial distribution~$f_0(x_0,p_0)$. Moreover, the FPE~\eqref{e:external_FPE} is of first order in time, reflecting the fact that the Langevin equations~\eqref{e:langevin_m} describe a Markovian process.
\par
If there are no external forces present, i.e., $\Force(t,x)\equiv 0$, then Eq.~\eqref{e:external_FPE} yields 
the following FPE for the momentum PDF $\phi(t,p)$ 
\be\label{e:nonrelativistic_FPE_nonlinear}
\f{\p \phi}{\p t} =
\f{\p}{\p p} \left[\ga(p)\, p\, \phi  + D(p) \f{\p \phi}{\p p} \right].
\ee
The stationary solution of Eq.~\eqref{e:nonrelativistic_FPE_nonlinear} is given by
\be \label{e:nonrelativistic_FPE_nonlinear_solution}
\phi_\infty(p)
=
\mcal{N}
\exp\biggl[-\int_{-p_*}^p \diff p'\;\f{\ga(p')}{D(p')}\, p'\biggr],
\ee
where $\mcal{N}$ is a normalization constant, and $p_*$ some arbitrary constant such that the integral in the exponential exists.
\par
The general form~\eqref{e:nonrelativistic_FPE_nonlinear_solution}  of the stationary solution implies that one may generate arbitrary momentum distributions (e.g., Maxwell, Bose, Fermi or power law distributions) by choosing the friction and noise amplitude functions $\ga(p)$ and $D(p)$ in a suitable manner~\cite{2000Eb,2006DuHiHa}. To briefly illustrate this, consider some normalized target PDF $\hat{\phi}(p)\ge 0$. We would like to fix the relation between $\ga$ and $D$ such that the stationary solution $\phi_\infty(p)$ coincides with $\hat{\phi}(p)$. Equating $\hat{\phi}(p)$ with $\phi_\infty(p)$ from Eq.~\eqref{e:nonrelativistic_FPE_nonlinear_solution}, taking the logarithm and differentiating with respect to $p$ we find the condition\footnote{If reexpressed in terms of the corresponding Ito or Stratonovich-Fisk friction coefficients $\ga_{*|\circ}(p)$ from Eqs.~\eqref{e:nonlinear_langevin_ito-b} and \eqref{e:nonlinear_langevin_sf-b}, then the derivative $D'(p)$ enters the lhs. of Eq.~\eqref{e:generalized_FDT}, which thus would take the form of a differential equation.}
\be\label{e:generalized_FDT}
\f{\ga(p)}{D(p)}\, p\equiv -\f{\diff}{\diff p}\log\hat{\phi}(p).
\ee
For instance, if a classical nonrelativistic Brownian particle is in thermal equilibrium with the surrounding heat bath, then  $\hat{\phi}$ is a Maxwell distribution
\be\label{e:_maxwell}
\hat{\phi}_\mrm{M}(p)=
\left(2\pi M/\gb\right)^{-1/2} 
\exp[-\gb\,p^2/(2M)]
\csp 
\gb:=(\kB \Temp)^{-1},
\ee
where $\Temp$ is the temperature of the heat bath and $\kB$ the Boltzmann constant. In this case, Eq.~\eqref{e:generalized_FDT} reduces to the 
the \emph{generalized Einstein fluctuation-dissipation  relation}~\cite{1994Kl,1995Klimontovich,2007Li}
\be\label{e:generalized_Einstein_relation}
D(p)=\ga(p)\;M\kB \Temp=\ga(p)\;M/\gb.
\ee
The fluctuation-dissipation relations~\eqref{e:generalized_FDT} and~\eqref{e:generalized_Einstein_relation} do fix only one of the two coefficients $\ga(p)$ and $D(p)$. Put differently, one is still free to adapt, e.g., the function $\ga(p)$ such that the stochastic process~\eqref{e:nonlinear_langevin} exhibits the correct relaxation behavior. This freedom is a main reason why the Langevin approach is successfully applicable to a wide range of thermalization processes~\cite{1994Kl}. Physically reasonable expressions for $\ga(p)$ may be deduced from kinetic theory~\cite{1924Ep,1971Ho,1982HoRaRa,1988Sv,2003Tr,2005HeRa} or microscopic Hamiltonian models that take into account the interactions as well as the statistical properties of the heat bath~\cite{1945Bo,1959Ma,1965FoKaMa,1973Zw,1985CoWeLi,1997Ha,2006DuHa}. Examples will be discussed in Section~\ref{s:nonrelativistic_microscopic_models}.
\par 
With regard to our subsequent discussion of relativistic Brownian motions,  it will be important to keep in mind that the nonlinear Langevin equations like Eq.~\eqref{e:nonlinear_langevin} provide a tool for constructing Brownian motion processes with arbitrary stationary velocity and momentum distributions~\cite{2000Eb,2006DuHiHa}.

\subsubsection{Nonrelativistic Ornstein-Uhlenbeck process}
\label{s:OUP}

The standard paradigm for a nonrelativistic Brownian motion process in the absence of external forces is the classical Ornstein-Uhlenbeck process~\cite{1930UhOr}, corresponding to constant coefficients
\be
D(P)\equiv D_0
\csp
\ga(p)\equiv \ga_0,
\ee
yielding, e.g., the Ito SDE
\bse\label{e:OUP_external}
\be
\dX&=&(P/M)\,\dt,\\
\dP&=&\label{e:OUP_external-b}
\Force(t,X)\,\dt-\ga_0\, P\,\dt + (2D_0)^{1/2}*\dB(t).
\ee
\ese
Note that in this particular case the choice of the discretization rule is not relevant when integrating the momentum equation~\eqref{e:OUP_external-b}, but a rule must be specified when considering nonlinear transformations $G(P)$; 
cf. remarks in Appendix~\ref{as:stochastic_calculus}.

\paragraph*{Free motion}
Considering free Brownian motions with $\Force(t,x)\equiv 0$ first, the solutions of Eqs.~\eqref{e:langevin_m} read explicitly
\bse\label{e:OUP_solution}
\be
X(t)&=&X(0)+\int_0^t \diff s\; P(s)/M,\\
P(t)&=&P(0)\,e^{-\ga_0 t}+(2D_0)^{1/2}e^{-\ga_0 t}\int_0^t e^{\ga_0 s}*\dB(s).
\ee
\ese
Combining the solution~\eqref{e:OUP_solution} with Eq.~\eqref{e:OUP_langevin_math_density}, one finds for the first two moments of the momentum coordinate~\cite{1930UhOr,1943Ch}
\bse
\be
\lan P(t)\ran&=&
P(0)\, e^{-\ga_0 t},\\
\lan P(t)^2\ran&=&
P(0)^2\,e^{-2\ga_0 t}+\f{D_0}{\ga_0}\; (1-e^{-2\ga_0 t}),
\ee
\ese
while the first centered moments of the position coordinate are obtained as
\bse\label{e:OUP_moments}
\be
\lan X(t)-X(0)\ran&\;=\;&
\f{P(0)}{\ga_0 M}\,(1-e^{-\ga_0 t}),\\
\lan [X(t)-X(0)]^2\ran&\;=\;&\notag
\f{2D_0 }{(\ga_0 M)^2}\,t  +
\left[\f{P(0)}{\ga_0 M}\right]^2 \left(1-e^{-\ga_0 t}\right)^2+
\notag\\
&&\qquad\qquad\quad\;\;
\label{e:OUP_moments_b}
\f{D_0}{\ga_0^3M^2}\,\left(-3+4e^{-\ga_0 t}-e^{-2\ga_0 t} \right).
\ee
\ese
The asymptotic spatial diffusion constant $\D_\infty$, not to be confused with the noise amplitude $D_0$, is usually defined by 
\be
2\D_\infty:=\lim_{t\to\infty} \f{1}{t}\;\lan [X(t)-X(0)]^2\ran.
\ee
From Eq.~\eqref{e:OUP_moments_b} we find for the Ornstein-Uhlenbeck process the classical result\footnote{A useful integral formula for the diffusion constant  for  nonlinear one-dimensional  Brownian motion processes  was  derived by Lindner~\cite{2007Li} recently, see Eq.~\eqref{e:lindner_diffusion} below.} 
\be\label{e:classical_diffusion_OUP-1} 
\D_\infty:=D_0/(\ga_0 M)^2.
\ee 
The FPE governing the momentum PDF $\phi(t,p)$ of the free Ornstein-Uhlenbeck process reads~\cite{Becker}
\be\label{e:FPE-nr}
\f{\p \phi}{\p t}
=
\f{\p}{\p p} \left(\ga_0 p\, \phi + D_0 \f{\p \phi}{\p p}\right).
\ee
Adopting the deterministic initial condition~\eqref{e:OUP_initial_conditions_b}, $\phi(0,p)=\gd(p-p_0)$, 
the time-dependent solution of Eq.~\eqref{e:FPE-nr} is given by~\cite{1930UhOr,Becker}
\be
\phi(t,p)
&=&\left\{\f{\ga_0}{2\pi D_0 [1-\exp(-2\ga_0 t)]}\right\}^{1/2}
\label{e:OUP_FPE_solution}
\exp\biggl\{
-\f{\ga_0[p-p_0 \exp(-\ga_0 t)]^2}{2D_0[1- \exp(-2\ga_0 t)]}
\biggr\}.
\ee
In the limit $t\to\infty$ this solution reduces to the stationary Gaussian distribution
\be\label{e:OUP_FPE_solution_stationary}
\phi_\infty(p)
=\left(\f{\ga_0}{2\pi D_0}\right)^{1/2} \exp\left(-\f{\ga p^2}{2D_0}\right).
\ee
For a given momentum distribution $\phi(t,p)$ of the Brownian particle, the corresponding velocity PDF $\psi(t,v)$ is defined by
\be\label{e:PDF_transformation}
\psi(t,v):=\f{\diff p}{\diff v}\;\phi(t,p(v)),
\ee
where $p=Mv$ in the nonrelativistic case. Hence, by imposing again the \emph{Einstein relation} from Eq.~\eqref{e:generalized_Einstein_relation}, which now reduces to $D_0= \ga_0 M\kB \Temp $, the stationary momentum PDF~\eqref{e:OUP_FPE_solution_stationary} is seen to be equivalent to Maxwell's  velocity distribution
\be\label{e:OUP_maxwell}
\psi_\mrm{M}(v)=
\left(\f{M}{2\pi \kB \Temp}\right)^{1/2} 
\exp\left(-\f{Mv^2}{2\kB \Temp}\right).
\ee
Moreover, by virtue of the Einstein relation $D_0= \ga_0 M\kB \Temp $, the asymptotic  diffusion constant from Eq.~\eqref{e:classical_diffusion_OUP-1} can be written in the form
\be\label{e:classical_diffusion_OUP-2}
\D_\infty:= \kB \Temp /(\ga_0 M).
\ee

\paragraph*{External force fields} 
The FPE describing the phase space density $f(t,x,p)$ of the stochastic process~\eqref{e:OUP_external} in an external force field $\Force(t,x)$ reads 
\be\label{e:OUP_external_FPE}
\f{\p f}{\p t} +\f{p}{M} \f{\p f}{\p x} + \Force(t,x)\;\f{\p f}{\p p} 
=
\f{\p}{\p p} \left(\ga_0 p f + D_0 \f{\p f}{\p p} \right).
\ee
Models of this type have been intensely studied during the past century, covering a wide range of application (see, e.g., Ref.~\cite{1990HaTaBo} for a review). However, for arbitrary time and position dependent force fields $\Force(t,x)$ it is generally very difficult, and in many cases even impossible, to find exact time-dependent solutions of the Fokker-Planck equation~\eqref{e:OUP_external_FPE}. In the simpler case of a time-independent, conservative force field $\Force(t,x)\equiv F(x)$ with confining\footnote{Conventionally, a potential $\Phi(x)$ is called \lq confining\rq\space if it increases sufficiently fast for $|x|\to\infty$ so that the phase space PDF $f$ is normalizable.} potential $\Phi(x)$, i.e.,
\be
F(x)=-\f{\p}{\p x} \Phi(x), 
\ee
one can determine the stationary solution attained in the limit $t\to \infty$. Imposing as above the Einstein relation $D_0= \ga_0 M\kB \Temp$, the stationary solution of Eq.~\eqref{e:OUP_external_FPE} is given by the Maxwell-Boltzmann distribution~\cite{Becker,Huang}
\be\label{e:OUP_external_FPE_solution}
f(x,p)
=\Z^{-1}
\exp\biggl\{-\gb\left [\f{p^2}{2M} + \Phi(x)\right]\biggr\}
\csp \gb:=(\kB \Temp)^{-1},
\ee
where the normalization constant $\Z$ is determined by Eq.~\eqref{e:OUP_normalization}.
\par
Another important class of applications concerns time periodic force fields, satisfying $\Force(t,x)=\Force(t+\Gd t,x)$ for some fixed period $\Gd t$. In this case it is sometimes possible to derive approximate asymptotic solutions of the FPE~\eqref{e:OUP_external_FPE} by considering the limit $t\to \infty$. These asymptotic solutions are usually also time periodic and can exhibit phase shifts. They may give rise to a number of interesting phenomena such as, e.g., stochastic resonance~\cite{1981Benzi,1983Benzi,1981NiNi,1989JuHaMa,1991JuHa,1993Ni,1998GaHaJuMa,1994NeSG,1999AnEtAl,2000LiSG,2004LiEtAl}. 
\par
From the purely mathematical perspective, SDEs define well-defined models of stochastic processes~\cite{KaSh91}; from the physicist's point of view, their usefulness for the description of real systems is \textit{a priori} an open issue. Hence, before directing our attention to the relativistic case, it is worthwhile to recall how nonrelativistic Langevin equations can be justified by means of microscopic models~\cite{1945Bo,1959Ma,1965FoKaMa,1985CoWeLi,1986Po,1988LeShPo,1997Ha,2006DuHa}.

\subsection{Microscopic models}
\label{s:nonrelativistic_microscopic_models}

When considering Langevin equations of the type~\eqref{e:langevin_m}, one may in principle distinguish between the two following tasks:
\begin{itemize}
\item[(a)]
One can postulate Langevin equations as phenomenological model equations, study the mathematical consequences and compare the resulting predictions  with experiments in order to (in)validate the theory. Adopting this approach, the parameters and the explicit functional form of the friction and noise amplitude functions have to be determined from experimental data~\cite{2006BlEtAl}.
\item [(b)]
Alternatively, one can try to motivate and derive Langevin equations from 
microscopic models. If successful, this approach yields explicit expressions for the friction and noise functions in terms of the microscopic model parameters. 
\end{itemize}
The remainder of this section addresses the latter problem, 
which has  attracted considerable interested over the past decades~\cite{1957Ko,1945Bo,1959Ma,1965FoKaMa,1973Zw,1981BY,1982HoRaRa,1985CoWeLi,1986Po,1988LeShPo,1997Ha,2006DuHa,2007ChKr}. Langevin equations provide an approximate stochastic description of the \lq exact\rq\space microscopic dynamics. Hence, in order to derive SDEs from, e.g., Hamilton mechanics one has to impose certain approximations. These approximations determine the range of applicability of the Langevin approach. Generally, one can pursue at least two different routes for deriving SDEs of the type~\eqref{e:langevin_m} from more precise models:
\begin{itemize}
\item[(1)] 
Starting from a Boltzmann-type equation~\cite{1981Er,Liboff,CercignaniKremer} or master equation~\cite{VanKampen} for the one-particle probability density of the Brownian particle, one can try to reduce these integro-differential equations to a Fokker-Planck equation by performing suitable approximations~\cite{1957Ko,1971Ho,1980Ha_2,1981Ha,1981BY,1982HoRaRa,1984HaEtAl,1988Sv,2003Tr,2005HeRa,2007ChKr}. Once the Fokker-Planck equation has been found, it is straightforward to write down a corresponding Langevin dynamics~\cite{Grigoriu,VanKampen}. The microscopic collision dynamics is then encoded in the scattering cross-sections appearing in the collision integral of the Boltzmann equation~\cite{Liboff,CercignaniKremer}.
\item[(2)]
Alternatively, one may start from a microscopic (e.g., Hamiltonian) model describing the interaction between Brownian particle and heat bath. After eliminating the heat bath degrees of freedom from the equations of motion for the Brownian particle~\cite{1945Bo,1959Ma,1965FoKaMa,1973Zw,1985CoWeLi,1986Po,1988LeShPo,1997Ha,2006DuHa}, one obtains a generalized Langevin equation which may be reduced to the form~\eqref{e:langevin_m} in certain limit cases. As a byproduct, the fluctuation-dissipation relations arise quite naturally within this approach upon assuming a probability distribution for the (initial) bath configuration. 
\end{itemize}
To briefly illustrate the second procedure by example, we next consider the oscillator  model~\cite{1945Bo,1959Ma,1965FoKaMa,1985CoWeLi,1986Po,1988LeShPo,1997Ha} and the elastic binary collision model~\cite{2006DuHa}. In contrast to the more frequently  studied harmonic oscillator model, the collision model from  Section~\ref{s:nonrelativistic_collision_model} can be extended to the relativistic case; cf. discussion in Section~\ref{s:relativistic_binary_collision_model}.

\subsubsection{Harmonic oscillator model}
\label{s:harmonic_oscillators}

The harmonic oscillator model presents the classical paradigm for constructing a generalized Langevin equation from a Hamiltonian model~\cite{1945Bo,1959Ma,1965FoKaMa,1985CoWeLi,1986Po,1988LeShPo,1990HaTaBo,1997Ha}. The Hamiltonian function upon which the derivation is based reads~\cite{1997Ha}
\be\label{e:harmonic_oscillator_hamiltonian}
H=
\f{P^2}{2M}+\Phi(X)+
\sum_{r}\left[\f{p_r^2}{2m_r}+
\f{m_r \go_r^2}{2}
\left(x_r-\f{c_r}{m_r\go_r^2} X\right)^2
\right].
\ee
Here $M$, $X$ and $P$ are the mass, position and momentum of the Brownian particle and $\Phi(x)$ is an external potential field; $x_r$ and $p_r$ denote the position and momentum of a heat bath particle with mass $m_r$, oscillator frequency $\go_r$ and coupling constant $c_r$. Equation~\eqref{e:harmonic_oscillator_hamiltonian} yields the following Hamilton equations of motions: 
\bse\label{e:HO_eom}
\be
\label{e:HO_eom_a}
\dot X&=&(P/M),
\qquad\; 
\dot P=F(X)+\sum_r c_r\left(x_r-\f{c_r}{m_r \go_r^2}X\right);\\
\label{e:HO_eom_b}
\dot x_r&=&(p_r/m_r),
\qquad 
\dot p_r=-m_r \go_r^2 x_r+c_r X
\csp
r=1,\ldots,N,
\ee
\ese
where $F(x)=-\diff \Phi(x)/\diff x$ is the conservative external force acting on the Brownian particle. As evident from Eqs.~\eqref{e:HO_eom}, Brownian particle and heat bath are coupled via linear forces in this model. By formally integrating Eqs.~\eqref{e:HO_eom_b} and inserting the solutions into Eq.~~\eqref{e:HO_eom_a}, one may eliminate the heat bath coordinates from Eqs.~\eqref{e:HO_eom_a}, yielding the exact \emph{generalized} Langevin equations~\cite{1985CoWeLi,1997Ha}
\bse\label{e:HO_langevin}
\be
\dot X&=&(P/M),\\
\dot{P}&=&F(X)-\int_{0}^t\diff s\;\nu(t-s)P(s)+L(t),
\ee
where, for given initial values $X(0),P(0)$, $\{x_r(0),p_r(0)\}$, the memory friction kernel $\nu(t-s)$ and the Langevin noise force $L(t)$ are given by~\cite{1997Ha}
\be
\nu(t-s)&:=&\label{e:HO_friction_kernel}
\f{1}{M}\sum_r \f{c_r^2}{m_r \go^2_r} \cos[\go_r(t-s)],\\
L(t)&:=&
\sum_r c_r\biggl\{
\left[x_r(0)-\f{c_r}{m_r \go_r^2}X(0)\right]
\cos(\go_r t)+
\f{p_r(0)}{m_r\go_r}\sin(\go_r t)
\biggr\}.\quad
\ee
\ese
In order to be able to characterize the properties of the noise force $L(t)$ by means of an fluctuation-dissipation relation, one still needs to impose a distribution for the 
initial conditions $\{x_r(0),p_r(0)\}$ of the bath variables. In principle, 
this initial distribution can be chosen arbitrarily. Of particular interest in canonical thermostatistics are equilibrium distributions of the Maxwell-Boltzmann type. In the case of the generalized Langevin equation~\eqref{e:HO_langevin}, a plausible choice for the initial bath distribution corresponds to the PDF
\be\notag
&&f_\bath(\{x_r(0),p_r(0)\}\,|\,X(0)=x_0\,)=
\Z^{-1}\times\\
&&\qquad\qquad\label{e:HO_initial_distribution}
\exp\biggl\{-\gb
\sum_r\biggl[
\f{p_r(0)^2}{2m_r}+
\f{m_r\go_r^2}{2} \left(x_r(0)-\f{c_r}{m_r\go_r^2} x_0\right)^2  
\biggr]
\biggr\},
\ee
where $\gb=(\kB \Temp)^{-1}$ is the inverse thermal energy, $\Temp$ the temperature, and $\Z$ the normalization constant. The initial position $X(0)=x_0$ of the Brownian particle enters in Eq.~\eqref{e:HO_initial_distribution} as an independent parameter, i.e.,  averages with respect to $f_\bath(\{x_r(0),p_r(0)\}\,|\,X(0)=x_0)$ are \emph{conditional} on the initial Brownian particle position $X(0)=x_0$. Averaging the stochastic force $L(t)$ with respect to $f_\bath$ from Eq.~\eqref{e:HO_initial_distribution}, one finds
\bse\label{e:HO_FDT}
\be
\lan L(t)\ran_\bath &=&0,\\
\lan L(t)L(s)\ran_\bath &=&M\kB \Temp \;\nu(t-s).\label{e:HO_FDT_b}
\ee
\ese
Equation~\eqref{e:HO_FDT_b} represents the fluctuation-dissipation relation for the generalized Langevin equation~\eqref{e:HO_langevin} given the initial bath distribution~\eqref{e:HO_initial_distribution}. The generalized Langevin equation~\eqref{e:HO_langevin} differs from Eqs.~\eqref{e:langevin_m} and \eqref{e:OUP_external} through the memory friction $\nu(t-s)$. The SDE~\eqref{e:OUP_external}, describing the Ornstein-Uhlenbeck process in an external force field, is recovered from Eqs.~\eqref{e:HO_langevin} in the limit case\footnote{The prefactor \lq$2$\rq\space is required in Eq.~\eqref{e:HO_limit} because of the convention \mbox{$\int_0^t\diff s\;\gd(t-s) P(s)= P(t)/2.$}}
\be\label{e:HO_limit}
\nu(t-s) = 2\ga_0\, \gd(t-s),
\ee
where $\ga$ is a constant friction coefficient. The limit case~\eqref{e:HO_limit} can be illustrated by rewriting the friction kernel~\eqref{e:HO_friction_kernel} in the more general form
\be\label{e:HO_friction_kernel_continuum}
\nu(t-s)=\int_0^\infty \diff \go\; C(\go)\, \cos[\go(t-s)].
\ee
By fixing the amplitude function $C(\go)$ as 
\be
C(\go)= \f{1}{M} \sum_r\f{c_r^2}{m_r \go^2_r}\;\gd(\go-\go_r),
\ee
one recovers the memory friction~\eqref{e:HO_friction_kernel} as a special case of Eq.~\eqref{e:HO_friction_kernel_continuum}. In order to obtain the limit case~\eqref{e:HO_limit} from Eq.~\eqref{e:HO_friction_kernel_continuum}, one can use the cosine-decomposition of the Dirac $\gd$-function
\be
\gd(t-s)
=\f{1}{2\pi}\int_{-\infty}^\infty \diff \go\; e^{i\go(t-s)}
=\f{1}{\pi}\int_0^\infty \diff \go\;  \cos[\go(t-s)].
\label{e:delta-function}
\ee
Hence, upon comparing Eqs.~\eqref{e:delta-function} and~\eqref{e:HO_friction_kernel_continuum}, the white noise limit~\eqref{e:HO_limit} corresponds to the particular choice
\be
C(\go)=(2\ga_0)/{\pi}.
\ee
\par
The harmonic oscillator model\footnote{The quantum mechanical generalization of the harmonic oscillator model represents a paradigm for quantum Brownian motions and has been studied, e.g., in \cite{1960Se,1961Se,1965FoKaMa,1983CaLe,1983GrTa,1984GrWeTa,1985CaLe,1987FoKa,2006HaIn}; see also the reviews by Grabert et al.~\cite{1988GrScIn} and H\"anggi and Ingold~\cite{2005HaIn}.} provides a useful microscopic justification for the Langevin equations~\eqref{e:langevin_m} and \eqref{e:OUP_external} of the Ornstein-Uhlenbeck process. Unfortunately, this model cannot be transferred to special relativity, as it is based on instantaneous harmonic interactions-at-distance which violate fundamental relativistic principles. Therefore, in the last part of this section we shall consider another microscopic model which is based on strictly localized elastic binary collisions and, thus, can be extended to special relativity.

\subsubsection{Elastic binary collision model}
\label{s:nonrelativistic_collision_model}

The one-dimensional elastic binary collision model \cite{2006DuHa,2008Du_PHD} is based on the idea that the stochastic motion of a Brownian particle (mass $M$) is caused by frequent elastic collisions with smaller heat bath particles (mass $m\ll M$).\footnote{Similar approaches are known from unimolecular rate theory, see, e.g., Section~V in~\cite{1990HaTaBo}. In the context of quantum Brownian motions, a quantum-mechanical version of the collision model was proposed and studied by Pechukas~\cite{1991Pe}, and Tsonchev and Pechukas~\cite{2000TsPe}.} As before, we denote the coordinates and momenta of the heat bath particles by $\{x_r,p_r\}$, where $r=1,\ldots,N$ and $N\gg 1$.

\paragraph*{Collision kinematics}
An elastic collision of the Brownian particle [velocity $V$, momentum $P=MV$, kinetic energy $E={P^2}/({2M})$] with a heat bath particle [velocity $v_r$, momentum $p_r=mv_r$, 
kinetic energy $\eps_r={p_r^2}/({2m}$] is governed by the energy-momentum conservation laws
\be\label{e:kinematics}
E+\eps_r=\hat{E}+\hat{\eps}_r\csp
P+p_r =\hat{P}+\hat{p}_r.
\ee
Here, hat-symbols refer to the state after the collision. Taking into account the kinematic conservation laws \eqref{e:kinematics}, we find that the
momentum gain $\Gd P_r$ of the Brownian particle per single collision is given by
\be\label{e:p-tilde-nonrel}
\Gd P_r:=\hat{P}-P=
-\f{2m}{M+m}\,P+\f{2M}{M+m} \,p_r.
\ee
To construct a Langevin-like equation from Eqs.~\eqref{e:kinematics} and \eqref{e:p-tilde-nonrel}, one considers the total momentum change $\gd P(t)= P(t+\gd t)-P(t)$ of the Brownian particle within the \lq mesoscopic\rq\space time interval $[t,t+\gd t]$, assuming that:
\begin{itemize}
\item collisions occurring within $[t,t+\gd t]$ can be viewed as independent events;
\item 
the time step $\gd t$ is sufficiently small, so that there occurs at most
only one collision between the Brownian particle and a specific heat bath
particle $r$ and that $|\gd P(t)/P(t)|\ll 1$ holds true; 
\item $\gd t$ is still sufficiently large, so that the total number of collisions within $\gd t$ is larger than $1$. 
\end{itemize}
These requirements can be fulfilled simultaneously only if $m\ll M$ holds.
With the above assumptions, one can approximate
\be\label{e:initial}
\gd P(t)\approx \sum_{r=1}^N \Gd P_r\; I_r(t,\gd t),
\ee
where $I_r(t,\gd t)\in\{0,1\}$  is the indicator function for a collision
with the heat bath particle~$r$ during the interval $[t,t+\gd t]$; i.e.
\be
I_r(t,\gd t)=\begin{cases}
1& \mrm{if\; a\; collision\; has\; occurred\; in}\;[t,t+\gd t],\\
0& \mrm{otherwise}.
\end{cases}
\ee
In the one-dimensional case, $I_r(t,\gd t)$ can be expressed in the form\footnote{The Heaviside-function $\Gt(x)$ is defined as the integral over the Dirac {$\gd$-function}, i.e.,  $\Gt(x):=0,x< 0$; $\Gt(0):=1/2$; \mbox{$\Gt(x):=1,x> 0$}. When considering higher space dimensions, the expression~\eqref{e:indicator} for the indicator function has to be modified accordingly, e.g., by taking into account the geometric shape of the Brownian particle.} 
\bse\label{e:indicator} 
\be
I_r(t,\gd t)
&=&
\Gt(X-x_r)\;\Gt(x'_r-X')
+
\Gt(x_r-X)\;\Gt(X'-x'_r),
\label{e:indicator_1}
\ee
where $X=X(t)$, $x_r=x_r(t)$ are the \lq initial\rq\space positions of the colliding particles at time $t$, and
\be\label{e:indicator_0}
X'=X+V\,\gd t \csp
x'_r=x_r+v_r\,\gd t
\ee
\ese
their projected positions\footnote{Of course, in the case of a collision the  position of the particles at time $t+\gd t$ will be different from the \lq projected\rq~ positions $X'$ and $x'$.} at time $t+\gd t$. The collision indicator from Eq.~\eqref{e:indicator}  is characterized by 
\be\label{e:indicator_properties}
\label{e:idempotent}
I_r(t,0)=0 
\csp
\left[I_r(t,\gd t)\right]^j =I_r(t,\gd t),
\qquad
j=1,2,\ldots.
\ee
The Taylor-expansion of $I_r$ at $\gd t=0$ reads~\cite{2006DuHa}
\bse\label{e:quasi_almost_langevin}
\be\label{e:indicator_taylor}
I_r(t,\gd t)&\approx&
\f{\gd t}{2}\;|v_r-V|\;\gd(x_r-X).
\ee
Combining Eqs.~\eqref{e:p-tilde-nonrel}, \eqref{e:initial} and \eqref{e:indicator_taylor} yields
\be
\label{e:binary_almost_langevin_full}
\gd P(t) &\approx& 
-2\left[\sum_{r=1}^N\f{m}{M+m} \;I_r(t,\gd t)\right] \,P(t) + 
2\sum_{r=1}^N \f{M}{M+m} \,p_r\;I_r(t,\gd t),
\ee
where, additionally, it was assumed that for each collision occurring within $[t,t+\gd t]$, the momentum of the Brownian particle before the collision is approximately equal to the \lq initial\rq\space value $P(t)$. In view of $m\ll M$, Eq.~\eqref{e:binary_almost_langevin_full} can be simplified further to give
\be\label{e:binary_almost_langevin}
\gd P(t)&\approx &
-2\left[\sum_{r=1}^N\f{m}{M} \;I_r(t,\gd t)\right] \,P(t) + 
2\sum_{r=1}^N \,p_r\;I_r(t,\gd t).
\ee
\ese
A comparison with the Langevin equation~\eqref{e:nonlinear_langevin} suggests that, heuristically, the first term on the rhs. of Eq.~\eqref{e:binary_almost_langevin} can be interpreted as a \lq friction\rq\space term, while the second term represents \lq noise\rq.
However, although looking quite similar to a  Langevin equation,  Eq.~\eqref{e:binary_almost_langevin} is still considerably more complicated than, e.g., the Langevin  equation~\eqref{e:nonlinear_langevin}. This is due to the fact that the collision indicators $I_r(t,\gd t)$ from Eq.~\eqref{e:indicator_taylor} depend not only on the Brownian particle's position  and velocity but also on the stochastic bath variables $\{x_r,v_r\}$. Nevertheless, it is possible to calculate the statistical properties of the momentum increments $\gd P(t)$ from Eqs.~\eqref{e:quasi_almost_langevin}, 
provided one specifies a distribution for the heat bath particles. 

\paragraph*{Bath distribution}
In principle, one can use Eqs.~\eqref{e:quasi_almost_langevin} to calculate the statistical moments $\lan (\gd P)^j \ran_\bath$ for an arbitrarily given heat bath PDF $f_\bath^N(\{x_r,p_r\})$. Here, we shall focus on the situation where the (infinitely large) heat bath is given by a quasi-ideal gas which is in thermal equilibrium with its environment. In this case, the one-particle PDF $f^1_\bath(x_r,p_r)$ is given by the spatially homogeneous Maxwell distribution  
\be\label{e:nonrelativistic_bath}
f_\bath^1(x_r,p_r)
&=&
{\left(2\pi m\kB \Temp\right)^{-1/2}}\; L^{-1}\;
\label{e:nonrelativistic_bath_a}
\exp[-{p_r^2}/(2m \kB \Temp)],
\ee
where $x_r\in [0,L]$ with $L$ being the one-dimensional container volume. Moreover, we will assume that:
\begin{itemize}
\item the heat bath particles are \emph{independently} and \emph{identically} distributed;
\item the distribution of the bath particles is not affected by the  collisions with the Brownian particle.
\end{itemize}
The above assumptions can be justified for a sufficiently large bath, if collisions between the bath particles rapidly reestablish a spatially homogeneous bath distribution.

\begin{figure}[t]
\center 
\vspace{0.5cm}
\includegraphics[width=9cm,angle=0]{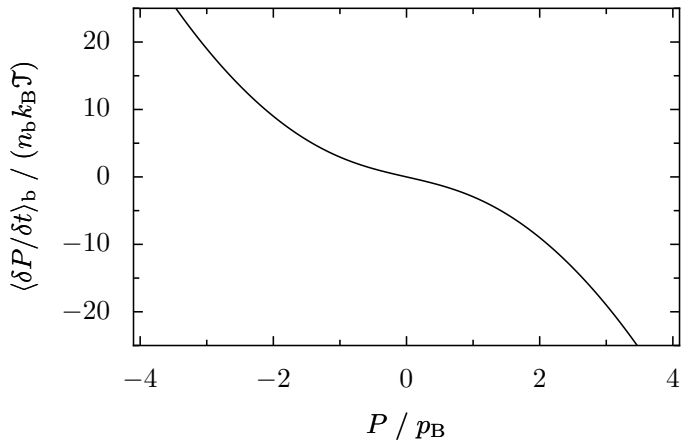}
\caption{\label{fig_nonrelativistic_drift} 
Nonrelativistic binary collision model. Mean drift force $\drift(P)=\lan \gd P(t)/\gd t \ran_\bath$ from Eq.~\eqref{e:binary_averaged_friction} with  $n_\bath=N/L$ denoting the number density of the heat bath particles and $\pB:=M(2\kB \Temp/m)^{1/2}$ the characteristic momentum of a Brownian particle (mass $M$), surrounded by heat bath particles of mass $m$ and temperature $\Temp$.}
\end{figure}

\paragraph*{Mean drift force}
We define the mean (momentum) drift as the average momentum change $\lan \gd P \ran_\bath$ over the interval $[t,t+\gd t]$, given the momentum value $P$ at time $t$.\footnote{More precisely, one should write the mean momentum drift in the form of a conditional expectation $\lan \gd P(t)\,|\,P(t)=p \ran_\bath$; however, for ease of notation we write $\lan \gd P \ran_\bath$ at this stage.} In the case of  Eq.~\eqref{e:binary_almost_langevin}, one finds~\cite{2006DuHa,2008Du_PHD}
\be
\lan \gd P(t) \ran_\bath
&=& 
-2N\,\left(\f{m}{M}\right) \;\lan I_r(t,\gd t)\ran_\bath \,P + 
2N\, \lan p_r\; I_r(t,\gd t)\ran_\bath.
\ee
To calculate the averages on the rhs. we note that, for a spatially uniform bath distribution as in Eq.~\eqref{e:nonrelativistic_bath}, one-particle expectation values of the form  $\lan G(x_r,v_r)\,I_r(t,\gd t)\ran_\bath$ can be calculated to first order in $\gd t$ as~\cite{2006DuHa}
\be
\lan G(x_r,v_r)\,I_r(t,\gd t)\ran_\bath
=\label{e:indicator_mean}
\f{\gd t}{2L}
\int_{-\infty}^{\infty}\!\!\!\diff v_r\;G(X,v_r)\;|v_r-V|\;
\psi_\bath(v_r).
\ee
Here, $\psi_\bath(v_r)$ denotes the one-particle velocity PDF of the heat bath particles, which in the case of Eq.~\eqref{e:nonrelativistic_bath} is given by the Maxwellian
\be
\label{e:binary_nonrelativistic_bath}
\label{e:maxwell_gas}
\psi_\bath(v_r)
&=&
\left(\vB^2 \pi\right)^{-1/2} 
\exp(-v_r^2/\vB^2)
\csp
\vB:=(2\kB \Temp/{m})^{1/2}.
\ee
By making use of Eq.~\eqref{e:indicator_mean}, one obtains for the mean drift of the collision model:\footnote{Higher moments and correlation functions may be calculated in a similar manner.}
\be
\lan \gd P(t) \ran_\bath 
&\approx&\notag
-2n_\bath\,\kB\Temp\; 
\biggl\{
\pi^{-1/2}\biggl(\f{P}{\pB}\biggr)
\exp\biggl[-\left(\f{P}{\pB}\right)^{2}\biggr]+ \\
&&\qquad\qquad\qquad\quad
\biggl[\biggl(\f{P}{\pB}\biggr)^2 +\f{1}{2}\biggr]
\;\mrm{erf}\biggl(\f{P}{\pB}\biggr)
\biggr\}
\;\gd t,
\label{e:binary_averaged_friction}
\ee
where $n_\bath=N/L$ is the number density of the heat bath particles,
\linebreak \mbox{$p_\mrm{B}:=M\vB=M(2\kB \Temp/{m})^{1/2}$} a characteristic thermal momentum scale, and the error function $\mrm{erf}(z)$ is defined by
\be\notag
\mrm{erf}(z):=\f{2}{\sqrt{\pi}}\int_0^z\diff x\;e^{-x^2}.
\ee
Figure \ref{fig_nonrelativistic_drift} depicts the mean drift force 
\be\label{e:mean_drift_force_nonrelativistic}
\drift(P):=\lan \gd P(t) /\gd t \ran_\bath,
\ee
obtained from Eq.~\eqref{e:binary_averaged_friction}.  The absolute value of this drift force grows linearly for small momentum values (Ornstein-Uhlenbeck regime) and quadratically for large momentum values. 
\par
In the remainder of this section, we are going to illustrate how one can use the result~\eqref{e:binary_averaged_friction} to obtain a  systematic procedure for approximating the model equations~\eqref{e:quasi_almost_langevin} by a nonlinear SDE of the  type~\eqref{e:nonlinear_langevin}.

\paragraph*{Langevin approximation}
Similar to the oscillator equation~\eqref{e:HO_langevin}, the
\linebreak Eq.~\eqref{e:binary_almost_langevin} for the momentum increments $\gd P(t)$ in the binary collision model is not yet a Langevin equation. SDEs of the type
\be\label{e:nonlinear_langevin_repeated}
\dP(t)=-\ga(P)\, P\,\dt + [2D(P)]^{1/2}\bullet\dB(t)
\ee 
are phenomenological model equations that provide a simplified description of the microscopic dynamics, which in the case of the collision model is more precisely described by Eq.~\eqref{e:binary_almost_langevin}. Therefore, to obtain a useful Langevin model, the coefficients $\ga(p)$ and $D(p)$  have to be chosen such that they yield the best possible approximation within this 
class of SDEs defined by Eq.~\eqref{e:nonlinear_langevin_repeated}.
Plausible general criteria for the \lq best approximation\rq\space can be formulated as follows: 
\par
The stochastic process described by Eq.~\eqref{e:nonlinear_langevin_repeated} should 
\begin{itemize}
\item  approach the correct stationary momentum distribution;
\item  exhibit the correct mean relaxation behavior.
\end{itemize}
The first criterion is equivalent to imposing the appropriate fluctuation-dissipation relation on the functions $\ga$ and $D$. For the elastic collision model considered here, the expected stationary momentum PDF of the Brownian particle is given by the Maxwell distribution
\be\label{e:binary_Maxwell}
\phi_\infty(p)= 
\left({2\pi M\kB \Temp}\right)^{-1/2}
\exp\bigl[-{p^2}/({2M \kB \Temp})\bigr]. 
\ee 
According to the discussion in Section~\ref{s:nonlinear_langevin_equations}, this implies that $\ga$ and $D$ must be coupled by the Einstein condition
\be
D(P)=\label{e:binary_langevin_coefficients_1b}
\ga(P)\,M\kB \Temp. 
\ee
The second (drift) criterion can be expressed mathematically as\footnote{$\lan \quad\,|\, P(t)=p\ran$ denotes by the conditional expectation with respect to the Wiener measure of the Brownian motion $B(t)$.} 
\be\label{e:mean_value_criterion}
\lan \f{\dP(t)}{\diff t}\;\biggl|\; P(t)=p\ran
&\overset{!}{=}&
\lan\f{ \gd P(t)}{\gd t}\;\biggl|\;P(t)=p\ran_\bath.
\ee
Taking into account the Einstein relation and Eq.~\eqref{e:conditional_mean_values-hk}, the lhs. of Eq.~\eqref{e:mean_value_criterion} is given by
\bse\label{e:C}
\be\label{e:C2b}
\lan \f{\dP(t)}{\diff t}\,\biggl|\, P(t)=p\ran
=
-[\ga(p)\, p-\ga'(p)\; M\,\kB \Temp],
\ee
where $\ga'(p):=\diff \ga(p)/\diff p$. The rhs. of Eq.~\eqref{e:mean_value_criterion} is obtained by substituting $P=p$ on the rhs. of Eq.~\eqref{e:binary_averaged_friction}, yielding, i.e.,
\be
\lan\f{\gd P(t)}{\gd t}\;\biggr|\;P(t)=p\ran_\bath=\drift(p),
\label{e:C2a}
\ee
\ese
where the mean drift force $\drift$ was defined in Eq.~\eqref{e:mean_drift_force_nonrelativistic}.
Hence, by virtue of Eqs.~\eqref{e:C}, we see that the drift criterion~\eqref{e:mean_value_criterion} is equivalent to the following ordinary differential equation (ODE) for $\ga(p)$:
\be\label{e:nonrelativistic_ODE}
-\ga(p)\, p\,+\ga'(p)\; M\,\kB\Temp =\drift(p).
\ee
With respect to the two criteria formulated above, the solution of this ODE gives the friction function $\ga$ that provides the \lq best\rq~Langevin approximation to the binary collision model. The initial condition for $\ga(p)$ must be specified such that the correct asymptotic behavior is obtained~\cite{2008Du_PHD}. Information about the collision model and the bath distribution is encoded in the mean drift force~$\drift(p)$. Evidently, the procedure leading to Eq.~\eqref{e:nonrelativistic_ODE} can be generalized to other interaction models/bath distributions as well -- provided the stationary distribution of the Brownian particle is known. Other types of interactions (e.g., nonelastic) would result in another function $\drift(p)$. A non-Maxwellian bath distribution would affect not only the rhs. of Eq.~\eqref{e:nonrelativistic_ODE} but also its lhs. due to a modified fluctuation-dissipation relation.
\par
\begin{figure}[t]
\center 
\includegraphics[height=4.2cm,angle=0]{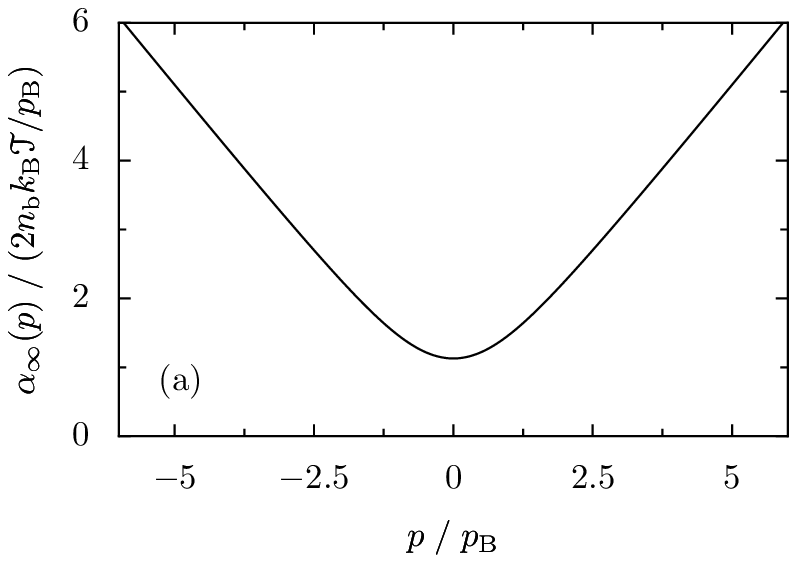}
\hspace{0.5cm}
\includegraphics[height=4.2cm,angle=0]{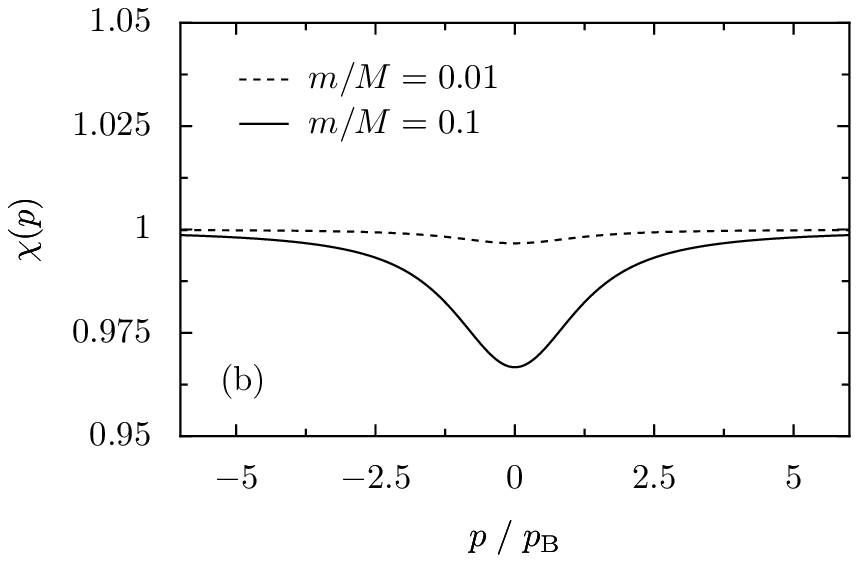}
\caption{\label{fig_nonrelativistic_friction} 
Nonrelativistic binary collision model. (a) Friction coefficient $\ga_\infty(p)=-\drift(p)/p$ from Eq.~\eqref{e:taylor_large_p} with  $n_\bath=N/L$ denoting the number density of the heat bath particles and $\pB:=M(2\kB \Temp/m)^{1/2}$ the characteristic momentum of a Brownian particle (mass $M$), surrounded by heat bath particles of mass $m$ and temperature $\Temp$. (b)~The ratio $\chi(p)$ from Eq.~\eqref{e:quality_test} as a  measure for the quality  of the approximation~$\ga_\infty(p)$.
}
\end{figure}

Unfortunately, it is usually very difficult or even impossible to find the exact analytical solution of the ODE~\eqref{e:nonrelativistic_ODE} for a realistic drift function $\drift(p)$. For practical purposes, one can obtain useful approximations, e.g., by considering the asymptotic behavior for $p\to \infty$ and $p\to 0$, respectively. In the case of the collision model, one finds that the approximation
\be\label{e:taylor_large_p}
\ga(p)\simeq-\drift(p)/p=:\ga_\infty(p),
\ee
which becomes exact for $p\to \infty$, is also applicable at small $|p|$-values, if $m\ll M$. This is illustrated in Fig.~\ref{fig_nonrelativistic_friction}~(b), which depicts the dimensionless ratio
\be\label{e:quality_test}
\chi(p):=[-\ga_\infty(p)\, p\,+\ga_\infty'(p)\, M\kB\Temp]/\drift(p).
\ee 
The function $\chi$ is an indicator for the quality of the approximative solution~$\ga_\infty(p)$, which is plotted in Fig.~\ref{fig_nonrelativistic_friction}~(a). As evident from the dotted curve in Fig.~\ref{fig_nonrelativistic_friction}~(b), for $m\ll M$ the function $\chi$ is close to unity even for small values $|p|$. This means that $\ga_\infty$ is a good approximation to the exact solution of Eq.~\eqref{e:nonrelativistic_ODE}. Thus, a Langevin equation of the type~\eqref{e:nonlinear_langevin_repeated} with $\ga_\infty(p)$ from Eq.~\eqref{e:taylor_large_p}  and \mbox{$D(p)=\ga_\infty(p)M\kB \Temp$} yields the correct stationary momentum distribution~\eqref{e:binary_Maxwell} and exhibits almost exactly the same mean relaxation behavior as Eq.~\eqref{e:quasi_almost_langevin}. In particular, such a nonlinear Langevin equation provides a considerably more accurate description of the Brownian motion in a gaseous heat bath than, e.g., a classical Ornstein-Uhlenbeck process with constant friction and noise coefficients, cf. Eq.~\eqref{e:OUP_external-b}. For instance, an Ornstein-Uhlenbeck (or Stokes-like) approximation is obtained by replacing $\ga_\infty(p)$ through its minimum value 
\be\label{e:stokes-nonrel} 
\ga_0&:=&\ga_\infty(0)=-\drift'(0)=
n_\bath\,\f{m}{M} \left(\f{8\kB \Temp}{\pi m}\right)^{1/2}
\ee
and fixing the Einstein relation $D(p)=\ga_0 M\,\kB \Temp=:D_0$.
Adopting these additional simplifications, the corresponding Ornstein-Uhlenbeck process can be expected to provide a useful description for slow Brownian particles at sufficiently low temperature values~$\Temp$. 
\par
Similar to the harmonic oscillator model from  Section~\ref{s:harmonic_oscillators}, the collision model provides useful insights into the approximations that must be made in order to obtain a Langevin equation from a microscopic model. Compared with the oscillator model, an advantage of the collision model is given by the fact that it can be extended to special relativity, cf.  Section~\ref{s:relativistic_binary_collision_model} below. 
\par
From a more general perspective, the above examples illustrate which objectives can(not) be achieved by phenomenological Langevin models that are based on Brownian motion processes.  Langevin equations of this type, and their corresponding Fokker-Planck equations, provide a simplified description of the underlying microscopic dynamics. The coefficient functions in the Langevin/Fokker-Planck equations allow one to construct stochastic processes that exhibit the  same asymptotic relaxation behavior  and approach the same stationary or asymptotic distribution like the actual physical process. Stationary distributions can often be inferred from thermostatistical (e.g., maximum entropy) principles, while the relaxation behavior must be deduced from the exact microscopic dynamics. In many cases, the resulting stochastic models     are sufficient for comparing with experimentally accessible data, but they may become inaccurate  for describing higher order correlations and/or the relaxation dynamics far from the asymptotic  state.

\subsection{Remarks and generalizations}

In this section we have focused on the most commonly considered examples of nonrelativistic Brownian motion processes, described by SDEs of the type~\eqref{e:langevin_m}. Such nonrelativistic Langevin equations have been studied extensively in various contexts over the past decades (see, e.g., the reviews~\cite{1994Kl,1990HaTaBo}). The list of successful applications covers a wide range of different areas including laser  physics~\cite{1994Kl,1995Klimontovich,1972KlKoLa}, optical lattices~\cite{1996MaElZo,2003Lu}, plasma physics~\cite{2003KhIvMoZh,2004KhIvMo,2003TrEbIgTk,2000ZaScTr}, high energy physics~\cite{1988Sv,2005HeRa}, biologically and chemically motivated population and reaction dynamics~\cite{1989Mi}, active Brownian motion  models~\cite{1993ScGr,1999EbScTi,2000ErEbScSc,1998ScEbTi,2001ScEbTi,2003Tr,2005EbDuErTr}, or theoretical and experimental studies of excitation and transition phenomena in nonlinear systems~\cite{2000EbErDuJe,2001EbLaUs,2001MaRiEV,2001DuEbErMa}.
\par
The stochastic processes defined by Eqs.~\eqref{e:langevin_m} share as a common feature that the underlying noise source is modeled by a standard Wiener process $B(t)$. In general, one can also consider other driving processes such as Poisson processes~\cite{KaSh91,Grigoriu} or L\'evy processes~\cite{1999JeMeFo,2000So,2005weron:016113,2008DeHoHa}, which may give rise to so-called anomalous super- or sub-diffusion effects; see, e.g., the reviews by Bouchaud and Georges~\cite{1990BoGe} and Metzler and Klafter~\cite{2000MeKl}.\footnote{Anomalous diffusion processes~\cite{2004sancho:250601,2006heinsalu:046133,2006goychuk:020101} exhibit a nonlinear growth of the mean square displacement, e.g., of the form to $t^\gk$ with $\gk>1$ and $\gk<1$ corresponding super- and sub-diffusion, respectively~\cite{1999JeMeFo,2000MeKl}.} Furthermore, one can abandon the assumption~\eqref{e:white_noise_b} of $\gd$-correlated \lq white\rq\space noise by considering stochastic processes that are driven by \lq colored\rq\space noise,  for example, by replacing  Eq.~\eqref{e:white_noise_b} with an exponential correlation
\be\label{e:coloured_noise}
\lan  \zeta(t)\, \zeta(s)\ran = \f{1}{\tau_\mrm{n}}\exp(-|t-s|/\tau_\mrm{n}),
\ee
where the parameter $\tau_\mrm{n}$ is the relaxation time of the driving noise $\zeta$. The mathematical analysis of processes driven by colored noise is considerably more complicated than in the case of $\gd$-correlated white noise (see Ref.~\cite{1995HaJu} for a review). The study of non-Gaussian and/or colored driving processes within a   relativistic framework still represents an open problem for the future. By contrast, the relativistic generalization of stochastic processes of the type~\eqref{e:langevin_m} has been subject of intense research in recent years, leading to the relativistic Brownian motion theory described in  Section~\ref{s:RBM_lab_time}.


\section{Relativistic equilibrium thermostatistics}
\label{s:Rel_Thermo}

The brief summary of nonrelativistic Brownian motion theory in the preceding section shows that equilibrium thermostatistics plays an important role in constraining the relation between friction and noise coefficients in Langevin equations by means of suitably chosen fluctuation-dissipation relations. Moreover, \textit{a priori} knowledge about the equilibrium properties of the heat bath is required if one wishes to derive (generalized) Langevin equations from microscopic models. Therefore, the present section intends to summarize relevant aspects of relativistic equilibrium thermostatistics in order to provide for the subsequent discussion of relativistic Langevin equations in Section~\ref{s:RBM_lab_time}. This part is structured as follows. 
\par 

Section~\ref{s:prelims} introduces some notation and discusses general properties of relativistic probability density functions.  Section~\ref{s:relativistic_equilibrium_distribution} focuses on the thermostatistics of stationary systems, since these will play the role of a heat bath later on. In this context, particular emphasis will be placed on the relativistic generalization of Maxwell's distribution for the following reason: Nonrelativistic Brownian motion models such as the classical Ornstein-Uhlen\-beck process are in obvious conflict with special relativity because they permit particles to move faster than the speed of light~$c$. Most directly, this can be  seen from the stationary velocity distribution, which  is a Maxwell velocity distribution and thus non-zero for velocities~\mbox{$|\bs v|>c$.}  The recent literature has seen some debate about the correct generalization of Maxwell's distribution in special  relativity~\cite{2002Ka,2006Le,1981HoScPi,1989HoShSc,2007DuHa,2007DuTaHa_2}. In Section~\ref{s:relativistic_equilibrium_distribution} we shall discuss recent molecular dynamics simulations~\cite{2007CuEtAl} which  favor a distribution that was proposed by J\"uttner~\cite{1911Ju} in 1911, i.e., six years after Einstein had formulated his theory of special relativity~\cite{1905Ei_SRT1,1905Ei_SRT2}. In addition to its relevance with regard to relativistic Brownian motion theory, the J\"uttner gas  also provides a useful model for illustrating the subtleties of relativistic thermodynamics. The latter topic is discussed separately in Appendix~\ref{s:relativistic_thermodynamics}, also addressing the controversy that has surrounded the Lorentz transformation laws of temperature and other thermodynamic quantities over the past 100 years~\cite{1907Ei,1908Pl,1963Ott,1965Ar,1965Ar_1,1965Bo,1965Ga,1965Su,1966La,1966La_1,1966Ki,1966Ar,1966Fr,1966Pa,1966Pa_2,1966Ro,1967Ga,1967La,1967Mo,1967No,1967Re,1967Wi,1967Wi_2,1968Ba,1968Ge,1968LaJo,1968Li,1968Na,1969Ba,1969Ha,1970Ge,1970LaJo,1970Yu,1971CaHo,1977Ag,1978Kr,1981La,1995Ko,1996LaMa,2005ArLoAn}.

\subsection{Preliminaries}
\label{s:prelims}
Section~\ref{s:notation} summarizes definitions and notational conventions.\footnote{For a more detailed introduction to special relativity we refer to Weinberg~\cite{Weinberg} or Sexl and Urbantke~\cite{SexlUrbantke}.} Subsequently, the \lq peculiar\rq\space Lorentz transformation behavior of one-particle phase space probability density functions (PDFs) will be addressed.

\subsubsection{Notation and conventions}
\label{s:notation}
In special relativity, an inertial frame $\Gs$ corresponds to a global Cartesian spacetime coordinate system. A spacetime event $\eve$ is labeled by a $(1+d)$-dimensional coordinate tuple \mbox{$\bar{x}=(x^\ga)=(ct,\bs x)=(t,x^1,\ldots, x^d)$} in $\Gs$, where $d$ is the number of space dimensions and, adopting natural units from now on, the speed of light $c=1$. Upper and lower Greek indices $\ga,\gb,\ldots$ take values $0,1,\ldots,d$, and we use Latin indices $i,k,\ldots\in\{1,\ldots,d\}$ for the spatial components. Vectors with upper indices are called contravariant. 
\par 
With respect to the Cartesian coordinate frame $\Gs$, the components $\eta_{\ga\gb}$ of the metric tensor of flat Minkowski spacetime are defined by~\cite{Weinberg} 
\be
\eta_{\ga\gb}=
\begin{cases}
-1 &\qquad\ga=\gb=0\\
+1 &\qquad\ga=\gb=1,\ldots, d\\
0  &\qquad\ga\ne\gb.
\end{cases}
\ee
By definition, the components of the covariant vector $(x_\ga)$ are obtained by contracting the contravariant vector $(x^\ga)$ with $\eta_{\ga\gb}$, i.e.,\footnote{Throughout, we will use the Einstein summation convention defined in Eq.~\eqref{e:a-einstein-convention}. 
} 
\be\label{e:a-einstein-convention}
x_\ga:=\sum_{\gb=0}^d\eta_{\ga\gb} x^\gb=:\eta_{\ga\gb} x^\gb
\qquad\Rightarrow\qquad
(x_\ga)=(-t,\bs x).
\ee
The tuples $(x^\ga)$ and $(x_\ga)$ will be called four-vectors, regardless of the number of space dimensions. The corresponding four-vector gradients are defined by
\bse
\be
(\p_\ga)&:=&\left(\parder{}{x^\ga}\right)
=\left(\parder{}{t},\parder{}{x^1},\ldots,\parder{}{x^d} \right)
=\left(\parder{}{t},\nabla\right),\\
(\p^\ga)&:=&\left(\parder{}{x_\ga}\right)
=\left(-\parder{}{t},\parder{}{x^1},\ldots,\parder{}{x^d} \right)
=\left(-\parder{}{t},\nabla \right).
\ee
\ese
The components $\eta^{\ga\gb}$ of the inverse metric tensor are determined by the condition
\bse
\be
x^\ga\overset{!}{=}\eta^{\ga\gb} x_\gb=\eta^{\ga\gb}\eta_{\gb\gc} x^\gc
\qquad\qquad
\fa (x^\ga),
\ee
or, equivalently, by
\be
\eta^{\ga\gb}\eta_{\gb\gc}\overset{!}{=} {\gd^\ga}_\gc,
\ee
\ese
where ${\gd^\ga}_\gc$ is the Kronecker $\gd$-symbol, yielding
\be
\eta^{\ga\gb}=\eta_{\ga\gb}.
\ee
The Minkowski spacetime distance between two events $\bar{x}_A=(x_A^\ga)=(t_A,\bs x_A)$ and \mbox{$\bar{x}_B=(x_B^\ga)=(t_B,\bs x_B)$} is defined by 
\be
d(\bar{x}_A,\bar{x}_B)^2
&:=&\notag
\eta_{\ga\gb}(x_A^\ga-x_B^\ga)(x_A^\gb-x_B^\gb)\\
&=&-(t_A-t_B)^2+(\bs x_A-\bs x_B)^2.
\ee 
By definition, the separation of two events is
\begin{itemize}
\item time-like, if $d(\bar{x}_A,\bar{x}_B)^2<0$;
\item light-like, if $d(\bar{x}_A,\bar{x}_B)^2=0$;
\item space-like, if $d(\bar{x}_A,\bar{x}_B)^2>0$.
\end{itemize}
Events with time-like separation can be causally connected by (a series of) signals travelling slower or equal to the speed of light. Events with light-like separation can be causally related only by undisturbed signals travelling at the speed of light.  Events with space-like separation are causally disconnected.
\par
The classical motion of a massive point-like particle through spacetime corresponds to a sufficiently smooth, time-like curve in $\Gs$, referred to as world-line. 
Consider a stationary observer $\Obs$, who is at rest in $\Gs$. It is natural that $\Obs$  parameterizes the particle motion using the coordinate time $t$ of $\Gs$, i.e., $\Obs$ describes the world-line as a curve $(x^0(t),x^i(t))$ with $x^0(t)=t$. In the vicinity of any point (event) on the particle's world-line, an infinitesimal proper time differential can be defined by
\be\label{e:a-proper-time}
\dtau :=(-\eta_{\ga\gb}\dx^\ga\dx^\gb)^{1/2}=(\dt^2-\dbx^2)^{1/2}=\dt\, (1-\bs v^2)^{1/2},
\ee
where $\bs v(t):=\dbx(t)/\dt$ is the particle velocity in $\Gs$. According to the postulates of special relativity, $\dtau$ is the time interval measured by an intrinsic clock, comoving with the particle, whereas $\dt$ is the coordinate time interval measured by a clock at rest in $\Gs$. The four-velocity $(u^\ga)$ of a massive particle is defined as the derivative of the world-line with respect to its proper time,
\be\label{e:a-4-velocity}
u^\ga:=\f{\dx^\ga}{\dtau}
\qquad\Rightarrow\qquad 
u_\ga u^\ga=-1.
\ee
For a point-like particle with rest mass $m>0$, the energy-momentum four-vector $(p^\ga)=(p^0,p^1,\ldots, p^d)=(\eps,\bs p)$ is defined by
\be\label{e:a-4-momentum}
p^\ga:= m u^\ga
\qquad\Rightarrow\qquad 
p_\ga p^\ga=-m^2.
\ee
Upon comparing with \eqref{e:a-proper-time}, one finds for a particle with velocity $\bs v$ in $\Gs$
\be
p^0=\eps=m\gc(\bs v),
\qquad
\bs p= m\bs \gc(\bs v)\,\bs v,
\qquad
\gc(\bs v):=(1-\bs v^2)^{-1/2}.
\ee

\paragraph*{Lorentz-Poincar\'e  transformations}
In special relativity affine-linear Lorentz-Poincar\'e transformations (LPTs) of the form
\bse\label{e:a-lpt}
\be\label{e:a-lpt-1}
\bar x'=\Gl \bar x +\bar a
\qquad\Leftrightarrow\qquad
x'^\ga={\Gl^\ga}_\gb x^\gb +a^\ga
\ee
describe the transition from an inertial frame $\Gs$ to another inertial frame $\Gs'$. The constant four-vector $a^\ga$ shifts the origins of time and space, while the constant Lorentz matrix $({\Gl^\ga}_\gb)$ may account for a spatial rotation, a change of orientation and/or a relative velocity between the two frames $\Gs$ and $\Gs'$~\cite{Weinberg}. The matrix components ${\Gl^\ga}_\gb$ are determined by the condition
\be\label{e:a-lpt-2}
d(\bar{x}_A',\bar{x}_B')^2 \overset{!}{=} d(\bar{x}_A,\bar{x}_B)^2 
\qquad\Leftrightarrow\qquad
{\Gl^\ga}_\gc {\Gl^\gb}_\gd\eta_{\ga\gb}\overset{!}{=} \eta_{\gc\gd},
\ee
\ese
which means that causal relations remain preserved during transitions between inertial systems. The LPTs~\eqref{e:a-lpt} form a group.\footnote{For group theoretical aspects of Lorentz transformations see, e.g., Refs.~\cite{Weinberg,SexlUrbantke,WeinbergQFT1}.} Of particular interest for our purpose, is the subgroup of \emph{proper} Lorentz transformations (LTs), defined by $a^\ga=0$ and  the additional constraints
\be\label{e:a-proper}
{\Gl^0}_0\ge 1
\csp
\det ({\Gl^\ga}_\gb)=+1.
\ee
The requirements~\eqref{e:a-proper} exclude time reversal and space inversion.
Examples are pure rotations
\be\label{e:a-rotations}
{\Gl^0}_0=1
\csp
{\Gl^i}_0={\Gl^0}_i=0
\csp
{\Gl^i}_j=R_{ij},
\ee 
where $(R_{ij})$ is a rotation matrix [i.e., $\det(R_{ij})=1$ and $R_{ij} R_{kj}=\gd_{ij}$], and Lorentz boosts~\cite{Weinberg}
\be\label{e:a-boost}
{\Gl^0}_0=\gc,
\qquad
{\Gl^i}_0={\Gl^0}_i=-\gc w^i,
\qquad
{\Gl^i}_j={\gd^i}_j+\f{w^iw^j}{\bs w^2}(\gc-1)
\ee
with velocity $\bs w=(w^1,\ldots,w^d)$ and Lorentz factor $\gc:=(1-\bs w^2)^{-1/2}$. To briefly illustrate the effect of a boost, consider a particle at rest in the spatial origin of $\Gs$ and, therefore, being described by the world-line $(x^\ga(t))\equiv (t,\bs 0)$ in $\Gs$. By applying the Lorentz boost~\eqref{e:a-boost} to $(x^\ga)=(t,\bs 0)$, we find
\be
x'^0= {\Gl^0}_0 x^0=\gc t=t'
\csp
x'^i= {\Gl^i}_0 x^0=-\gc w^i t=-w^i t',
\ee
which means that the particle travels at constant velocity $\bs w'=-\bs w$ through~$\Gs'$; i.e., $\Gs'$ moves with velocity $\bs w$ relative to $\Gs$. The inverse of the transformation matrix~\eqref{e:a-boost} is obtained by replacing $\bs w$ with $-\bs w$. 
\par
From Eq.~\eqref{e:a-lpt} and the definition~\eqref{e:a-4-momentum} of the four-momentum, one finds the relativistic energy-momentum transformation law
\be\label{e:a-momentum-lpt}
p'^\ga={\Gl^\ga}_\gb p^\gb.
\ee
Combining Eqs.~\eqref{e:a-momentum-lpt} and \eqref{e:a-lpt-2}, one can verify the well-known mass-shell condition
\be\label{e:a-mass-shell}
m^2=\eps^2-\bs p^2=\eps'^2-\bs p'^2=m'^2,
\ee
which means that the rest mass $m$ is a Lorentz invariant. In  particular, the mass shell condition~\eqref{e:a-mass-shell} implies that Eq.~\eqref{e:a-momentum-lpt} is equivalent to a restricted \emph{nonlinear} transformation $\bs p \mapsto \bs p'=\bs p(\bs p')$, given by
\be\label{e:a-p-transformation}
p'^i(\bs p)={\Gl^i}_0(m^2+\bs p^2)^{1/2}+{\Gl^i}_jp^j.
\ee

\subsubsection{Probability densities in special relativity}
\label{s:probality_densities}
With regard to the subsequent discussion, it is worthwhile to address a few subtleties concerning the definition and transformation behavior of probability density functions (PDFs) in special relativity~\cite{1969VK,2001DeRiLe}.

\paragraph*{Relativistic one-particle phase space distributions} 
To start with, we consider the \emph{one-particle} phase space PDF $f(t,\bs x,\bs p)\ge 0$, where $(t, \bs x, \bs p)$ are measured with respect to the inertial \lq lab\rq\space frame $\Gs$. For a relativistic many-particle system  with conserved particle number $N$ (e.g., a gas of identical particles), the function $f$ can be defined operationally as follows~\cite{1969VK}:
\par 
An observer $\Obs$, who is at rest in $\Gs$ and observes the system at $\Gs$-time $t$,  finds
$$N\, f(t,\bs x,\bs p)\;\diff^d x\diff^d  p$$
 particles in the $2d$-dimensional phase space interval $[\bs x,\bs x+\diff \bs x]\times [\bs p,\bs p+\diff \bs p]$. Assuming that the dynamics of each particle is described by functions $\bs X_r(t)$ and $\bs P_r(t)$ in $\Gs$, the \emph{fine-grained} one-particle PDF $f$ of the $N$-particle system is defined by\footnote{Since the kinetic momentum $\bs P_r$ is uniquely linked to the velocity $\bs V_r(t):=\diff \bs X_r(t)/\diff t$, the definition~\eqref{e:PDF_microscopic_definition_1} formalizes the idea of classifying the particle curves $\bs X_r(t)$ according to their positions and time derivatives at time $t$.} 
\bse\label{e:PDF_microscopic_definitions}
\be\label{e:PDF_microscopic_definition_1}
f(t,\bs x,\bs p)=
N^{-1}\sum_{r=1}^N\gd(\bs x-\bs X_r(t))\;\gd(\bs p-\bs P_n(t)).
\ee
From this definition it is evident that $f$ satisfies the $t$-\emph{simultaneous} normalization condition
\be\label{e:normalization_t}
1=\int \d^d x \diff^d p\;f(t,\bs x,\bs p).
\ee 
Note that this integral is taken along the hyperplane \lq\lq $t$=constant\rq\rq\space in $\Gs$.
\par
Alternatively, when considering, e.g., the random motion of a single Brownian particle in a fluctuating medium, the quantity  $f(t,\bs x,\bs p)\;\diff^d x\diff^d p$ can be interpreted as the probability of finding the Brownian particle at lab time $t$ in \mbox{$[\bs x,\bs x+\diff \bs x]\times [\bs p,\bs p+\diff \bs p]$}. In the latter case, it is usually assumed that a potential trajectory is realized with a certain \textit{a priori} probability. Mathematically, this idea is implemented by introducing latent variables $\go$ in order to label the potential trajectories\footnote{For example, if the particle dynamics in $\Gs$ is described by differential equations of the form $\diff\bs X(t)/\diff t=\bs V[\bs X(t),\bs P(t)],\;\diff\bs P(t)/\diff t=\bs K[\bs X(t),\bs P(t)]$ with given (well-behaved) functions $\bs V:\R^d\times \R^d\to\R^d$ and $\bs K:\R^d\times \R^d\to\R^d$, then a trajectory is uniquely determined by specifying the values $\bs X(t_0)=\bs x_0$ and $\bs P(t_0)=\bs p_0$ at some instant $t_0$ in $\Gs$. In this case, one could choose $\go=(\bs x_0,\bs p_0)$. More generally, $\go$ could also label different realizations of some background field which affects the particle dynamics.} by writing  $\bs X(t;\go)$ and $\bs P(t;\go)$. The assignment of \textit{a priori} probabilities is equivalent to specifying a PDF $\Phi(\go)$ on the set of the latent variables~$\{\go\}$. In this case, the phase space density $f$ in $\Gs$ is defined by~\cite{1969VK,2001DeRiLe} 
\be\label{e:PDF_microscopic_definition_2}
f(t,\bs x,\bs p)=
\int \diff\go\; \Phi(\go)\;
\gd(\bs x-\bs X(t;\go))\;\gd(\bs p-\bs P(t;\go)),
\ee
\ese
and this $f$ is again subject to the normalization condition~\eqref{e:normalization_t}.
\par
\begin{figure}[t]
\centering
\includegraphics[width=9.7cm,angle=0]{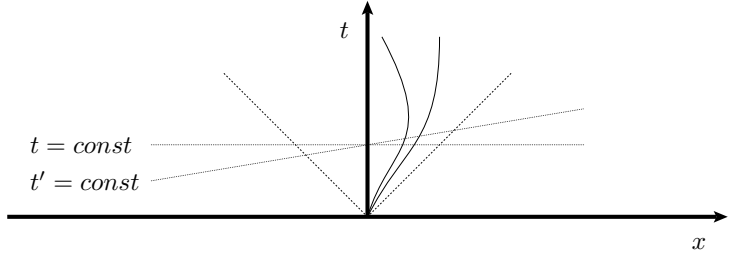}
\caption{
\label{fig:MovingFrame}
The solid curves represent two world-lines starting at the same spacetime point (dashed lines indicate the forward lightcone)  The one-particle phase space PDF $f(t,\bs x,\bs p)$ measures the number of world-lines that pierce through the hyperplane \lq\lq $t$=constant\rq\rq\space within the interval  
$[\bs x,\bs x+\diff \bs x]$  while having momentum values in the range 
$[\bs p,\bs p+\diff \bs p]$. Similarly,  the PDF $f'(t',\bs x',\bs p')$ measures the number of world-lines piercing through the hyperplane \lq\lq $t'$=constant\rq\rq\space within the interval  
$[\bs x',\bs x'+\diff \bs x']$  while having momenta in the range 
$[\bs p',\bs p'+\diff \bs p']$. 
 }
\end{figure}

Equations~\eqref{e:PDF_microscopic_definitions} refer explicitly to the inertial  rest frame $\Gs$ of the observer~$\Obs$. Now consider a second observer~$\Obsm$ at rest in another inertial frame $\Gs'$ that moves with constant velocity $\bs w\ne 0$ relative to~$\Gs$. Employing an analogous operational definition as $\Obs$, the moving observer $\Obsm$~will measure another distribution $f'(t',\bs x',\bs p')$ and it arises the question how the two functions $f'(t',\bs x',\bs p')$ and $f(t,\bs x,\bs p)$ are related to each other. 
In the nonrelativistic theory, the change from one inertial system to another does not affect the time coordinate; hence, one can use the standard transformation laws for PDFs in that case [see, e.g., Eq.~\eqref{e:PDF_transformation}]. By contrast, the situation becomes more complicated in the relativistic theory, because now the definition of $f$ and $f'$ is based on an observer-dependent notion of simultaneity:  The measurements of $\Obs$ and $\Obsm$ refer to the two different hyperplanes \lq\lq $t$=constant\rq\rq\space and  \lq\lq  $t'$=constant\rq\rq~in Minkowski space, respectively. 
This is illustrated in Fig.~\ref{fig:MovingFrame}.
\par
In an insightful paper~\cite{1969VK} published in 1969, van Kampen proved  that the one-particle phase space PDF $f$ transforms as a Lorentz scalar, i.e.,\footnote{One can find several insufficient \lq proofs\rq~of Eq.~\eqref{e:pdf_invariance} in the literature,  cf. the discussion in~\cite{1969VK,2001DeRiLe}. In this context it is sometimes claimed, erroneously, that the phase space element $\diff^d x\diff^d p$ is a Lorentz scalar; in Section~2 of their paper, Debbasch et al.~\cite{2001DeRiLe} demonstrate that this is not true in general.}
\be\label{e:pdf_invariance}
f(t,\bs x,\bs p)=f'(t',\bs x',\bs p'),
\ee
where $(t,\bs x,\bs p)$ and $(t',\bs x',\bs p')$ are connected by Lorentz transformations. His (first) proof of Eq.~\eqref{e:pdf_invariance} merely uses an assumption about the uniqueness of particle trajectories and a reparameterization of the particles trajectories in terms of their invariant proper times (cf. Section 3 in~\cite{1969VK}). As a consequence, Eq.~\eqref{e:pdf_invariance} represents a generic kinematical result and applies to a broad class of interaction models. Furthermore, van Kampen  showed that\footnote{The result~\eqref{e:same_normalization} is nontrivial due the fact that the integrals refer to different hyperplanes \lq\lq $t$=constant\rq\rq\space and  \lq\lq  $t'$=constant\rq\rq~in Minkowski spacetime, respectively. The  proof of Eq.~\eqref{e:same_normalization} follows from Gauss' theorem, by making use of the fact that the absolute particle velocities are bounded by the speed of light $c=1$, cf. Eqs. (34) and (35) in~\cite{1969VK}.}
\bse
\be\label{e:same_normalization}
\int \diff^d x'\diff^d p'\;f(t',\bs x',\bs p')=\int \diff^d x\diff^d p\;f(t,\bs x,\bs p),
\ee 
implying that the function $f'$ satisfies the $t'$-simultaneous  normalization condition
\be\label{e:normalization_t_prime}
1=\int \d^d x' \diff^d p'\;f(t',\bs x',\bs p').
\ee 
\ese
We next summarize a few consequences of Eq.~\eqref{e:pdf_invariance}. 

\paragraph*{Density-current vector} 
Equation~\eqref{e:pdf_invariance} implies that the quantities 
\bse\label{e:4-density}
\be\label{e:4-density-a}
\gr(t,\bs x)&=&\int \diff^d p\, f(t,\bs x,\bs p)\\
\bs j(t,\bs x)&=&\int \diff^d p\;  f(t,\bs x,\bs p)\;\bs v,
\ee
\ese
where $\bs v=\bs p/\eps=\bs p/p^0$ is the particle velocity, can be combined into a density-current four-vector field $\bar{j}(\bar x)=(j^\ga)=(\gr,\bs j)$. As any covariant vector field, $\bar j$ transforms under a Lorentz transformation~\mbox{$\bar x\mapsto\bar x'=\Gl \bar x$} as
\be
\bar{j}'(\bar x')=\Gl \bar j(\Gl^{-1} \bar x')=\Gl \bar j(\bar x).
\ee
The four-vector character of $(j^\ga)$ becomes particularly evident by rewriting Eq.~\eqref{e:4-density} in the form
\be\label{e:4-density-covariant}
j^\ga(t,\bs x)=\int \f{\diff^d p}{p^0}\; f(t,\bs x,\bs p)\;p^\ga,
\ee
where
\be\label{e:invariant_measure}
\f{\diff^d p}{p^0}
=\f{\diff^d p'}{p'^0}
\ee
is invariant under Lorentz  transformations~\cite{CercignaniKremer,1969VK}. Furthermore, as shown in Section 4 of van Kampen's paper~\cite{1969VK}, $j^\ga$ satisfies the continuity equation
\be\label{e:4-density-conservation}
\p_\ga j^\ga=\f{\p}{\p t}\gr +\nabla \cdot \bs j=0, 
\ee
stating the conservation of particle number or probability, in agreement with Eq.~\eqref{e:same_normalization}. The kinematical proof of Eq.~\eqref{e:4-density-conservation} as given in~\cite{1969VK} does not require 
knowledge about the equations of motions, but uses only the existence of the particle trajectories.

\paragraph*{Energy-momentum density tensor} 
Equation~\eqref{e:4-density-covariant} can be generalized to define a symmetric \emph{energy-momentum (density) tensor} field by~\cite{CercignaniKremer,1969VK}
\be\label{e:stress-tensor-covariant}
\gt^{\ga\gb}(t,\bs x)=
\int \f{\diff^d p}{p^0}\; f(t,\bs x,\bs p)\;p^\ga p^\gb.
\ee
In particular, the \lq\lq ${00}$\rq\rq-component gives the (kinetic) energy density per particle. In contrast to the particle number  conservation law~\eqref{e:4-density-conservation}, the conservation of the energy-momentum tensor is a dynamical property which requires knowledge about the equations of motions (see example in the next section). 
\par
In principle, \lq\lq higher  moment tensors\rq\rq~can be defined in a similar manner: 
\be\label{e:higher-stress-tensor-covariant}
\hat\gt^{\ga\gb\gc\ldots}(t,\bs x)=
\int \f{\diff^d p}{p^0}\; f(t,\bs x,\bs p)\;p^\ga p^\gb p^\gc\ldots.
\ee
However, conventional thermodynamic and hydrodynamic analysis~\cite{1976Is,1981Is,1987Is} usually focuses on relations involving $j^\ga$ and $\gt^{\ga\gb}$.

\paragraph*{Marginal momentum distributions} 
The zero-component $\gr(t,\bs x)$  of the density-current vector $\bar j(\bar x)$, defined in Eq.~\eqref{e:4-density-a}, can be interpreted as the marginal distribution of the particle positions. Similarly, one may define marginal momentum distributions with respect to $\Gs$ and $\Gs'$ by
\bse
\be
\phi(t,\bs p)&=&\int\diff^d x\, f(t,\bs x,\bs p),\\
\phi'(t',\bs p')&=&\int\diff^d x'\, f'(t',\bs x',\bs p').
\ee
\ese
Again, $\phi$ and $\phi'$ refer to different hyperplanes in Minkowski spacetime, respectively. Thus, in general, it is impossible to calculate $\phi'$ from $\phi$ alone or \textit{vice versa}. There exist, however, a few exceptions such as, e.g., a system of freely moving, identical particles (i.e., no interactions, no external fields, no walls). In any inertial frame, 
such a collection of particles is described by a time-dependent\footnote{We assume that at least two particles have different velocities.} one-particle phase space PDF $f$ satisfying~\cite{1969VK}
\be
f(t,\bs x,\bs p)=f(0,\bs x-\bs v t,\bs p),
\ee
where $\bs v=\bs p/(m^2+\bs p^2)^{1/2}$ is the velocity. Moreover, since the individual particle momenta do not change, the marginal momentum distribution must be time-independent in any inertial frame. In particular, in this case  -- and only in this case -- it is true that each particle observed by $\Obs$ as having momentum~$\bs p$ is observed by $\Obs'$ as having momentum $\bs p'$, i.e.,~\cite{1969VK}
\be
\phi(\bs p)\diff^d p=\phi'(\bs p')\diff^d p',
\ee
where $\bs p'(\bs p)$ is the restricted Lorentz transformation from Eq.~\eqref{e:a-p-transformation}. Taking into  account that $\diff^d p'= (p'^0/p^0)\diff^d p$, cf. Eq.~\eqref{e:invariant_measure}, one thus obtains
\be\label{e:pdf_invariant_free}
(m^2+\bs p^2)^{1/2}\,\phi(\bs p)=(m^2+\bs p'^2)^{1/2}\,\phi'(\bs p').
\ee
It should be stressed again that this formula holds true only in the case of an unconfined, non-interacting systems; it is \emph{not} valid anymore in the presence of a confinement (see example in Section~\ref{s:relativistic_equilibrium_distribution}).

\paragraph*{Multi-particle distributions}
The discussion in the remainder mostly concerns one-particle distributions. Nonetheless, we mention that Eq.~\eqref{e:pdf_invariance} can be generalized to the case of $N$-particle phase space PDFs $f_N$, yielding~\cite{1969VK}
\be
f_N(t_1,\bs x_1,\bs p_1;\ldots;t_N,\bs x_N,\bs p_N)
=
f_N'(t_1',\bs x_1',\bs p_1';\ldots;t_N',\bs x_N',\bs p_N'),
\ee
where for $n=1,\ldots,N$ the coordinates $(t_n',\bs x_n',\bs p_n')$ and $(t_n,\bs x_n,\bs p_n)$ are connected by a Lorentz-transformation, and $f_N$ is the \emph{multiple-time} probability density for the lab observer $\Obs$ to observe particle 1 at time $t_1$ near  $(\bs x_1,\bs p_1)$, \emph{and} particle 2 at time $t_2$ near  $(\bs x_2,\bs p_2)$, etc..
\par
The above results clarify the transformation behavior of PDFs in special relativity, but they do not yet answer the question as to \emph{which} PDF provides the correct description for a given physical system as, e.g., a relativistic gas in equilibrium. The latter problem will be addressed in the next part.

\subsection{Stationary systems with confinement}
\label{s:relativistic_equilibrium_distribution}

Of particular relevance in equilibrium thermodynamics are confined systems that can be described by an isotropic, stationary PDF $f(t,\bs x,\bs p)=\vphi(\bs x,\bs p)$ in a specific inertial frame $\Gs$. A typical example is an equilibrated gas, enclosed in a container box which is at rest in the lab frame~$\Gs$. In standard Brownian motion theory such systems often play the role of the heat bath.

\subsubsection{General remarks}

\paragraph*{Lab frame} 
In the lab frame $\Gs$, a spatially homogeneous gas is described by a PDF of the form
\bse
\be\label{e:PDF_stationary}
\vphi(\bs x,\bs p)=V^{-1}\; \Ind(\bs x;\Volume)\;\phi(\bs p),
\ee 
where $V=\gl_d(\Volume)$  is the rest volume number (i.e., the Lebesgue measure~\cite{1951Halmos}) of the spatial container region $\Volume\subset\R^d$ in $\Gs$, and $\Ind(\bs x;\Volume)$ denotes the indicator function of the box, i.e., 
\be\label{e:indicator_definition}
\Ind(\bs x;\Volume):=
\begin{cases}
1,&\qquad \bs x\in\Volume,\\
0,&\qquad \bs x\not\in\Volume.
\end{cases}
\ee
\ese
To be more specific, we consider a cubic container of length $L$ such that $\Volume=[-L/2,L/2]^d$. In this case, $\Vol=\gl_d(\Volume)=L^d$  and the indicator function can be expressed as
\be\label{e:volume_indicator}
\Ind(\bs x;\Volume):=\prod_{i=1}^d\Gt(L/2+x^i)\,\Gt(L/2-x^i).
\ee
Isotropy and stationarity of the gas in the lab frame $\Gs$ require the marginal momentum distribution to be rotationally invariant,  $\phi(\bs p)=\hat{\phi}(|\bs p|)$, yielding for the four-current $(j^\ga)=(\gr,\bs j)$ in~$\Gs$
\bse\label{e:current_volume_indicator}
\be
\gr(t,\bs x)&=&V^{-1}\;\Ind(\bs x;\Volume),\\
\bs j(t,\bs x)&=&\bs 0.
\ee
\ese
It is obvious that this current satisfies the continuity equation~\eqref{e:4-density-conservation}, $\p_\ga j^\ga=0$. 
Furthermore, due to the isotropy of the momentum distribution $\phi$ in~$\Gs$, the energy-momentum tensor~\eqref{e:stress-tensor-covariant} becomes diagonal with components  given by\footnote{As usual, we denote by $\bs p \cdot \bs v$ the ordinary Euclidean scalar product of two $d$-dimensional vectors $\bs p=(p^i)$ and $\bs v=(v^i)$, i.e., $\bs p \cdot \bs v:=\sum_{i=1}^d p^i v^i$. Moreover, we abbreviate $\bs p^2:=\bs p\cdot \bs p=p_ip^i=\sum_{i=1}^d p^i p^i.$}
\bse\label{e:stress-tensor-covariant-gas}
\be
\gt^{\ga\gb}(t,\bs x)
=
V^{-1}\,\Ind(\bs x;\Volume)
\begin{cases} 
\energy,    &\qquad\ga=\gb=0,\\
\vir /d ,    &\qquad\ga=\gb=1,\ldots, d,\\
0,& \qquad\ga\ne\gb.
\end{cases}
\ee
Here, we have defined the one-particle mean values
\be
\energy&:=&\int {\diff^d p}\;\phi(\bs p)\;\eps,\\
\vir &:=& \int \diff^d p \; \phi(\bs p)\; \bs p \cdot \bs v.
\label{e:virial_temperature}
\ee
\ese 
with $\eps(\bs p)=(\bs p^2+m^2)^{1/2}$ denoting the energy of a gas particle, and $\bs v(\bs p)= \bs p/\eps$ the velocity. 
\par
It is worthwhile to calculate the four-divergence of the energy momentum  tensor:
\bse\label{e:em-divergence}
\be
\p_\ga\gt^{\ga \gb}= 
(Vd)^{-1} \vir
\begin{cases}
0,&\qquad\gb=0,\\
\p_i\Ind(\bs x;\Volume), &\qquad\gb=i,
\end{cases}
\ee
where, in the case of a cubic box $\Vol:=\gl_d(\Volume)=[-L/2,L/2]^d$, we find
\be
\p_i\Ind(\bs x;\Volume)
&=&\notag
\left[\gd(L/2+x^i)-\gd(L/2-x^i)\right]\times\\
&&\qquad\label{e:em-divergence-b}
\prod_{j\ne i}\Gt(L/2+x^j)\,\Gt(L/2-x^j);
\ee
\ese
i.e., \emph{the boundaries are sources of stress}~\cite{1969VK_2}.\footnote{Usually, it is assumed that the  momentum conservation violation of the gas, $\p_\ga \gt^{\ga\gb}_\mrm{gas}\ne 0 $, is compensated for by the energy-momentum tensor $\gt^{\ga\gb}_\mrm{conf}$ of the confinement (environment), i.e., 
$\p_\ga( \gt^{\ga\gb}_\mrm{gas} +\gt^{\ga\gb}_\mrm{conf})\equiv 0$; a similar problem occurs in continuum models of the electron, cf.~\cite{1986CaJi}.}
\par
To illustrate the meaning of the energy-momentum tensor $\gt^{\ga\gb}$, consider the mean (integrated) energy-momentum vector
\bse\label{e:em-vector-t}
\be
\lan p^\gb\ran_t:=
\int \diff^d x \diff^d p\; f(t,\bs x,\bs p)\; p^\gb.
\ee
This quantity can be rewritten in terms of $\gt^{\ga\gb}$ as follows:
\be\label{e:em-vector-t-b}
\lan p^\gb\ran_t
= 
\int \diff^d x\; \gt^{0 \gb}(t,\bs x)
=
\int_t \diff\gs_\ga\; \gt^{\ga \gb}(t,\bs x).
\ee
The directed surface element normal to the hyperplane \lq\lq $t$=constant\rq\rq\space in $\Gs$-coordinates is given by $(\diff \gs_\ga)=(\diff^d x,\bs 0)$; cf. Appendix~\ref{appendix:surface_integrals}.
Thus, a $d$-dimensional spatial integration in $\Gs$ is equivalent to a  surface integral over the hyperplane \lq\lq $t$=constant\rq\rq in $(1+d)$-dimensional  Minkowski space. For the energy-momentum tensor~\eqref{e:stress-tensor-covariant-gas} one finds explicitly
\be\label{e:em-vector-t-c}
\lan p^\gb\ran_t=
\begin{cases}
\energy,&\qquad \gb=0,\\
0,&\qquad \gb\ne0.
\end{cases}
\ee
\ese
It is important to note that $\langle p^\gb\rangle_t$ is a \emph{non-local} quantity, as it represents a sum over components of the energy-momentum tensor at different spacetime points. As emphasized by Gamba~\cite{1967Ga} and discussed below, this aspect becomes relevant if one considers the question how a moving observer $\Obsm$ could, in principle, determine~$\langle p^\gb\rangle_t$. 

\paragraph*{Moving frame}
Consider an observer $\Obsm$ who is at rest in a frame $\Gs'$ that moves at velocity $\bs w=(w,0,\ldots,0)$ along the $x^1$-axis of $\Gs$. Denoting the spacetime coordinates and four-momentum in~$\Gs$ by $(x^\ga)=(t,x^1,\ldots,x^d)$ and $(p^\ga)=(\eps,p^1,\ldots,p^d)=(\eps,\bs p) $ and those in $\Gs'$ by $(x'^\ga)=(t',x'^1,\ldots,x'^d)$ and $(p'^\ga)=(\eps',p'^1,\ldots,p'^d)=(\eps',\bs p')$, we obtain from Eqs.~\eqref{e:a-lpt-1}, \eqref{e:a-boost}  and~\eqref{e:a-p-transformation} the explicit Lorentz transformations
\begin{align}\notag
t'&=\gc (t-wx^1)
&
t&=\gc (t'+wx'^1)
&
\\\label{e:w-boost}
x'^1&= \gc (-wt+x^1)
&
x^1&= \gc (wt'+x'^1)
&\\\notag
\eps'&=\gc (\eps -w p^1)
&
\eps&=\gc (\eps' +w p'^1)
&
\\\notag
p'^1&=\gc (-w\eps + p^1)
&
p^1&=\gc (w\eps' + p'^1)
&
\\\notag
x'^j&= x^j
&
p^j&= p'^j
&
j&=2,\ldots,d,
\end{align}
where $\gc=\gc(w)=(1-w^2)^{-1/2}$ is the Lorentz factor; $\eps=(m^2+\bs p^2)^{1/2}$ and $\eps'=(m^2+\bs p'^2)^{1/2}$ denotes the relativistic energy, respectively. From Eqs.~\eqref{e:w-boost} one can obtain an explicit expression for the phase space PDF $f'$ measured by the moving observer. If the observer moves at three-velocity $\bs w=(w,0,\ldots,0)$ through the lab frame, then from her point of view the container  box moves at velocity
\be\label{e:velocity_inversion}
\bs w'=(w',0,\ldots,0)=(-w,0,\ldots,0)=-\bs w
\ee
through $\Gs'$. By inserting Eqs.~\eqref{e:w-boost} and \eqref{e:velocity_inversion} into Eq.~\eqref{e:pdf_invariance} for a stationary lab distribution of the form~\eqref{e:PDF_stationary}, one obtains the PDF $f'$ in the moving frame $\Gs'$ as
\be
f'(t',\bs x',\bs p')
&=&\notag
V^{-1}\;\Ind(\gc (-w't'+x'^1),x'^2\ldots,x'^d;\Volume)\;
\times\\
&&\qquad\qquad\qquad
\phi(\gc (-w'\eps' + p'^1),p'^2,\ldots,p'^d).
\label{e:boosted-f}
\ee
While the phase space PDF $f$ was stationary in~$\Gs$, the associated PDF $f'$ is not stationary in~$\Gs'$ due to the motion of the container box.\footnote{In fact, this also happens in the nonrelativistic case.} Considering the representation~\eqref{e:volume_indicator} of the indicator function~$\Ind$ in $\Gs$, one can introduce a corresponding  quantity $\Ind'$ in $\Gs'$ by
\be
\Ind'(t',\bs x';\Volume)
&:=&\notag
\Ind(\gc(-w't'+x'^1),x'^2,\ldots,x'^d;\Volume)\\
&=&\notag
\Gt(L/2+\gc (x'^1-w't'))\;\Gt(L/2-\gc (x'^1-w't'))\times\\
&&\qquad\notag
\prod_{i=2}^d\Gt(L/2+ x'^i)\;\Gt(L/2-x'^i)\\
&=&\notag
\Gt(L'/2+(x'^1-w't'))\;\Gt(L'/2-(x'^1-w't'))\times\\
&&\qquad
\prod_{i=2}^d\Gt(L/2+x'^i)\;\Gt(L/2-x'^i),
\label{e:scalar_indicator}
\ee
where $L'=L/\gc$ is the Lorentz-contracted length in $x'^1$-direction.
By means of Eq.~\eqref{e:scalar_indicator}, one can express the components of the conserved four-current $(j'^\ga)=(\gr',\bs j')=(\gc \gr,-\gc \bs w \gr)$ in $\Gs'$ as [cf. Eq~\eqref{e:current_volume_indicator}]
\bse
\be
\gr'(t',\bs x')&=&V'^{-1}\;\Ind'(t',\bs x';\Volume)\\
\bs j'(t',\bs x')&=&V'^{-1}\;\Ind'(t',\bs x';\Volume)\;\bs w'= \gr'\bs w',
\ee
\ese
with $V':=V/\gc$ denoting the Lorentz-contracted box volume in the moving frame~$\Gsm$. By inserting the expression~\eqref{e:scalar_indicator} into Eq.~\eqref{e:boosted-f}, one may perform a volume integration along the hyperplane \lq\lq $t'$=constant\rq\rq~to obtain the marginal momentum density in the moving frame as~\cite{1969VK}
\bse\label{e:pdf_invariance_box}
\be\label{e:pdf_invariance_box-a}
\phi'(\bs p')
&:=&\
\int\diff ^dx'\;f'(t',\bs x',\bs p') 
=\gc^{-1}\; \phi(\bs p(\bs p')).
\ee
This result was already obtained by Dirac~\cite{1924Di} in 1924; it is sometimes written in the equivalent  form~\cite{1969VK}\footnote{Experimentally, the momentum PDF $\phi(\bs p)$ can be determined by collecting the momentum values $\bs p_r(t)$ of the $N$ gas particles, measured at the same instant of time $t$ in $\Gs$, into a histogram. Similarly, the function $\phi'(\bs p')$ can be determined from a histogram of the momentum values $\bs p_r'(t')$, measured at the same time~$t'$ in~$\Gs'$.}
\be\label{e:pdf_invariance_box-b}
{\phi'(\bs p')}/{V'}={\phi(\bs p)}/{V}.
\ee
\ese
As anticipated above, Eqs.~\eqref{e:pdf_invariance_box}  differ from the corresponding result~\eqref{e:pdf_invariant_free} for an unconfined system of non-interacting particles. At first sight, it may seem surprising that the presence of the box alters the transformation properties of the momentum distribution. The underlying physical explanation is that the observations by $\Obs$ and $\Obs'$ are not synchronous and that in the time between their observations some particles collide with the container walls.\footnote{Cf. discussion in Section 6 of van Kampen's paper~\cite{1969VK}.} 
\par
The marginal momentum distributions present another example of non-locally defined quantities. In the history of relativity, such quantities have been a source of considerable confusion.\footnote{Gamba~\cite{1967Ga} gives a detailed list of examples, including the Lorentz contraction of length: \lq\lq Only rarely is it pointed out clearly that the measurements of the two observers ... \emph{do not refer to the same set of events}.  The \lq ends\rq\space of the rod being taken as contemporary for both observer at rest and moving observer, are in fact different points in the four-dimensional (absolute) spacetime. Once this is clearly indicated, the accepted definition [of length] is not particularly harmful.\rq\rq} Pitfalls may be avoided by keeping in mind that~\cite{1967Ga}:
\par
\emph{In general, non-locally defined quantities are \underline{not} connected by Lorentz transformations if they refer to different subsets (e.g., hyperplanes) in Minkowski spacetime. Therefore, a complete measuring instruction for a nonlocal observable must include a statement about the set of events over which the quantity is defined.}
\par
To illustrate this, we may consider the non-locally defined  vector
%
%
\bse
\be
\lan p'^\gb\ran_{t'}
&:=&
\int \diff^d x'\diff^d p'\; f'(t',\bs x',\bs p')\;p^\gb
=
\int_{t'} \diff\gs'_\ga\; \gt'^{\ga \gb}(t',\bs x'),
\label{e:em-vector-t-prime}
\ee
representing a surface integral along the hyperplane \lq\lq $t'$=constant\rq\rq.\footnote{As discussed in more detail in Appendix~\ref{appendix:surface_integrals}, for a surface integral along the hyperplane \lq\lq $t'$=constant\rq\rq~in~$\Gs'$ one finds in $\Gs'$-coordinates
$$
\int _{t'} \diff \gs'_\ga \gt'^{\ga \gb\ldots}(t',\bs x')
=\int \diff^dx'\,\gt'^{0 \gb\ldots}(t',\bs x').
$$
}
Although the average four-momentum vector  $\langle p'^\gb\rangle_{t'}$ from  Eq.~\eqref{e:em-vector-t-prime} looks quite similar to $\langle p^\gb\rangle_{t}$ from Eq.~\eqref{e:em-vector-t}, the crucial  difference is given by the fact that these two objects refer to two distinct hyperplanes, respectively, as signaled by the different labels $t$ and $t'$. For the energy-momentum  tensor~\eqref{e:stress-tensor-covariant-gas},  the rhs. of Eq.~\eqref{e:em-vector-t-prime} can be evaluated to give~\cite{1970JoLa_1}
\be\label{e:em-vector-t-prime-explicit}
\lan p'^\gb\ran_{t'}
=
\begin{cases}
\gc(\energy+w^2 \vir /d),  &\qquad \gb=0,\\
-\gc w(\energy+  \vir /d), &\qquad \gb= 1,\\
0,                      &\qquad \gb>1.
\end{cases}
\ee
\ese
Upon comparing this result with Eq.~\eqref{e:em-vector-t-c}, we note that 
$\langle p^\gb\rangle_{t}$ and $\langle p'^\gb\rangle_{t'}$ are indeed \emph{not} related by a Lorentz transformation, i.e.,
$$
\lan p'^\ga\ran_{t'}
\ne 
\langle p'^\ga\rangle_t={\Gl^\ga}_\gb\, \langle p^\gb\rangle_{t}\;.
$$ 
This inequality is, in fact, caused by the non-vanishing divergence of the energy-momentum vector in the presence of a container, cf. Eq.~\eqref{e:em-divergence}.
The actual Lorentz transformation $\langle p'^\gb\rangle_t$ of $\langle p^\gb\rangle_{t}$ is obtained by keeping the underlying hypersurface \lq\lq $t$=constant\rq\rq~fixed~\cite{1967Ga,1970Yu}, i.e.,
\bse
\be
\lan p'^\gb\ran_t
:= 
{\Gl^\gb}_\mu\int \diff^d x\; \gt^{0 \mu}(t,\bs x)
=
{\Gl^\gb}_\mu\int_t \diff\gs_\ga\; \gt^{\ga \mu} 
= 
\int_t \diff\gs_\ga'\; \gt'^{\ga \gb},
\ee
producing the expected result
\be
\lan p'^\gb\ran_{t}
=
\begin{cases}
\gc\energy,&\qquad \gb=0,\\
-\gc w\energy,&\qquad \gb= 1,\\
0, &\qquad \gb>1.
\end{cases}
\ee
\ese
With regard to experimental observations this means that, in order to determine $\langle p'^\gb\rangle_{t}$, a moving observer $\Obs'$ must first reconstruct the momentum values on the hyperplane~\lq\lq $t$=constant\rq\rq~from her data before being able to compare with the averages $\langle p^\gb\rangle_{t}$ of a lab observer $\Obs$.
\par
This gives rise to a bit of a \textit{dilemma} when attempting to formulate a relativistic thermodynamic theory that is based on global system averages such as, e.g., the total internal energy, momentum, etc.:
\begin{itemize} 
\item 
Following Planck~\cite{1908Pl} and Einstein~\cite{1907Ei}, one can build a theory based on observer-simultaneously defined quantities, as those in Eq.~\eqref{e:em-vector-t-prime-explicit}, but this leads to a loss of covariance because different observers would use different hyperplanes depending on their velocities.
\item
Alternatively, one can define thermodynamic quantities with respect to a specific, fixed hyperplane to obtain a manifestly covariant formalism~\cite{1970Yu}. In this case, however, the question arises as to which hyperplane is the most appropriate one (cf. discussion below).
\end{itemize}
These difficulties can, in principle, be avoided by formulating a 
thermodynamic field theory~\cite{1940Ec,1976Is,1981Is,1986ShMuRu,1987Is,1997KrMu,1999-lrr-1,2008Mu}  in terms of local tensorial quantities as, e.g., $j^\ga$ and $\gt^{\ga\gb}$. In view of the fact that traditional nonrelativistic thermodynamics~\cite{1974Ca,Becker} intends to describe a many-particle system by means of a few macroscopic control parameters, one could argue that such a field theoretic approach is already somewhat closer to hydrodynamics.

\subsubsection{J\"uttner gas}
\label{s:juettner_gas}

The above results apply to an arbitrary stationary momentum PDF  $\phi(\bs p)$ in the lab frame $\Gs$. With regard to Brownian motion theory, one is particularly interested in thermal equilibrium distributions. When postulating relativistic Langevin equations~\cite{1997DeMaRi,2005Zy,2005DuHa,2005DuHa_2}, these distributions must be known in advance in order to correctly specify 
the relativistic fluctuation-dissipation relation. 
Similarly, thermal equilibrium distributions are 
required as an input, if one wishes to derive Langevin-type equations from microscopic models, cf. Section~\ref{s:nonrelativistic_microscopic_models}.\footnote{Knowledge of the relativistic equilibrium distributions is essential for the correct interpretation of experimental observations in high energy and astrophysics~\cite{1998ItKoNo,2006DiDrSh,2006RaGrHe,2006HeGrRa}. Examples include thermalization processes in heavy ion collision experiments~\cite{2006HeGrRa,2006RaGrHe} and ultra-relativistic plasma beams~\cite{2006DiDrSh,2008BrEtAl}, or the relativistic Sunyaev-Zel'dovich (SZ) effect~\cite{1998ItKoNo}, describing the distortion of the cosmic microwave background (CMB) radiation spectrum due to the interaction of CMB photons with hot electrons in clusters of galaxies~\cite{1972SuZe,1984SZNature,1993SZNature}. The predicted strength of the spectral distortions and the cosmological parameters inferred from the SZ effect depend on the assumed velocity distribution of the relativistic electrons~\cite{1998ItKoNo}.}

At the beginning of the last century it was commonly accepted that a dilute (quasi-ideal) gas in equilibrium is described by the Maxwellian velocity PDF~\cite{1867Ma}
\bse\label{e:maxwell}
\be\label{e:maxwell-v}
\psi_\M({\bs  v};m,\gb,d)=
[\gb m/(2\pi)]^{d/2} 
\exp(-\gb m \bs v^2/2),
\ee
or, equivalently, by the one-particle momentum distribution 
\be\label{e:maxwell-p}
\phi_\M({\bs  p};m,\gb,d)=
[\gb/(2\pi m)]^{d/2} 
\exp[-{\gb\bs p^2}/({2 m})],
\ee
\ese
where $m$ is the rest mass of a gas particle, $\bs v=\bs p/m \in \mathbb{R}^d$ the nonrelativistic velocity, and $\Temp=(\kB \gb)^{-1}$ the temperature. After  Einstein~\cite{1905Ei_SRT1,1905Ei_SRT2} had formulated his theory of special relativity  in 1905, Planck and others noted immediately that the distribution~\eqref{e:maxwell} is in conflict with the fundamental relativistic postulate that velocities cannot exceed the light speed~$c$. In 1911 a solution to this problem was put forward by  J\"uttner~\cite{1911Ju}, who proposed to replace Maxwell's PDF by
\bse\label{e:juttner}
\be
\label{e:juttner-a}
\psi_\mrm{J}({\bs  v};m,\gb,d)
=\Z_d^{-1} m^{d}\gamma(\bs v)^{2+d}\exp[-\gb m\gamma(\bs v)]\;\Gt(1-|\bs v|),
\ee
yielding for the relativistic momentum  $\bs p=m\bs v\gc(\bs v)$ the PDF
\be
\label{e:juttner-c}
\phi_\J(\bs p;m,\gb)= \Z_d^{-1} \exp[-\gb (m^2+\bs p^2)^{1/2}].
\ee
\ese
Similar to the Maxwell distribution~\eqref{e:maxwell}, J\"uttner's distribution~\eqref{e:juttner} refers to a lab frame~$\Gs$ where the vessel, enclosing the gas, is at rest. For space dimensions $d=1,2,3$, the normalization constant
\bse
\be
\Z_d(m,\gb)=
\int\diff^dp\;\exp[-\gb (m^2+\bs p^2)^{1/2}]
\ee
can be expressed as~\cite{1911Ju}
\be
\Z_1 &=& 2m\; K_1(\gb m),\\
\Z_2 &=& 2 \pi m^2  \exp(-\gb m) (1 + \gb m)/(\gb m)^2,\\
\Z_3 &=& 4\pi m^3\; K_2(\gb m)/(\gb m),
\ee
\ese
with $K_n(z)$ denoting modified Bessel functions of the second kind~\cite{AbSt72}. The average energy per particle is obtained by logarithmic differentiation
\bse\label{a-e:Juttner-energy}
\be
\EW{\eps}_d=-\f{\p}{\p \gb}\ln \Z_d,
\ee
yielding
\be\label{a-e:Juttner-energy-b}
\EW{\eps}_1&=&m\; \f{K_0(\gb m)+K_2(\gb m)}{2K_1(\gb m)},\\
\EW{\eps}_2&=&\f{2}{\gb}+\f{m^2 \gb}{1 + m \gb},\\
\EW{\eps}_3&=&\f{3}{\gb}+m\;\f{K_1(\gb m)}{K_2(\gb m)},
\label{a-e:Juttner-energy-d}
\ee
\ese
and exhibiting the limiting behavior
\bse
\be
&&\lim_{\gb\to \infty} \EW{\eps}_d =m, \\
&&\lim_{\gb\to 0} \gb \EW{\eps}_d  =d.
\ee
\ese
Moreover, one can show that, for arbitrary space dimensions $d$ the expectation value $\EW{\bs p\cdot\bs v}$ is independent of the mass $m$; more precisely
\be\label{e:single_virial}
\EW{\bs p\cdot\bs v}=\EW{\bs p^2/\eps}=d/\gb,
\ee
which allows one to regard $\EW{\bs p\cdot\bs v}$ as a statistical thermometer (cf. discussion below).

\paragraph*{Maximum (relative) entropy principle}
J\"uttner~\cite{1911Ju} originally derived the distribution~\eqref{e:juttner} from a maximum entropy principle~\cite{1970JoLa_2,2007DuTaHa_2,2008De} by postulating that, for a sufficiently large particle number $N\gg 1$, the one-particle equilibrium distribution in phase space $\vphi(\bs x,\bs p)$ be a maximizer of the Boltzmann-Gibbs-Shannon (BGS) entropy\footnote{The BGS entropy $\ent[\phi|h^{-d}]$ from Eq.~\eqref{e:entropy_principle-a} is a relative entropy~\cite{1976Ochs,1976Ochs_2,1951KuLe,1978We,1991We} with respect to the Lebesgue measure (normalized by the constant $h^{d}$) on the $2d$-dimensional \emph{one}-particle phase space $\{(\bs x,\bs p)\}$; i.e., by writing Eq.~\eqref{e:entropy_principle} one has already fixed a specific reference measure~\cite{2007DuTaHa_2,1991We}. } 
\bse\label{e:entropy_principle}
\be\label{e:entropy_principle-a}
\ent[\phi|h^{-d}]=
-\kB \int_\Volume\diff^dx\int\diff^dp\; \vphi(\bs x,\bs p) \ln[ h^d \vphi(\bs x,\bs p)],
\ee
while satisfying the constraints
\be
\label{e:juttner_constrain_1}
1&=&\int_\Volume\diff^dx\int\diff^dp\; \vphi(\bs x,\bs p),\\
\lan \eps\ran&=&\int_\Volume\diff^dx\int\diff^dp\; \vphi(\bs x,\bs p)\;\eps,
\label{e:juttner_constrain_2}
\ee
\ese
for a given mean energy value per particle $\lan \eps\ran=\HamEn/N$. Pursuing a similar approach, J\"uttner later also derived the corresponding expressions for relativistic quantum gases obeying Bose and Fermi statistics~\cite{1928Ju}.\footnote{Alternative derivations of the J\"uttner distribution are discussed by Pauli~\cite{1921Pa},  Synge~\cite{1957Sy}, Matolcsi et al.~\cite{1996MaKrSz}, and Debbasch~\cite{2008De}.}
\paragraph*{Microcanonical approach} 
An alternative way~\cite{1921Pa,1996MaKrSz,2008De} of justifying the J\"uttner distribution~\eqref{e:juttner} is based on the microcanonical ensemble. To illustrate this, consider a dilute system of weakly interacting relativistic particles enclosed in a vessel $\Volume\subset \R^d$ of volume $V=\gl_d(\Volume)$ which is at rest in the lab frame $\Gs$. Assume that, with respect to~$\Gs$, this system can -- in some approximation -- be described by a truncated Hamiltonian of the form
\bse\label{e:MC_relativistic}
\be\label{e:relativistic_gas_hamiltonian}
H(\bs x_1,\ldots \bs p_N)=\sum_{s=1}^N(m^2+\bs p_s^2)^{1/2}=\HamEn,
\ee
complemented by the condition of elastic reflections at the boundaries of the vessel. By writing~\eqref{e:relativistic_gas_hamiltonian} one demands  that the energy stored in the interaction is negligible compared with the energy carried by the particles, so that $\HamEn$ is approximately conserved. The corresponding microcanonical $N$-particle distribution is given by the PDF
\be\label{e:MC_relativistic_PDF}
f_N^\MC(\bs x_1,\ldots, \bs p_N) =\go^{-1}
\begin{cases}
\gd(\HamEn-H(\bs x_1,\ldots \bs p_N)),&\qquad \bs x_s\in\Volume,\\
0,&\qquad \bs x_s\not\in\Volume.
\end{cases}
\ee
The normalization constant $\go=\go(\HamEn,V,N)$ can be obtained from the integrated phase volume $\Go=\Go(\HamEn,V,N)$ by performing a  differentiation with respect to $\HamEn$, i.e.,
\be
\Go&:=&\int \diff \Gc_N\; \Gt(\HamEn-H(\bs x_1,\ldots \bs p_N)),\\
\go&:=&\int \diff \Gc_N\;\gd(\HamEn-H(\bs x_1,\ldots \bs p_N))
=\f{\p \Go}{\p\HamEn},
\ee
where, for identical particles, the integration measure in $N$-particle phase space is defined by
\be
\diff\Gc_N:=
(N!h^{dN})^{-1}\;
\begin{cases}
\diff^dx_1\cdots\diff^dx_N\diff^dp_1\cdots\diff^dp_N ,
&\qquad \bs x_s\in\Volume,\\
0,&\qquad \bs x_s\not\in\Volume.
\end{cases}
\ee
\ese
The one-particle momentum PDF corresponding to $f_N^\MC$ from Eqs.~\eqref{e:MC_relativistic_PDF} is defined by
\be
\phi_{(N)}(\bs p)
&:=&\notag
\EW{N^{-1}\sum_{s=1}^N\gd(\bs p-\bs p_N)}^\MC\\
&=&\int \diff\Gc_N\;f_N^\MC(\bs x_1,\ldots, \bs p_N)\;
N^{-1}\sum_{s=1}^N\gd(\bs p-\bs p_N).
\ee
As discussed in Pauli's early review~\cite{1921Pa} and more recently also by Matolsci et al.~\cite{1996MaKrSz}, in the limit of an infinite particle number $N\to\infty$, the PDFs $\phi_{(N)}$ converge to the J\"uttner distribution $\phi_\J$ from Eq.~\eqref{e:juttner-c}. 
\par
Quite generally, a microcanonical PDF of the form~\eqref{e:MC_relativistic_PDF} gives rise to a number of rigorous \emph{equipartition theorems}~\cite{Becker,Huang,2007Ca} such as, e.g.,\footnote{In this case, the summation convention is abrogated on the lhs. of Eq.~\eqref{e:MC_equipartition}; i.e., Eq.~\eqref{e:MC_equipartition} holds separately for each $s=1,\ldots,N $ and $i=1,\ldots,d$.}
\be\label{e:MC_equipartition}
\EW{p^i_s \f{\p H}{\p p^i_s}}^\MC
=\f{\Go}{\go}
=\left[\f{\p}{\p  \HamEn} \ln \Go \right]^{-1}
=:\kB \Temp^\MC(\HamEn,N),
\ee
for each $s=1,\ldots,N $ and $i=1,\ldots,d$. The rhs. of Eq.~\eqref{e:MC_equipartition} defines ~\cite{Becker,Huang,2007Ca} the usual microcanonical temperature of the gas in the lab frame. In the case of the Hamiltonian~\eqref{e:relativistic_gas_hamiltonian}, the equipartition theorem~\eqref{e:MC_equipartition} is consistent with Eq.~\eqref{e:single_virial} upon identifying $\gb=(\kB \Temp^\MC)^{-1}$ in the limit $N\gg 1$.

\begin{figure}[b]
\centering\vspace{0.5cm}
\includegraphics[width=5.6cm,angle=0]{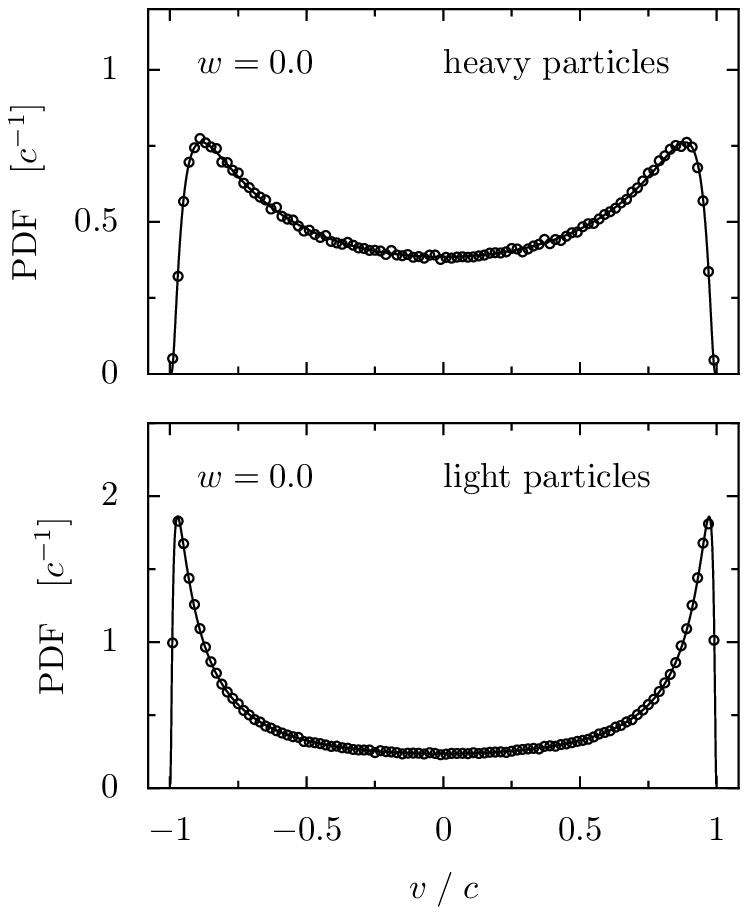}
\hspace{1cm}
\includegraphics[width=5.6cm,angle=0]{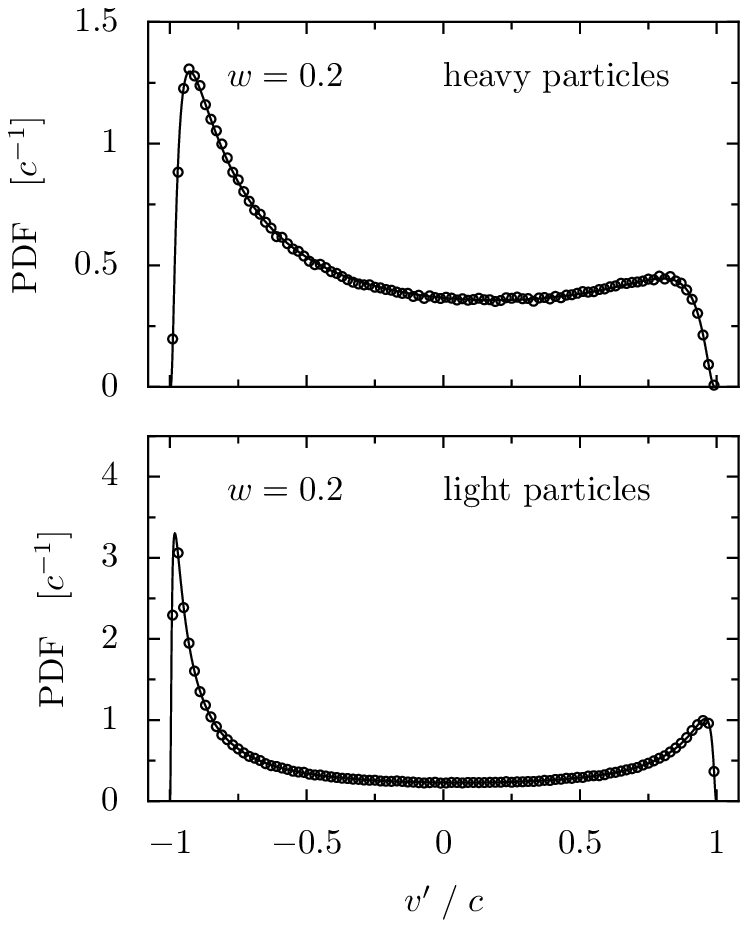}
\caption{
\label{fig01PRL} 
\emph{Left column:} Equilibrium velocity PDFs as measured by an observer with velocity $w=0$ in the lab frame $\Sigma$. The numerically obtained one-particle velocity PDFs ($\circ$) are based on a fully relativistic, deterministic 1D  molecular dynamics algorithm~\cite{2007CuEtAl} with $N_1=1000$ light particles of mass $m_1$ and $N_2=1000$ heavy particles with mass $m_2=2m_1$. Considering elastic interactions between the gas particles and elastic reflections at the boundaries, the mean energy per particle in the lab frame,   $\bar{\eps}=[\sum_{r=1}^{N_1}m_1\gc(v_r)v_r+\sum_{s=1}^{N_2}m_2\gc(v_s)v_s]/(N_1+N_2)$, remains conserved during the simulation. The solid curves in the upper and lower panel correspond to J\"uttner functions~(\ref{e:juttner-a}) with different particle masses but same  inverse temperature parameter $\gb=0.701\, (m_1c^2)^{-1}$. The latter can be determined from the value $\bar{\eps}=2.5 m_1c^2$ used in the simulation by the procedure described in~\cite{2007CuEtAl}.
\emph{Right column:} Equilibrium velocity PDFs as measured by a moving observer $\Gs'$ who travels at constant velocity $w=0.2c$ relative to the lab frame~$\Sigma$. Parameter values and initial conditions are the same as those in the left column. In contrast to the diagrams with $w=0$, the PDFs for $w=0.2$ are based $\Gs'$-simultaneously measured velocities. The solid curves represent the theoretically expected, Lorentz-boosted  J\"uttner function  $\psi_\J'(v';m,\gb,w)$ from Eq.~\eqref{e:moving}. The very good agreement with the simulation data corroborates the validity of the J\"uttner distribution. In particular, since in the stationary (equilibrium) state both particle species are characterized by the same $\gb$-value, the definition $\Temp:=(\kB \gb)^{-1}$ yields a meaningful statistical temperature concept.
}
\end{figure}

\paragraph*{Computer experiments} 
Although the J\"uttner distribution~\eqref{e:juttner} became widely accepted~\cite{1921Pa,1957Sy,1963Is,1971TeWe,1980DG,2008De},
several modifications have been discussed in the literature~\cite{1981HoScPi,1989HoShSc,2002Ka,2005Sc,2005SiLi,2006Le,2007DuHa,2007DuTaHa_2}. The validity of J\"uttner's Eqs.~\eqref{e:juttner} was recently confirmed in fully relativistic one-dimensional (1D) molecular dynamics simulations~\cite{2007CuEtAl}, see Fig.~\ref{fig01PRL},  as well as in (semi-)relativistic two and three-dimensional simulations~\cite{2006AlRoMo,2008MoGhBa}. By restricting the dynamics to one space dimension, it is possible  simulate localized elastic particle interactions in a relativistically consistent manner without needing to introduce fields. This is due to the fact that, in the 1D case, the outgoing momenta  $(\hat{p}_A,\hat{p}_B)$ of two colliding particles $A$ and $B$ are uniquely determined by the momentum values $({p}_A,{p}_B)$ before the collision by means of the relativistic energy momentum conservation laws~\cite{2007DuHa}
\bse\label{e:collision_kinematics}
\be
\eps(m_A,p_A)+\eps(m_B,p_B)&=&\eps(m_A,\hat p_A)+\eps(m_B,\hat p_B),\\
p_A+p_B&=&\hat{p}_A+\hat{p}_B,
\ee
\ese 
where $\eps(m,p)=(m^2+p^2)^{1/2}$. In higher space dimensions, additional assumptions about the spatio-temporal structure of the collision processes need to be made~\cite{1949WhFe,1963CuJoSu,1965DaWi,1966DaWi,1978Ko_b,1984MaMuSu}.
In order to be fully consistent with the requirements of special relativity, relativistic  interactions $d>1$ space dimensions must be formulated in terms of fields. Direct simulation of the field dynamics is numerically expensive and, therefore,  practically infeasible in most cases. Alternatively, one can use simplified semi-relativistic models  such as, e.g., effective hard-sphere models where the interaction radius is defined with respect the center-of-mass frame of the colliding particles. Generally, it can be expected that such simplified models yield satisfactory results in the low density regime~\cite{2006AlRoMo}, but they may lead to inconsistencies at high densities, e.g., when three-body encounters become relevant.

In addition to (in)validating theoretically predicted distribution functions,  numerical simulations can be useful for illustrating the meaning of concepts such as \lq temperature\rq\space and \lq thermal equilibrium\rq\space in special relativity. Figure~\ref{fig01PRL}  shows results of a 1D relativistic gas simulation as described in~\cite{2007CuEtAl}. The simulated gas consists of $N_1$ \lq light\rq\space particles (mass $m_1$) and $N_2$ \lq heavy\rq\space particles (mass $m_2>m_1$). Particle collisions in the gas are governed by the conservation laws~\eqref{e:collision_kinematics} and collisions with the boundaries  are elastic in the lab frame (i.e., $p\to-p$ in $\Gs$). As evident from  Fig.~\ref{fig01PRL}, the numerically obtained stationary velocity distributions are very well matched by 1D J\"uttner functions~\eqref{e:juttner} with different masses but \emph{same} parameter $\gb$. Hence, the J\"uttner distribution not only agrees with the numerical data, it also yields a well-defined concept of \lq temperature\rq\space in the lab frame: Intuitively, the temperature $\Temp$ is thought to be an intensive quantity that equilibrates to a common value if two or more systems are brought into contact with each other (i.e., may exchange different forms of energy). In the example considered here, it is natural to regard the particle species as two different subsystems that may exchange energy via elastic collisions. After a certain relaxation time, the combined system approaches a \lq thermodynamic equilibrium state\rq, where each subsystem is described by the same asymptotic, two-parametric velocity PDF $\psi_\J(v;m_i,\gb)$, differing only via the rest masses~$m_i$. The commonly shared distribution parameter $\gb$ may thus be used to \emph{define} a relativistic (statistical) equilibrium temperature in $\Gs$ by 
\be
\Temp:=(\kB \gb)^{-1},
\ee
in agreement with the interpretation of $\gb$ in the nonrelativistic case.
\par
However, for this concept of temperature to be meaningful, a restriction of the accessible spatial volume is required -- be it by means of periodic boundary conditions, by imposing reflecting walls or other types of  confinements~\cite{1971Ho_2}.\footnote{The critical role of the boundary conditions in relativistic systems has been emphasized by Sinyukov~\cite{1983Si} and van Kampen~\cite{1968VK,1969VK}. Loosely speaking, if a many-particle system has reached a universal stationary equilibrium state, then each particle \lq knows\rq\space about the presence of the confinement/walls because equilibration typically requires momentum reversal at the walls, e.g., in order to maintain a uniform density and a well-balanced average collision frequency. In particular, if the walls are considered to be stationary objects then they single out a preferred frame of reference. The relevance of the boundary conditions is even more obvious in quantum mechanics/statistics due to their direct effect on the energy spectra and, thus, on the density of states~\cite{1928Ju}.} Otherwise, it cannot be expected that a many-particle system approaches a universal stationary equilibrium state  independent of the specific initial conditions.
\par
The simulations can also be used to determine the equilibrium velocity distributions as seen from another frame $\Sigma'$, moving with velocity $w$ relative to the lab frame~$\Sigma$, see Fig.~\ref{fig01PRL}. In contrast to the diagrams for $w=0$, the numerical data points for  $w=0.2$ are obtained by measuring the particle velocities $v_s'$ simultaneously with respect to $\Sigma'$. The solid curves in the right column of Fig.~\ref{fig01PRL} correspond to the PDF
\be\label{e:moving}
\psi_\J'(v';m,\gb,w)=\gc(w)^{-1}
\frac{m\,\gamma(v')^{3}}{ \Z_1}
\exp[-\gb \gc(w) m\gamma(v')\; (1+w v')],
\ee
obtained by making use of transformation law~\eqref{e:pdf_invariance_box}. The excellent agreement between the simulation data and Eq.~\eqref{e:moving} confirms the validity of Eq.~\eqref{e:pdf_invariance_box} and, thereby, also the scalar Lorentz transformation behavior of the phase space PDF $f(t,\bs x,\bs p)$~\cite{1969VK}.
Moreover, for the model under consideration, one may state more precisely: Two relativistic gas components are in \lq thermodynamic equilibrium\rq\space for any observer if their one-particle velocity PDFs are given by generalized J\"uttner functions~\eqref{e:moving} with same parameters $\gb$ \emph{and} $w$. It is only in this case that the net energy-momentum transfer between the different gas components in the container vanishes~(a more general version of this criterion is discussed in Ref.~\cite{1968VK}).
\par
\begin{figure}[t]
\centering
\vspace{0.5cm}
\includegraphics[width=6.7cm,angle=0]{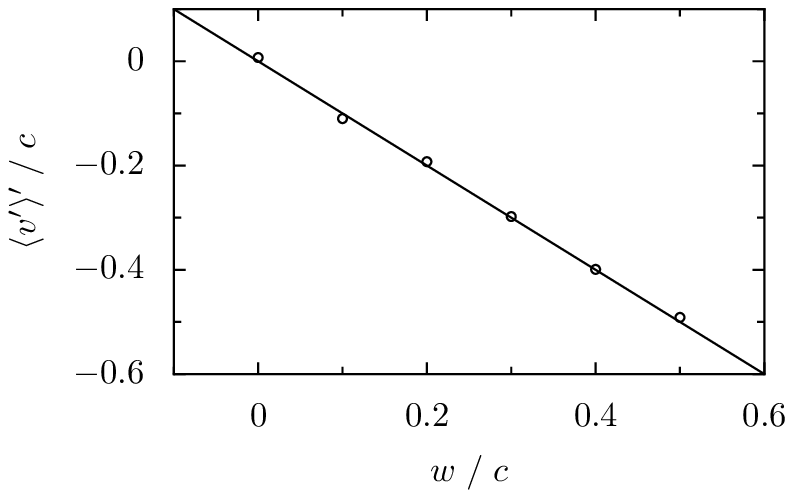}
\hspace{0.4cm}
\includegraphics[width=6.5cm,angle=0]{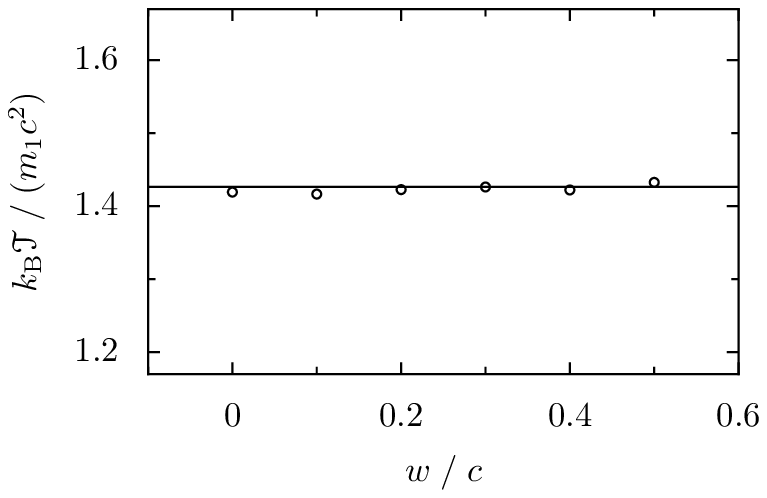}
\caption{
\label{fig03_extra} 
Measured mean particle velocity ($\circ$, left digram) and estimated temperature ($\circ$, right diagram) based on Eqs.~\eqref{e:thermometer+velo} as a function of the observer velocity $w$. Solid lines correspond to the theoretically expected values, respectively, using the same simulation parameters and initial conditions as in Fig.~\ref{fig01PRL}.
}
\end{figure}

The numerical simulations illustrate that $\Temp=(\kB\gb)^{-1}$ yields a useful statistical temperature concept in the lab frame $\Gs$, defined by the boundary (container box). The question as to which temperature value is measured by a moving observer $\Obs '$ cannot be uniquely answered -- the answer depends on the respective definitions (i.e., measurement conventions) for heat, internal energy, etc. as   employed by $\Obs'$. These aspects are discussed in more detail in the Appendix~\ref{s:relativistic_thermodynamics}. At this point, it suffices to mention that $\Temp:=(\kB \gb_\J)^{-1}$ can also be determined from $t'$-simultaneously sampled velocity data (right column of Fig.~\ref{fig01PRL}) by means of the expectation value\footnote{Equation~\eqref{e:thermometer} is obtained by combining the microcanonical equipartition theorem for a Hamiltonian $H=\sum_{i=1}^{N_1}E(m_1,p_i)+ \sum_{j=1}^{N_2}E(m_2,p_j)$ with the Lorentz invariance of the relativistic phase space PDF $f$.}~\cite{1967La}
\bse\label{e:thermometer+velo}
\be
\label{e:thermometer}
\kB \Temp=m\gc(w)^3\left\langle {\gc(v')\;(v'+ w)^2} \right\rangle_{t'},
\ee
where 
\be
w=-\langle v'\rangle_{t'}
\ee
\ese
is the mean velocity of the gas measured by the moving observer. One can test the validity of Eq.~\eqref{e:thermometer} explicitly by using simulation data obtained for different values of $w$, see Fig.~\ref{fig03_extra}. Hence, in this sense,  Eqs.~\eqref{e:thermometer+velo} defines a Lorentz invariant statistical gas thermometer -- when adopting this thermometer a moving body appears   neither hotter nor colder~\cite{2007CuEtAl}.\footnote{The mean value from Eq.~\eqref{e:thermometer} can be used to measure the rest temperature, which plays a central role in van Kampen's~\cite{1968VK} approach to relativistic thermodynamics. Evidently, upon multiplying Eq.~\eqref{e:thermometer} by factors $\gc(w)^\ga$, $\ga\ne 0$ one can construct thermometers that measure \lq other\rq\space temperatures; e.g., $a=-1$ would correspond to Planck's~\cite{1908Pl} formulation of relativistic thermodynamics and $a=1$ to proposals made by Eddington~\cite{1923Ed} and Ott~\cite{1963Ott}.}   

We conclude this part with another general remark: 
The above discussion has focused on the case where an \lq ordinary\rq\space thermal equilibrium state is approached, corresponding to a stationary \lq exponential\rq\space one-particle momentum distribution of the J\"uttner type~\cite{1911Ju,1928Ju}. In general, one could also imagine non-equilibrium scenarios that give rise to (quasi-)stationary momentum  distributions which differ from the J\"uttner function (see, e.g, \cite{2002Ka,2005Ka,2006Ka,2005SiLi,2006Si,2007LiJiZh,2004Be}). When attempting to develop a thermodynamic formalism for such non-standard distributions the entropy functional needs to be modified appropriately~\cite{2007Ca}. However, our subsequent discussion of relativistic Langevin equations will concentrate on the case where the heat bath, which surrounds the Brownian particle, is described by a spatially homogeneous J\"uttner function.


\section{Relativistic Brownian motion processes in phase space}
\label{s:RBM_lab_time}

The present section describes the generalization of the Langevin theory of Brownian motions to the framework of special relativity~\cite{1997DeMaRi,1998DeRi,2001BaDeRi,2001BaDeRi_2,2004Zy,2005Zy,2005DuHa,2005DuHa_2,2006DuHa,2007DuHa,2007DuTaHa_2,2006Fa,2007FrLJ,2007AnFr,2007Fi,2007Li,2008ChDe,2007ChDe_1,2007ChDe_2,2008Ba_1,2008Ba_2}. More precisely, we will consider stochastic differential equations (SDEs) that describe Markov processes in relativistic one-particle phase space.\footnote{A main motivation for considering processes $(\bs X(t),\bs P(t))$  rather than \lq pure\rq\space position processes $\bs X(t)$ is that 
in, in the relativistic case,  it is impossible to define nontrivial relativistic Markov processes $\bs X(t)$ in Minkowski spacetime; cf.  {\L}opusza{\`n}ski~\cite{1953Lo}, Dudley (Theorem~11.3 in~\cite{1965Du}),  Hakim (Proposition~2 in~\cite{1968Ha}), and also the discussion in Section~\ref{s:relativistic_diffusion}.} Relativistic Langevin equations present a useful tool for modeling the dynamics of relativistic particles in a random environment. Examples include the analysis of thermalization effects in relativistic plasmas~\cite{2006HeGrRa,2006HeGrRa_2,2006RaGrHe,2008RaHe} and astrophysical systems~\cite{2006DiDrSh,2007Za}. Here, we would like to provide a comprehensive introduction to the underlying mathematical theory.
\par
From a conceptual perspective, one can distinguish two complementary approaches towards modeling relativistic stochastic processes in one-particle phase space: The first approach is based on postulating  evolution equations for the (transition) probability density of the stochastic process, as e.g., integro-differential master-type equations or Fokker-Planck-type partial differential equations. Within a relativistic framework, the condition of subluminal particle propagation imposes stringent constraints on the structure of the evolution equations. These aspects were studied in the early pioneering works of  {\L}opusza{\`n}ski~\cite{1953Lo}, Schay~\cite{1961Sc} and  Hakim~\cite{1968Ha}. Alternatively, one may start by postulating SDEs as  phenomenological model equations and subsequently derive the evolution equations for the associated (transition) probability densities. Here, we are going to pursue the latter route for mainly two reasons: We will be able to  directly compare with the nonrelativistic case discussed in  Section~\ref{s:nonrelativistic} and, more importantly, the physical origin  of the stochastic dynamics --  the  interaction with a complex  background medium --  is \lq less hidden\rq\space within the Langevin approach.\footnote{For a detailed analysis of the complementary approach we refer to Hakim~\cite{1968Ha}.}
\par
According to our knowledge, Debbasch et al.~\cite{1997DeMaRi} were the first to propose relativistic Langevin equations as models for relativistic Brownian motions. More precisely, these authors constructed a relativistic Ornstein-Uhlenbeck process (ROUP) by postulating an additive Gaussian white noise force in the rest frame of the bath while adapting the friction such that the stationary momentum distribution is given by a  J\"uttner function.  During the past decade, various properties of the ROUP were studied by Debbasch and Rivet~\cite{1998DeRi}, Barbachoux et al.~\cite{2001BaDeRi,2001BaDeRi_2}, and Zygad{\l}o~\cite{2005Zy}.  As will be illustrated below, the ROUP represents a special limit case of a larger class of relativistic Langevin models~\cite{2006DuHiHa,2007AnFr,2008ChDe,2007ChDe_1,2007ChDe_2}, which also includes the relativistic Brownian motion (RBM) process proposed in ~\cite{2005DuHa,2005DuHa_2}.  Processes belonging to this class share the common feature  that their stationary momentum distribution is given by a J\"uttner function, but they may exhibit a significantly different relaxation behavior, resulting e.g. in a different temperature dependence of the spatial diffusion constant~\cite{2007AnFr}.
\par
The present section is structured as follows. Section~\ref{s:labframe} introduces the axiomatic Langevin approach, i.e., suitable SDEs and fluctuation-dissipation relations are postulated in order to provide a simplified model of the complex interaction between the Brownian particle and its environment (heat bath).
 Specific example processes will be analyzed in Section~\ref{s:RBM_examples}.   Section~\ref{s:asymptotic_MSD} discusses  the temperature dependence of the asymptotic mean square displacement for three different relativistic Brownian motion processes~\cite{2007Li,2007AnFr}. It is demonstrated that the diffusion constant is sensitive to variations of the friction coefficients. This implies that measurements of the diffusion constant may reveal information about the underlying microscopic interactions. In Section~\ref{s:moving_observer}  relativistic Brownian motion processes will be analyzed from the viewpoint of a moving observer.  The  relativistic generalization of the binary collision model from Section~\ref{s:nonrelativistic_collision_model} is discussed in  Section~\ref{s:relativistic_binary_collision_model}, representing a simple microscopic model for relativistic Brownian motions.

\subsection{Relativistic Langevin and Fokker-Planck equations}
\label{s:labframe}
\label{s:concepts}

When considering Langevin equations as models of Brownian motion, one implicitly assumes 
that it is possible and reasonable to separate the degrees 
of freedom of  the Brownian particle from those of the environment (heat bath). Adopting this point of view, one can specify two distinguished frames of reference:  the lab frame $\Gs$, defined as the inertial restframe of the heat bath, and the inertial frame $\Gs_*$ that is comoving with the Brownian particle at a given instant of time.\footnote{Apart from an irrelevant shift of the origin, the inertial lab frame $\Gs$ is uniquely determined by the requirement that the mean velocity of the heat bath particles, which is assumed to be constant in any inertial frame, vanishes in $\Gs$. Similarly, the instantaneously comoving frame $\Gs_*$ is determined by the condition that the $\Gs_*$-velocity of the Brownian particle is equal to zero at the given instant of time. Generally, we assume that the time coordinates $t$ and $t_*$ can be measured, e.g., by using atomic clocks that are at rest in $\Gs$ or  $\Gs_*$, respectively.} To start with, we consider the question how one can model the dynamics of a relativistic Brownian particle in the lab frame~$\Gs$.

\subsubsection{Relativistic Langevin equations: general construction principles}

Roughly speaking, a relativistic Brownian motion process is a stochastic process whose absolute velocity $|\bs V(t)|=|\diff \bs X/\dt|$ does not exceed the light speed  $c=1$ at any time; i.e., the process must satisfy the condition
\be\label{e:subluminal}
|\bs V(t)|
= \f{|\bs P|}{P^0}
= \f{|\bs P|}{(M^2+ \bs P^2)^{1/2}}\le 1
\qquad\qquad \fa t,
\ee
where $M>0$ denotes the rest mass of the Brownian particle.
Of particular interest here are processes can be modeled by means of SDEs  similar to the nonrelativistic Langevin equations~\eqref{e:langevin_m}. The basic idea for constructing such processes is to couple the noise sources (e.g., Wiener processes) only to the relativistic momentum  $\bs P =(P^i)$, $i=1,\ldots,d$, which can take values in whole $\R^d$. By doing so, the condition~\eqref{e:subluminal} is automatically fulfilled. However, before one can actually write down specific SDEs for the relativistic momentum components $P^i$, an additional question needs to be addressed, namely~\cite{1968Ha,2006De}:

\paragraph*{The choice of the time parameter}
A fundamental assumption (postulate) of nonrelativistic Galilean physics is the existence of a universal time~$t$. Therefore, within the nonrelativistic Langevin theory, it is quite natural to identify this universal time~$t$ with the time parameter of the stochastic driving process, which is often taken to be a multi-dimensional Wiener process~ $\bs B(t)$.
By contrast, in special relativity the notion of time becomes frame-dependent and it is important to specify in advance which time parameter is used to quantify the fluctuations of the stochastic driving process.
\par 
When considering the stochastic motion of a relativistic Brownian particle, two characteristic time parameters can be distinguished: 
The coordinate time~$t$ of the inertial lab frame~$\Gs$,  which may be interpreted as the proper time of the heat bath, and the proper time~$\tau$ of the Brownian particle. In principle, either of the two parameters could be used to formulate SDEs for the spatial components of the particle momentum, $\bs P =(P^i)$. However, within the conventional Langevin picture of Brownian motion, one usually considers friction and noise as externally imposed forces that act upon the Brownian particle and reflect the fluctuations within the heat bath. Therefore, it seems more natural to characterize the statistical properties of the noise source in terms of the lab time~$t$. However, as we shall discuss in Section~\ref{s:propertime} one can reparameterize a given lab-time Langevin equation in term of the associated  proper-time.\footnote{Within this lab-time approach, the proper time becomes a stochastic quantity, and one could, for example, ask for the probability of finding  at lab time $t$ the particle's proper time in the interval $[\tau,\tau+\dtau]$. Conversely, if adopting the proper time $\tau$ as the primary deterministic evolution parameter, one could ask for the probability to find the particle at proper time $\tau$ within the spacetime interval $[t,t+\dt ]\times [\bs x,\bs x+\dbx]$ with respect to the lab frame. The proper-time approach was studied in Refs.~\cite{1965Ha,1968Ha,2005OrHo} and most recently in \cite{2008DuHaWe}.}

\paragraph*{Lab-time Langevin equation} 
Within the \emph{lab-time approach}~\cite{1997DeMaRi,1998DeRi,2001BaDeRi,2001BaDeRi_2,2004Zy,2005Zy,2005Ri,2005DuHa,2005DuHa_2} one aims at constructing  $t$-parameterized  $2d$-dimensional stochastic processes $\left\{\bs X(t), \bs P(t)\right\}$ with respect to the lab frame $\Gs$, where the position coordinates $\bs X=(X^i)$ and the spatial momentum coordinates $\bs P=(P^i)$ are connected by the standard relativistic differential relation
\bse\label{e:RLE}
\be\label{e:RLE_position}
\dX^i(t)
=(P^i/{P^0})\; \dt,
\ee
with $P^0(t)=E(t)=(M^2+ \bs P^2)^{1/2}$ denoting the relativistic energy of the particle and $V^i(t)=P^i/P^0$ its velocity components in $\Gs$. Stochasticity is implemented into the dynamics by coupling the momentum components $P^i(t)$ to an external noise source via the SDEs\footnote{E.g., for the post-point rule \lq\lq$\odot=\bullet$\rq\rq, Eqs.~\eqref{e:RLE} constitute a special case of the general SDE~\eqref{e:HK-SDE_d}, upon identifying $n=2d$, $\bs Y=(X^1,\ldots,X^d,P^1,\ldots,P^d)$ and
$
A^i={P^i}/{P^0} 
,\;
{C^i}_j=0,\;
A^{i+d}=-{a^i}_j P^j
,\;
{C^{i+d}}_j={c^i}_j
$ for $i=1,\ldots,d$.} 
\be\label{e:RLE_momentum}
\dP^i(t)
= \Force^i \dt-{a^i}_j\;P^j\; \dt +{c^i}_j\odot \diff B^j(t),
\ee
\ese
where 
\begin{itemize}
\item 
$\Force^i$ denotes a deterministic external force (e.g., Lorentz force), $-{a^i}_j\;P^j$ is a phenomenological friction force, and the last term represents noise;
 \item 
$\odot\in\{*,\circ,\bullet\}$ signals the discretization rule (i.e., the stochastic integral definition, cf. Appendix~\ref{as:stochastic_calculus});
\item
in general, the functions $\Force^i$, ${a^i}_j$ and ${c^i}_j$ can depend on $(t,\bs X, \bs P)$;
\item
the noise source is usually modeled by the $d$-dimensional standard Wiener\footnote{In principle, one could also consider other driving process (as, e.g., Levy or Poisson processes) but we are not aware of such studies.} process~$\bs B(t)=(B^1(t),\ldots,B^d(t))$ in a relativistic context.
\end{itemize}

Equations~\eqref{e:RLE} govern the spatial components of the four-vectors $(X^0,\bs X)$ and $(P^0,\bs P)$. One can still add an equation for time component $X^0$ by setting
\bse\label{e:RLE_zero}
\be \label{e:RLE_time}
\dX^0(t)
=(P^0/{P^0})\; \dt=\dt,
\ee
The equation for the energy component  $P^0$ can be derived from  Eq.~\eqref{e:RLE_momentum} by applying the (backward) Ito  formula, cf. Appendix~\ref{as:stochastic_calculus}, to the mass-shell condition $P^0(t)=(M^2+\bs P^2)^{1/2}$, yielding 
\be
\dP^0(t)
&=&\notag
\biggl\{
(\Force_i- a_{ij}P^j)\f{P^i}{P^0}+
\gl_\odot\;\f{D_{ij}}{2}\biggl[\f{\gd^{ij}}{P^0}- \f{P^i P^j} {(P^0)^3}\biggr]
\biggr\}\dt +\\
&&\;\label{e:RLE_energy}
\f{P^i}{P^0}\; c_{ir}\odot \diff B^r(t),
\ee
\ese
where$$
\Force_i:=\Force^i
\csp
D_{ij}:=D^{ij}:={c^i}_r {c^j}_r
\csp
c_{ir}:={c^i}_r,
$$
and, depending on the discretization rule,
$$
\gl_*=1
\csp
\gl_\circ=0
\csp
\gl_\bullet=-1.
$$
The $\gl_\odot$-term in Eq.~\eqref{e:RLE_energy} reflects the modification of differential calculus for Ito and backward-Ito SDEs, cf. Appendix~\ref{as:stochastic_calculus}.
\par
Finally, we still note that the proper-time process associated with Eqs.~\eqref{e:RLE} is defined by
\be
\dtau(t)=(M/P^0)\,\dt.
\ee
In Section~\ref{s:propertime}, we shall discuss how the lab-time Langevin equations~\eqref{e:RLE} can be reparameterized in terms of $\tau$.

\subsubsection{Fokker-Planck equations}
\label{s:relativistic_FPE}
In the case of the Ito rule \lq\lq$\odot=*$\rq\rq\space in Eq.~\eqref{e:RLE_momentum}, the corresponding Fokker-Planck equation (FPE) is given by [cf. Eq.~\eqref{e:HK-FPE_d}] 
\bse\label{e:rel-FPE}
\be\label{e:a-FPE-high-1-ito}
\left(\f{\p}{\p t} +\f{p^i}{p^0}\f{\p}{\p x^i}\right) f_*
&=&
\f{\p}{\p p^i}\left[ 
(-\Force^i+{a^i}_j p^j )f_* +\f{1}{2}
 \f{\p}{\p p^k}({c^i}_r {c^k}_r f_*)\right], 
\ee
in the case of the Stratonovich-Fisk mid-point rule \lq\lq$\odot=\circ$\rq\rq\space by
\be\label{e:a-FPE-high-1-sf}
\left(\f{\p}{\p t} +\f{p^i}{p^0}\f{\p}{\p x^i}\right) f_\circ
&=&
\f{\p}{\p p^i}\left[ 
\left(-\Force^i+{a^i}_j p^j - 
\f{1}{2} {c^k}_r \f{\p}{\p p^k} {c^i}_r    
\right) f_\circ +
\f{1}{2}
 \f{\p}{\p p^k}({c^i}_r {c^k}_r  f_\circ)\right]
 \qquad
\qquad
\ee
and in the case of the post-point (backward Ito) rule \lq\lq$\odot=\bullet$\rq\rq\space by
\be\label{e:a-FPE-high-1-pp}
\left(\f{\p}{\p t} +\f{p^i}{p^0}\f{\p}{\p x^i}\right) f_\bullet
&=&
\f{\p}{\p p^i}\left[ 
\left(-\Force^i+{a^i}_j p^j - 
{c^k}_r \f{\p}{\p p^k} {c^i}_r    
\right) f_\bullet +
\f{1}{2}
 \f{\p}{\p p^k}({c^i}_r {c^k}_r  f_\bullet)\right]
  \qquad
\qquad
\ee
with $f_\odot(t,\bs x,\bs p)$ denoting the phase space PDF of the relativistic Brownian particle in $\Gs$, and $p^0=(M^2+\bs p^2)^{1/2}$, respectively. 
\par
Deterministic initial data $\bs X(0)=\bs x_0$ and $\bs P(0)=\bs p_0$ corresponds to the localized initial condition 
\be
f_\odot(0,\bs x,\bs p)=\gd(\bs x-\bs x_0)\;\gd(\bs p-\bs p_0).
\ee 
\ese
Due to the linearity of Eqs.~\eqref{e:rel-FPE}, more general solutions can be obtained by integrating the solutions of Eqs.~\eqref{e:rel-FPE} over a non-localized initial distribution 
$f_0(\bs x_0,\bs p_0)$. However, it should be noted that, within a relativistic framework, it is very difficult to determine non-localized lab-isochronous initial data by means of experimental measurements.

\subsubsection{Free motions in an isotropic bath and Einstein relations}
Similar to the nonrelativistic case, physical constraints on the coefficient functions ${a^i}_j$ and ${c^i}_j$ come from symmetry properties of the heat bath and from the requirement that Eqs.~\eqref{e:RLE} must reproduce the correct stationary distribution and the correct relaxation behavior. For example, in the absence of external force-fields, $\Force^i\equiv 0$, and  if the heat bath is stationary, isotropic and homogeneous in the lab frame $\Gs$, the coefficient matrices take the simplified diagonal form
\be\label{e:a-isotropic}
{a^i}_j=\ga\, {\gd^i}_j
\csp
{c^i}_j=(2D)^{1/2}\,{\gd^i}_j.
\ee
Under the stated assumptions, the functions $\ga$ and $D$ depend only on the Brownian particles' \emph{absolute} momentum -- or, equivalently, on its relativistic energy $p^0=(M^2+\bs  p^2)^{1/2}$ i.e., $\ga(\bs p)=\hat{\ga}(p^0)$ and $D(\bs p)=\hat{D}(p^0)$. 
In this case, the relativistic Langevin equations~\eqref{e:RLE} simplify to
\bse\label{e:REL_bath_3D_diagonal}
\be 
\dX^i(t)&=&({P^i}/{P^0})\;\dt,\\
\dP^i(t)&=&-\hat{\ga} P^i \dt+(2\hat{D})^{1/2} \odot \dB^i(t),
\ee
\ese
and the Fokker-Planck equations~\eqref{e:rel-FPE} take the form
\bse\label{e:a-FPE-high-2}
\be 
\left(\f{\p}{\p t} +\f{p^i}{p^0}\f{\p}{\p x^i}\right) f_*
&=&
\f{\p}{\p p^i}\left[ 
\hat{\ga} p^i f_* +
\f{\p}{\p p^i}(\hat{D} f_*)\right],
\\
\left(\f{\p}{\p t} +\f{p^i}{p^0}\f{\p}{\p x^i}\right) f_\circ
&=&
\f{\p}{\p p^i}\left[ 
\hat{\ga} p^i f_\circ +
\hat{D}^{1/2}\f{\p}{\p p^i} ( \hat{D}^{1/2}f_\circ)\right],
\\
\left(\f{\p}{\p t} +\f{p^i}{p^0}\f{\p}{\p x^i}\right) f_\bullet
&=&
\f{\p}{\p p^i}\left[ 
\hat{\ga} p^i f_\bullet +
\hat{D}\f{\p}{\p p^i} f_\bullet\right],
\ee
\ese
An additional constraint on the functions $\ga(p)=\hat{\ga}(p^0)$ and $D(p)=\hat{D}(p^0)$ arises from thermostatistical considerations: If the motion of the Brownian particle is restricted to a finite volume $\Volume\subset\R^d$ and if the heat bath is in a thermal equilibrium state at temperature $\Temp=(\kB  \gb)^{-1}$, then the expected stationary solution $f_\infty(\bs x,\bs p)$ of Eqs.~\eqref{e:rel-FPE} is a spatially homogeneous J\"uttner distribution~\cite{1911Ju}\footnote{Cf. the discussion in Section~\ref{s:juettner_gas}.},
\be\label{e:RBM-juettner}
f_\infty(\bs x,\bs p)=\mcal{N}\; \exp[-\gb(M^2+\bs p^2)^{1/2}]\;
\Ind(\bs x;\Volume),
\ee
with $\Ind(\bs x;\Volume)$ being the indicator function of the accessible volume $\Volume$ as defined in Eq.~\eqref{e:indicator_definition}. 
By inserting Eq.~\eqref{e:RBM-juettner} into the Fokker-Planck equations~\eqref{e:rel-FPE} one finds that, depending on the discretization rule, the functions $\ga(p)=\hat{\ga}(p^0)$ and $D(p)=\hat{D}(p^0)$ must satisfy the
generalized fluctuation-dissipation  relations~\cite{1997DeMaRi,2005DuHa,2005DuHa_2} 
\bse\label{e:Einstein_relativistic}
\be 
\odot=*:\qquad
0 &\equiv& 
\hat{\ga}(p^0)\; p^0-\gb\; \hat{D}(p^0)+\hat{D}'(p^0) ,\\
\odot=\circ:\qquad
0 &\equiv& 
\hat{\ga}(p^0)\; p^0-\gb\; \hat{D}(p^0)+\hat{D}'(p^0)/2 ,\\
\odot=\bullet:\qquad
0 &\equiv& 
\hat{\ga}(p^0)\; p^0 - \gb\; \hat D(p^0).
\label{e:Einstein_relativistic_c}
\ee
\ese
where $\hat{D}'(p^0):=\diff\hat{D}(p^0)/\diff p^0$. 
Equations~\eqref{e:Einstein_relativistic} are also referred to as the \emph{relativistic Einstein relations}. 
In particular, by comparing Eq.\eqref{e:Einstein_relativistic_c} with the nonrelativistic Einstein relation~\eqref{e:generalized_Einstein_relation}, we note that the mass has been replaced with energy $p^0$ in the relativistic case.

\subsection{One-dimensional examples and mean square displacement}
\label{s:RBM_examples}

In this part, we will consider one-dimensional example processes ($d=1$). Their generalization to higher space dimensions is straightforward but analytic calculations become more tedious. We first summarize the SDEs for the energy process $P^0(t)$ and the velocity process $V(t):=P(t)/P^0(t)$ in  Section~\ref{s:energy_equation}. Subsequently, analytical and numerical results for the asymptotic diffusion constants of specific example processes will be discussed~\cite{2007Li,2007AnFr}.

\subsubsection{Discretization rules, energy and velocity equations}
\label{s:energy_equation}

Considering an isotropic thermalized heat bath in $d=1$ space dimensions and the post-point discretization, Eq.~\eqref{e:REL_bath_3D_diagonal} takes form
\bse\label{e:RLE_post}
\be\notag
\diff X(t) &=&(P/P^0)\;\dt,\\
\label{e:RLE_post-b}
\diff P(t) &=&-\ga_\bullet(P)\, P\;\dt +[2D(P)]^{1/2}\bullet \diff B(t),
\ee
where $B(t)$ is characterized by the Gaussian distribution~\eqref{e:OUP_langevin_math_density} and $\ga_\bullet(p)=\hat{\ga}_\bullet(p^0)$ and $D(p)=\hat{D}(p^0)$. Similar to the nonrelativistic case, one can replace Eq.~\eqref{e:RLE_post-b} by the equivalent Stratonovich-Fisk SDE\footnote{See also the corresponding discussion by H\"anggi~\cite{1980Ha}, and H\"anggi and Thomas (page 293 of Ref.~\cite{1982HaTh}).}
\be\label{e:RLE_mid} 
\diff P(t)&=&
-\ga_\circ(P)\, P\, \dt+[2D(P)]^{1/2} \circ \diff B(t)\\ 
\notag
\ga_\circ(p)&:=&\ga_\bullet(p)-D'(p)/(2p),
\ee
or by the equivalent Ito SDE 
\be\label{e:LE_relativistic_Ito 1D}
\diff P(t)&=&
-\ga_*(P)\, P\, \dt+[2D(P)]^{1/2} *\diff B(t)\\ 
\ga_*(p)&:=&\ga_\bullet(p)-D'(p)/p, 
\notag 
\ee
\ese
where $D'(p):=\diff D(p)/\diff p$. Compared with Eqs.~\eqref{e:RLE_post-b} and~\eqref{e:RLE_mid}, the Ito form~\eqref{e:LE_relativistic_Ito 1D} is most convenient for numerical simulations.
\par
Imposing that the solution of corresponding FPE~\eqref{e:a-FPE-high-2} be given by a one-dimensional J\"uttner function of the form~\eqref{e:RBM-juettner}, the 
friction and noise coefficients must satisfy the  Einstein relations~\eqref{e:Einstein_relativistic}. In terms of the friction coefficients $\ga_\odot(p)$, the Einstein relations can also be rewritten as
\bse
\be
0
&\equiv & \label{e:Einstein_relativistic_hk}
\ga_\bullet(p)\; p^0-\gb D(p)
\\
0
&\equiv& \label{e:Einstein_relativistic_mid}
\ga_\circ(p)\; p^0 -\gb D(p)+  D'(p)\; p^0/(2p)
\\
0
&\equiv& \label{e:Einstein_relativistic_ito}
\ga_*(p)\; p^0-\gb D(p)+  D'(p)\; p^0/p.
\ee
\ese
From  Eq.~\eqref{e:RLE_energy} one obtains  for the relativistic energy process $P^0(t):=(M^2+P^2)^{1/2}$ the following SDE
\be
\diff P^0(t)
&=&
\notag 
\left\{-\hat{\ga}_\odot(P^0)\,P^0 \left[1-\left(\f{M}{P^0}\right)^2\right] +\gl_\odot
\f{\hat{D}(P^0)}{P^0}\left(\f{M}{P^0}\right)^2\right\}\diff t + \\
&&
\left\{2\,\hat{D}(P^0)
\left[1-\left(\f{M}{P^0}\right)^2\right]\right\}^{1/2}\odot \dB(t),
\ee
where $\gl_*=1$, $\gl_\circ=0$, and $\gl_\bullet=-1$. Furthermore, by defining for $P(V)=MV\;(1-V^2)^{-1/2}$ new coefficients
$$
\tilde{\ga}(V):=\ga(P(V))
\csp
\tilde{D}(V):=D(P(V))
$$
and applying the (backward) Ito  formula, cf. Appendix~\ref{as:stochastic_calculus}, to the relativistic velocity formula $V(t)=P/(M^2+\bs P^2)^{1/2}$, one finds the following SDE for the velocity process
\be\notag 
\diff V(t)&=&
\left[
-\tilde{\ga}(V)\;(1-V^2)
-\gl_\odot\;\left(\f{3\tilde{D}}{M^2}\right) (1-V^2)^2
\right]\;V\,\dt +
\\
&&\label{e:velocity_formula}
\left[\left(\f{2\tilde{D}}{M^2}\right)\;
(1-V^2)^{3}\right]^{1/2}\odot \dB(t).
\ee
As discussed in the next section, this equation can be used to calculate the asymptotic mean square displacement.

\subsubsection{Asymptotic mean square displacement}
\label{s:asymptotic_MSD}
A primary objective within any Brownian theory is to determine the asymptotic diffusion constant $\D_\infty$, corresponding to the plateau values in Fig.~\ref{f:MSD_comparison}. For a one-dimensional diffusion process $X(t)$ with velocity $V(t)$, the asymptotic diffusion constant $\D_\infty$ is defined by
\bse
\be
\D_\infty=\lim_{t\to\infty}\lan [X(t)-X(0)]^2\ran/(2t),
\ee
where the spatial displacement is given by
\be
X(t)-X(0)=\int_0^t\ds \;V(s).
\ee
\ese
The asymptotic diffusion constant $\D_\infty$ may be expressed in terms of the velocity correlation function $\lan V(t)V(s)\ran$ by virtue of
\be
\D_\infty
&=&\notag
\lim_{t\to\infty}\;
\f{1}{2}\,\f{\diff}{\dt}\lan [X(t)-X(0)]^2\ran\\
&=&\notag
\lim_{t\to\infty}\;
\f{1}{2}\,\f{\diff}{\dt}
\int_0^t\ds \int_0^t\ds' \;\lan V(s)V(s')\ran\\
&=&
\lim_{t\to\infty}
\int_0^t\ds \;\lan V(t)V(s)\ran.
\ee
Assuming that the velocity process $V(t)$ is (approximately) stationary, which means that $\lan V(t)V(s)\ran=\lan V(t-s)V(0)\ran$ holds (at least in good approximation), and substituting $u=t-s$, we recover Kubo's formula
\be
\D_\infty
&=&\label{e:diffusion_kubo}
\lim_{t\to\infty}
\int_0^t\diff u \;\lan V(u)V(0)\ran.
\ee
As recently discussed by Lindner~\cite{2007Li}, for a one-dimensional Langevin equation of the form
\be
\dV(t)&=& - a_\bullet(V)\,V\,\dt + [2\,b(V)]^{1/2}\bullet\dB(t) 
\ee 
with symmetric coefficient functions, $a_\bullet(v)=a_\bullet(-v)$ and $b(v)=b(-v)$, the Kubo formula~\eqref{e:diffusion_kubo} gives rise to the following integral representation for the asymptotic diffusion constant:
\be\label{e:lindner_diffusion}
\D_\infty=
\f{\int_0^{v_+} \diff y\; e^{U(y)}
\left[\int_{y}^{v_+}\diff x\; e^{-U(x)}\, x/b(x)\right]^2}
{\int_0^{v_+} \diff z\; e^{-U(z)}/b(z)}.
\ee 
Here, $v_+\in[0,\infty]$ represents the upper bound for the velocity range, and
\bse\label{e:velocity_potential}
\be\label{e:velocity_potential-a}
U(v):=\int_0^v\diff w\;\mu_*(w)/b(w)
\ee
is an effective velocity potential with Ito drift
\be\label{e:velocity_potential-b}
\mu_*(v)&:=&a_*(v)\,v=a_\bullet(v)\,v - b'(v).
\ee
\ese
In general, the formula~\eqref{e:lindner_diffusion} has to be integrated numerically, but for the first two models from Section~\ref{s:rbm_examples} the integrals may also be evaluated analytically. The $d$-dimensional generalization of Eq.~\eqref{e:lindner_diffusion} was recently derived by Angst and Franchi~\cite{2007AnFr}. 

\subsubsection{Examples}
\label{s:rbm_examples}
We discuss three specific one-dimensional relativistic Langevin models whose stationary momentum distributions are J\"uttner functions $\phi_\J(p)\propto\exp[-\gb(p^2+M^2)^{1/2}]$ with heat bath temperature  $\Temp =(\kB \gb )^{-1}$. In this case, the relativistic Einstein relation~\eqref{e:Einstein_relativistic} implies that only one of the two functions $\ga_\bullet(p)$ and $D(p)$ can be chosen arbitrarily.

\paragraph*{Constant noise amplitude}
As a first example, we consider the so-called 'Relativistic Ornstein-Uhlenbeck process' (ROUP), proposed by Debbasch et al.~\cite{1997DeMaRi,1998DeRi} and also studied by Zygad{\l}o~\cite{2005Zy}. The ROUP is defined by the choice 
\bse
\be\label{e:ROUP_friction}
\ga_\bullet(p)=\ga_c\; {M}/{p^0},
\ee 
where $\ga_c>0$ is a constant friction  parameter. From the relativistic Einstein relation~\eqref{e:Einstein_relativistic}, one then finds
\be\label{e:ROUP_Einstein}
D(p)\equiv {\ga_\bullet(p)\, p^0}\,{\gb^{-1}}
={\ga_c M}\,{\gb^{-1}}
=:D_c,
\ee
i.e., the ROUP corresponds to the limit case of constant noise amplitude. The associated Langevin equation reads 
\be
\diff P(t) 
&=&-\ga_c\,\f{M}{P^0}\, {P}\; \dt +
\left(\f{2\ga_c M}{\gb}\right)^{1/2}\bullet \diff B(t)\\
&=&-\ga_c\,\f{M}{P^0}\, {P}\; \dt +
\left(\f{2\ga_c M}{\gb}\right)^{1/2}* \diff B(t).
\label{e:ROUP_RLE}
\ee
The discretization rule is irrelevant here, because the noise amplitude \mbox{$D_c={\ga_c M}/{\gb}$} does not depend on the momentum $P$ for this particular case. However, the rules of stochastic calculus have to be specified, if one wishes to write down the SDE for the associated velocity process $V(t):=P/P^0$. Defining the useful abbreviation
\be 
\chi:=\gb M=M/(\kB \Temp),
\ee
one finds
\be
\diff V(t) &=&\notag
-\ga_c\, \biggl[ 
(1-V^2)^{3/2}- \f{3}{\chi}(1-V^2)^{2}\biggr]\,V\; \dt +\\
&&
\qquad
\left[\f{2\ga_c}{\chi}\;(1-V^2)^3\right]^{1/2}\bullet \diff B(t)\\
 &=&\notag
-\ga_c\, \biggl[ 
(1-V^2)^{3/2}+\f{3}{\chi}(1-V^2)^{2}\biggr]\,V\; \dt +\\
&&
\qquad
\left[\f{2\ga_c}{\chi}\;(1-V^2)^3\right]^{1/2}* \diff B(t).
\ee
\ese
Using the Ito form, as required in Eqs.~\eqref{e:velocity_potential-b}, we see that the ROUP~\eqref{e:ROUP_RLE} corresponds to 
\bse
\be 
\mu_*(v)&=&\ga_c \left[(1-v^2)^{3/2}+
\f{3}{\chi}\,(1-v^2)^{2}
\right]\,v,\\
b(v)&=&
\f{\ga_c}{\chi}\,(1-v^2)^{3}
\ee
\ese
with an upper velocity bound $v_+=c=1$ in Eq.~\eqref{e:lindner_diffusion}.
In this case the general integral formula~\eqref{e:lindner_diffusion} for the asymptotic diffusion constant can be evaluated analytically by making use of the identity~\eqref{e:bessel_integral_identity} 
\be\label{e:bessel_integral_identity}
(-1)^\nu\f{\diff^\nu}{\diff \chi^\nu} K_0(\chi)&=&\int_0^1\diff v\; \exp\biggl(-\f{\chi}{\sqrt{1-v^2}}\biggr)\;(1-v^2)^{-(\nu+2)/2},
\ee
where, for~\mbox{$\nu=0,1,2\ldots$}, $K_\nu(z)$ denotes the modified Bessel function of the second kind~\cite{AbSt72}. Remarkably, one  recovers for the ROUP the \lq classical\rq~ result~\cite{2007AnFr}, cf. Eq.~\eqref{e:classical_diffusion_OUP-2},
\be\label{e:diffusion_ROUP_theory}
\mcal{D}_\infty^{\mrm{ROUP}}=(\ga_c \chi)^{-1}={\kB\Temp}/(M\ga_c)
\ee 
for all parameter values $(\ga_c,\Temp,M)$.

\paragraph*{Constant friction coefficient in the  backward-Ito SDE} 
An alternative relativistic Brownian motion (RBM) model~\cite{2005DuHa,2005DuHa_2} corresponds to the special case of a constant friction function $\ga_\bullet(p)\equiv \ga_\dagger$ in the backward-Ito SDE~\eqref{e:RLE_post}. In this case,  the relativistic Einstein relation~\eqref{e:Einstein_relativistic} yields the momentum dependent noise amplitude 
\bse\label{e:RBM_DH}
\be\label{e:RBM_Einstein}
D(p)={\ga_\dagger p^0}\,{\gb^{-1}}.
\ee
The relativistic (backward) Ito Langevin equations of this model read explicitly
\be\label{e:RBM_RLE}
\diff P(t) 
&=&
-\ga_\dagger\,P\; \dt +\left(\f{2\ga_\dagger P^0}{\gb}\right)^{1/2}\bullet \diff B(t)\\
&=&\label{e:rbm_dunkel_ito_p}
-\ga_\dagger\left(\f{\gb P^0-1}{\gb P^0}\right)\,P\; \dt +
\left(\f{2\ga_\dagger P^0}{\gb}\right)^{1/2} * \diff B(t).
\ee
The corresponding SDEs for the velocity process $V(t)=P/P^0$ are given by
\be
\diff V(t) 
&=&\notag
-\ga_\dagger\, \biggl[ (1-V^2) - \f{3}{\chi}(1-V^2)^{3/2}\biggr]\,V\;\dt +\\
&&
\qquad\left[\f{2\ga_\dagger}{\chi}\;(1-V^2)^{5/2}\right]^{1/2}
\bullet \diff B(t)\\
&=&\notag
-\ga_\dagger\, \biggl[ (1-V^2)+ 
\f{2}{\chi}(1-V^2)^{3/2}\biggr]\,V\;\dt + \\
&&\label{e:RBM_DH_velo_ito}
\qquad\left[\f{2\ga_\dagger}{\chi}\;(1-V^2)^{5/2}\right]^{1/2}
* \diff B(t).
\ee
\ese
With regard to numerical simulations, the Ito form is more convenient, cf. Appendix~\ref{as:stochastic_calculus}. 
\par
Various properties of the RBM process~\eqref{e:RBM_DH} have been analyzed by Fa~\cite{2006Fa}, Lindner~\cite{2007Li}, Fingerle\footnote{Fingerle~\cite{2007Fi} discusses a fluctuation theorem for this process; see also Cleuren et al.~\cite{2008ClEtAl}.}~\cite{2007Fi}, and Angst and Franchi~\cite{2007AnFr}. In particular, from Eq.~\eqref{e:RBM_DH_velo_ito} we see that this model corresponds to 
\bse
\be
\mu_*(v)&=&\ga_\dagger \left[(1-v^2)+
\f{2}{\chi}\,(1-v^2)^{3/2}
\right]\,v,\\
b(v)&=&\,
\f{\ga_\dagger}{\chi}\,(1-v^2)^{5/2},
\ee
\ese
with an upper velocity bound $v_+=c=1$ in Eq.~\eqref{e:lindner_diffusion}. As for the ROUP, the integral formula~\eqref{e:lindner_diffusion} can be calculated analytically by making use of the identity~\eqref{e:bessel_integral_identity} and  one then finds that\footnote{Equation~\eqref{e:lindner_diffusion_rbm} is an equivalent, more compact representation of Lindner's result Eq.~(10) in~\cite{2007Li}.}
\be\label{e:lindner_diffusion_rbm}
\mcal{D}_\infty^{\mrm{RBM}}= (\ga_\dagger \chi)^{-1}\;\f{K_0(\chi)}{K_1(\chi)}.
\ee
At low temperatures $\gb:=(\kB \Temp)^{-1} \to\infty$, Eq.~\eqref{e:lindner_diffusion_rbm} reduces to the well-known classical result $\mcal{D}_\infty^{\mrm{class}}= {\kB\Temp}/({M\ga_\dagger})$, cf. Eq.~\eqref{e:classical_diffusion_OUP-2}. In the opposite limit of very high temperatures, i.e., for $\gb M\ll 1$, one finds a logarithmic dependence~\cite{2007AnFr} 
\be\label{e:log_asymptotics}
\mcal{D}_\infty^{\mrm{RBM}}=(\ga_\dagger M)^{-1}
\left\{-\gc_\eps+\ln(2/\chi) +\mcal{O}[(\gb M)^2]\right\},
\ee 
where $\gc_\eps\simeq 0.577216$ is the Euler constant. However, it should be kept in mind that, due to the increasing importance of particle annihilation/creation at high energies, classical non-quantum theories become invalid in the high temperature limit $\chi=\gb M\ll 1$, and, therefore, the asymptotic expansion~\eqref{e:log_asymptotics} appears to be of limited practical use.
\par
By comparing with the ROUP, we observe that \mbox{$\mcal{D}_\infty^{\mrm{RBM}} \le \mcal{D}_\infty^{\mrm{ROUP}}$} holds true for same values of the friction coefficients $\ga_c=\ga_\dagger$, cf. Fig.~\ref{f:MSD}. Intuitively, this can be explained by the fact that, for the ROUP, the absolute value of the friction force is bounded by $\ga_c M$, cf. Eq.~\eqref{e:ROUP_RLE}, whereas the friction force is unbounded for the RBM model~\eqref{e:RBM_RLE}, thereby suppressing spatial diffusion more strongly in the latter case. 

\paragraph*{Constant friction coefficient in the Ito SDE} 
The RBM process defined by Eq.~\eqref{e:RBM_RLE} is characterized by a constant friction coefficient $\ga_\dagger$, when adopting the post-point discretization rule $(\bullet)$. Another model, referred to as RBM(I) hereafter, is obtained by considering a constant friction coefficient $\ga_*$ in the Ito Langevin equation
\bse
\be\label{e:RBM2_RLE}
\diff P(t) =-\ga_*\,P\; \dt +\left[\f{2\ga_*}{\gb^2}(1+\gb P^0)\right]^{1/2}* \diff B(t),
\ee
where the noise amplitude is chosen such that the Einstein relation~\eqref{e:Einstein_relativistic_ito} is satisfied. The Ito SDE of the associated velocity process $V(t):=P/P^0$ reads 
\be
\diff V(t) &=&\notag
-\ga_*\, \biggl[(1-V^2)
+\f{3}{\chi}(1-V^2)^{3/2}
+\f{3}{\chi^2} (1-V^2)^{2}
\biggr]\,V\;\dt +\\
&&\label{e:RBM2_RLE-velo}
\qquad
\left\{\f{2\ga_*}{\chi}\;
\left[(1-V^2)^{5/2}+ \chi^{-1} (1-V^2)^{3}\right]\right\}^{1/2}* \diff B(t).
\ee
\ese
In this case, we have
\bse
\be
\mu_*(v)&=& \ga_*\, \biggl[(1-v^2)
+\f{3}{\chi}(1-v^2)^{3/2}
+\f{3}{\chi^2} (1-v^2)^{2}
\biggr]\,v,\\
b(v)&=&
\f{\ga_*}{\chi}
\left[(1-v^2)^{5/2}+{\chi^{-1}} (1-v^2)^{3}\right].
\ee
\ese
From these equations the velocity potential is obtained as
\be\notag
U(v)=\ln\left[
\f{\chi+1}
{\chi(1-v^2) +(1-v^2)^{3/2}}
\right]-\chi [1-(1-v^2)^{-1/2}],
\ee
yielding for the asymptotic diffusion constant:
\be\label{e:diffusion_RBM_ito}
\mcal{D}_\infty^{\mrm{RBM(I)}}=
\left[\ga_* K_1(\chi)\right]^{-1}
\int_0^1\diff v\;
\f{e^{-\chi(1-v^2)^{-1/2}}}{\chi(1-v^2) +(1-v^2)^{3/2}}.
\ee
The remaining integral can be evaluated numerically. As illustrated in Fig.~\ref{f:MSD}, the theoretical predictions from Eqs.~\eqref{e:lindner_diffusion_rbm}, \eqref{e:diffusion_ROUP_theory} and \eqref{e:diffusion_RBM_ito} are in good agreement with the numerically obtained estimates of the asymptotic diffusion constant. 
\par
\begin{figure}[t]
\center 
\vspace{0.5cm}
\includegraphics[angle=0]{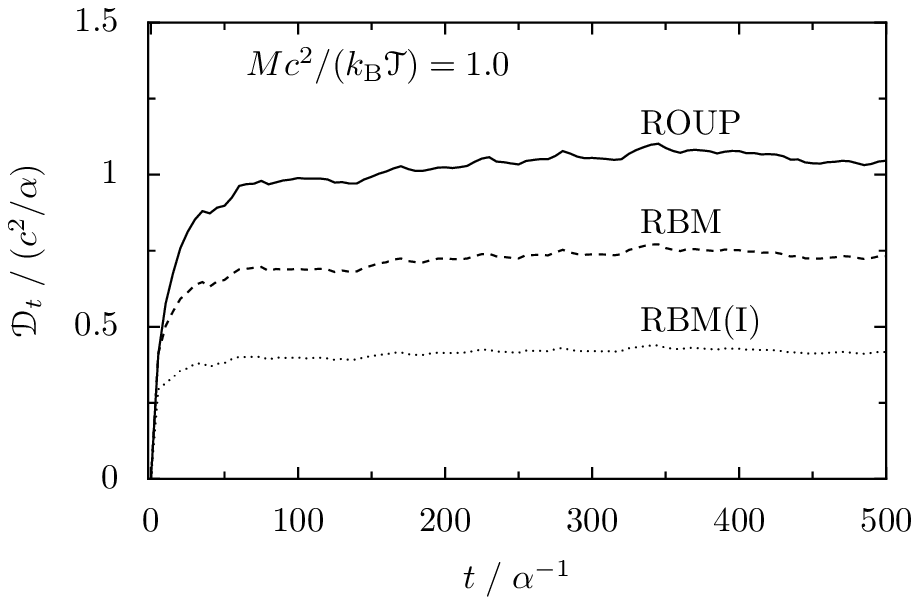}
\caption{\label{f:MSD_comparison} 
Time evolution of the spatial mean square displacement $\D_t:=\lan [X(t)-X(0)]^2\ran/(2t)$ for the ROUP~\cite{1997DeMaRi} model (solid line) from Eq.~\eqref{e:ROUP_RLE}, the RBM~\cite{2005DuHa} model (dashed) from Eq.~\eqref{e:RBM_RLE} and the RBM(I) model (dotted) from Eq.~\eqref{e:RBM2_RLE} at same temperature-mass ratio $\chi^{-1}:=\kB \Temp/(Mc^2)=1$. The plots are based on a simulation with $N=1000$ trajectories, initial conditions $X(0)=0$, $P(0)=0$ for each trajectory, and discretization time step $\Delta t=10^{-4}\;\ga_{c/\dagger}^{-1}$.
}
\end{figure}
\begin{figure}[h]
\centering 
\vspace{0.5cm}
\includegraphics[width=9.5cm,angle=0]{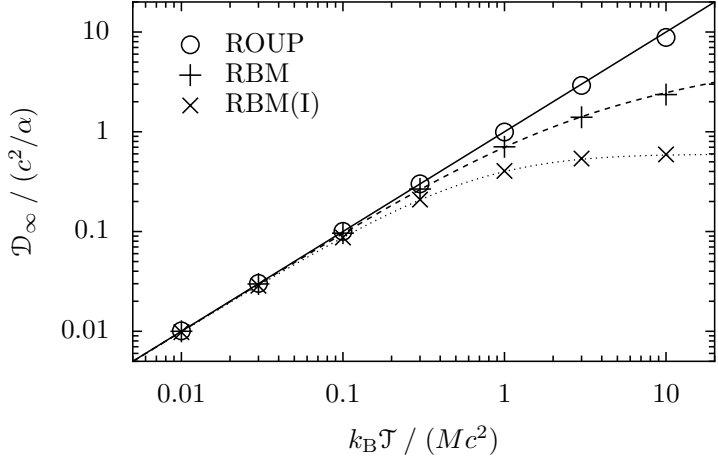}
\caption{\label{f:MSD} 
Temperature dependence of the asymptotic diffusion constant $\D_\infty$ for the ROUP~\cite{1997DeMaRi} from Eq.~\eqref{e:ROUP_RLE}, the RBM model~\cite{2005DuHa} from Eq.~\eqref{e:RBM_RLE}, and the RBM(I) model from Eq.~\eqref{e:RBM2_RLE}. Symbols '$\circ$', '$+$' and '$\times$' represent the results of computer simulations  ($N=100$ trajectories, initial conditions $X(0)=0$, $P(0)=0$ for each trajectory, discretization time step $\Delta t=10^{-4}$ in units of $\ga_{c/\dagger/*}^{-1}$), obtained by averaging over the numerically determined values $\{\D_{100}, \D_{110},\D_{120},\ldots,\D_{500}\}$. Solid, dashed and dotted lines correspond to the theoretical predictions from Eqs.~\eqref{e:diffusion_ROUP_theory}, \eqref{e:lindner_diffusion_rbm} and \eqref{e:diffusion_RBM_ito}, respectively.
}
\end{figure}

The three model processes~\eqref{e:ROUP_RLE},~\eqref{e:RBM_RLE} and ~\eqref{e:RBM2_RLE} give rise to the same stationary momentum PDF $\phi_\J(p)$, but their respective relaxation behavior differs strongly. This is illustrated in Fig.~\ref{f:MSD_comparison}, which depicts the time evolution of the spatial mean square displacement divided by time,
\be
\D_t:=\lan [X(t)-X(0)]^2\ran/(2t),
\ee
for all three models at same temperature $\Temp$. The curves in Fig.~\ref{f:MSD_comparison} were calculated numerically from Eqs.~\eqref{e:ROUP_RLE}, \eqref{e:rbm_dunkel_ito_p} and~\eqref{e:RBM2_RLE}, respectively, using an algorithm similar to those described in~\cite{2007Li,2007AnFr}, see also Appendix~\ref{appendix:numerics}.

The ROUP and the two RBM models considered in this part represent special limit cases of the general Langevin equation~\eqref{e:RLE_post} with arbitrarily chosen friction coefficient functions. Nonetheless, they yield useful insights: As evident from Fig.~\ref{f:MSD}, at moderate-to-high temperatures the diffusion constant can vary significantly for different friction models. For realistic systems, the exact functional shape (i.e., energy dependence) of the friction function $\ga$ is determined by the microscopic interactions.  This result implies that simultaneous measurements of the temperature and the diffusion constants can reveal information about the underlying microscopic forces. Below, in Section~\ref{s:relativistic_binary_collision_model}, we will outline a general procedure for deducing more realistic friction coefficients $\ga$ from microscopic models.

\subsection{Proper-time reparameterization}
\label{s:propertime} 

The relativistic Langevin equations considered thus far are parameterized in terms of the time-coordinate $t$ of the lab frame $\Gs$. We shall now discuss how these equations can be re-parameterized in terms of the proper-time $\tau$. For this purpose, we start from the Ito Langevin equation\footnote{For simplicity, we assume that $\bs B (t)=(B^j(t))$  is $d$-dimensional, implying  that $C^i_{~j}$ is a square matrix. However, all results still hold if $\bs B(t)$ has a different dimension.}
\bse\label{e:RLE_time-change-ito}
\be
\dX^\ga(t)
&=&\label{e:RLE_position_time-change-ito}
(P^\ga/{P^0})\; \dt,\\
\dP^i(t)
&=& \label{e:RLE_momentum_time-change-ito}
A^i\; \dt +{c^i}_j *\diff B^j(t),
\ee
In general, the functions $A^i$ and ${C^i}_j$ may depend on $(t,X^i,P^i)$. Upon applying Ito's formula~\cite{KaSh91, protter} to the mass-shell condition $P^0(t)=(M^2+\bs P^2)^{1/2}$,  Eq.~\eqref{e:RLE_momentum_time-change-ito} yields the following equation for the relativistic energy
\be
\dP^0(t)
=
\biggl\{
A_i \f{P^i}{P^0}+
\f{D_{ij}}{2}\biggl[\f{\gd^{ij}}{P^0}- \f{P^i P^j} {(P^0)^3}\biggr]
\biggr\}\dt + 
\f{P^i}{P^0}\; c_{ir}*\diff B^r(t)
,\label{e:RLE_0-b}
\ee
\ese
where $A_i:=A^i$, $D_{ij}:=D^{ij}=c^i_r c^j_r:=\sum_r c^i_r c^j_r$ and $c_{ir}:={c^i}_r$.
The Fokker-Planck equation for the associated phase space PDF $f(t,\bs x,\bs p)$ reads
\be\label{e:FPE-time-change}
\biggl(\f{\p}{\p t} +\f{p^i}{p^0}\f{\p}{\p x^i}\biggr) f
=
\f{\p}{\p p^i}\biggl[-A^i f +\f{1}{2}
\f{\p}{\p p^k} \bigl(D^{ik} f \bigr)\biggr].
\ee
As before, we consider deterministic initial data $\bs X(0)=\bs x_0$ and $\bs P(0)=\bs p_0$, which translates into $f(0,\bs x,\bs p)=\gd(\bs x-\bs x_0)\;\gd(\bs p-\bs p_0)$. 

\par
We are interested in rewriting Eqs.~\eqref{e:RLE_time-change-ito}  in terms of the proper-time $\tau$, defined by 
\be
\dtau(t)=(1-\bs V^2)^{1/2}\dt
\csp
\tau(0)=0.
\ee 
The proper-time differential may also be expressed as
\bse\label{e:proper-time-heuristics}
\be\label{e:proper-time-heuristics-a}
\dtau(t)=(M/P^0)\,\dt.
\ee
The inverse of the function $\tau(t)$ will be denoted by $\hat X^0(\tau) = t(\tau)$ and represents the time coordinate of the particle in the frame $\Sigma$ parameterized by the proper time~$\tau$. Our goal is to find SDEs for the reparameterized processes $\hat{X}^\ga(\tau):= X^\ga(t(\tau))$ and $\hat{P}^\ga(\tau)= P^\ga(t (\tau))$ in~$\Gs$. Heuristically, the SDEs can be derived from the relation
\be
\dB^j(t)
\;\simeq\;
\sqrt{\dt}
\;=\;
\left({\hat{P}^0}/{M}\right)^{1/2}\sqrt{\dtau}
\;\simeq\;
\left({\hat{P}^0}/{M}\right)^{1/2}*\diff\hat{B}^j(\tau),
\quad
\label{e:proper-time-heuristics-b}
\ee
\ese
where $\hat{B}^j(\tau)$ is a standard Wiener process with time-parameter $\tau$.
Equation~\eqref{e:proper-time-heuristics-b} can be justified rigorously by applying theorems for the time-change of (local) martingale processes~\cite{protter,2008DuHaWe}. Inserting Eqs.~\eqref{e:proper-time-heuristics} in Eqs.~\eqref{e:RLE} one finds
\bse\label{e:RLE-tau}
\be
\diff \hat X^\ga(\tau)&=&(\hat P^\ga/M)\,\dtau,\\
\diff\hat P^i(\tau) &=&
\hat A^i\; \dtau + {\hat c^i}_{~j} *\diff \hat B^j(\tau),
\ee
where the transformed coefficients are given by 
\be
\hat A^i &: =&   
(\hat P^0/M)\; A^i(\hat X^0, \hat{\bs X}, \hat {\bs  P}),\\
\hat c^i_{~j} &:=& 
(\hat P^0 / M)^{1/2}\; 
c^i_{~j}(\hat X^0, \hat{\bs X}, \hat{\bs P}).
\ee
Analogous to Eq.~\eqref{e:RLE_0-b}, the reparameterized energy equation is obtained as
\be
\diff\hat{P}^0(\tau)
=
\biggl\{
\hat{A}_i \f{\hat{P}^i}{\hat{P}^0}+
\f{\hat{D}_{ij}}{2}\biggl[\f{\gd^{ij}}{\hat{P}^0}- \f{\hat{P}^i \hat{P}^j} {(\hat{P}^0)^3}\biggr]
\biggr\}\dt + 
\f{\hat{P}^i}{\hat{P}^0}\; \hat{c}_{ir}*\diff \hat{B}^r(\tau)
,\label{e:RLE_0-b-tau}
\ee
\ese
where now ${\hat D}^{ik}:={\hat{c}^i}_r {\hat{c}^k}_r$. The FPE for the associated probability density $\hat{f}(\tau, x^0, \bs x, \bs p)$ reads
\be\label{e:FPE-tau}
\biggl(\f{\p}{\p \tau} +\f{p^\ga}{M}\f{\p}{\p x^\ga}\biggr) \hat{f}
=
\f{\p}{\p p^i}\biggl[-\hat{A}^i \hat{f} +\f{1}{2}
\f{\p}{\p p^k} \bigl({{\hat D}^{ik}}\hat{f}\bigr)\biggr]
\ee
We note that now $\hat{f}\;\diff x^0\diff^d x \diff^d p$ gives probability of finding the particle at proper-time~$\tau$ in the $(2d+1)$-dimensional interval $[t,t+\dt]\times [\bs x,\bs x+\diff \bs x]\times [\bs p,\bs p+\diff \bs p]$ in the inertial frame $\Sigma$.
\par
\begin{figure}[t]
\centering
\includegraphics[width=8.5cm]{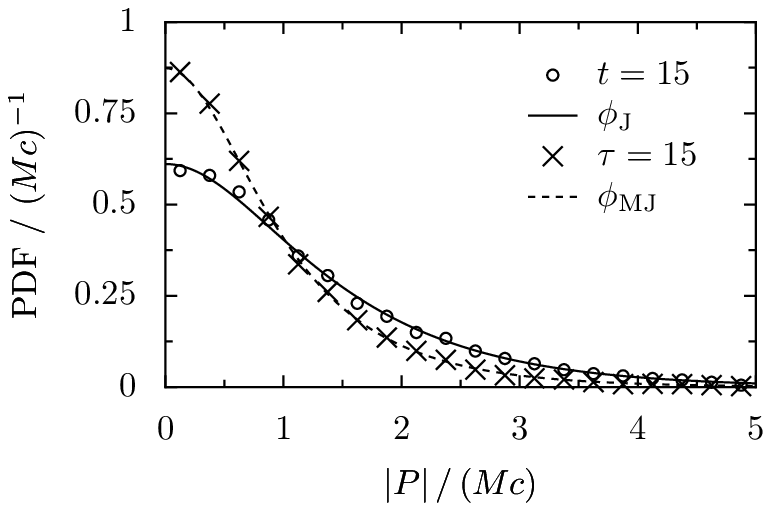}
\caption{\label{fig01}
\lq Stationary\rq\space probability density function (PDF) of the absolute momentum $|P|$ measured at time $t=15$ ($\times$) and proper-time $\tau=15$ ($\circ$), respectively, from $10000$ sample trajectories of the one-dimensional $(d=1)$ relativistic Ornstein-Uhlenbeck process~\cite{1997DeMaRi}, corresponding to coefficients $D(p^0)=D_c=const$ and $\ga(p^0)=\gb D_c/p^0 $ in Eqs.~\eqref{e:RLE_time-change-ito} and \eqref{e:constraints-time-change}. Simulation parameters: $\dt=0.001$, $M=c=\gb=D_c=1$.  The solid and  dashed lines show the corresponding standard and modified J\"uttner functions $\phi_\J\propto \exp(-\gb p^0)$ and $\phi_\MJ\propto \exp(-\gb p^0)/p^0$, respectively.
}  
\end{figure}

An interesting consequence of this reparameterization can be illustrated by considering the free motion in a thermalized heat bath, which is stationary, isotropic and position independent in the lab frame $\Gs$. As discussed above, in this case a plausible ansatz reads
\bse\label{e:constraints-time-change}
\be
A^i=-\hat{\ga}(p^0)\,p^i \csp
{c^i}_j=[2\hat{D}(p^0)]^{1/2}\;{\gd^i}_j,
\ee
where the friction and noise coefficients $\hat\ga$ and $\hat D$ depend on the energy $p^0=(M^2+\bs p^2)^{1/2}$ and satisfy the relativistic Einstein relation 
\be\label{e:einstein-time-change}
0\equiv
\hat \ga(p^0)\, p^0 + \hat D'(p^0) -\gb D(p^0).
\ee
\ese
Then, the stationary momentum distribution of the lab-time FPE~\eqref{e:FPE-time-change} is given by a J\"uttner function~\cite{1911Ju,2007CuEtAl}, i.e.,
\bse
\be
f_\infty:=\lim_{t\to\infty } f \propto 
\f{\exp(-\gb p^0)}{\mcal{Z}_\mrm{J}}
=\phi_\J(\bs p)
\ee
Remarkably, however, in this case the stationary solution $\hat{f}_\infty$ of the  corresponding proper-time FPE~\eqref{e:FPE-tau} is given by the modified J\"uttner function~\cite{2007DuTaHa_2}
\be
\hat f_\infty:=\lim_{\tau\to\infty } \hat{f} \propto 
\f{\exp(-\gb p^0)}{\mcal{Z}_\mrm{J}\;p^0}
=:\phi_\MJ(\bs p).
\ee
\ese
This can be confirmed by direct numerical simulation of Eqs.~\eqref{e:RLE_time-change-ito}. An example is given in Fig.~\ref{fig01}, which depicts the numerically obtained PDF of the absolute momentum $|P|$ for the one-dimensional ROUP~\cite{1997DeMaRi}. The numerical PDFs were calculated from $10000$ sample trajectories, by measuring the momentum  at constant lab-time $t$ ($\times$) and constant proper-time $\tau$ ($\circ$), respectively. The solid and  dashed lines show the corresponding standard and modified J\"uttner functions, respectively.  The physical explanation for the difference between $f_\infty$ and $\hat{f}_\infty$ is that measurements at $t=const$ and $\tau=const$ are non-equivalent even if both $\tau,t\to \infty$.
\par
The modified J\"uttner distribution $\phi_\MJ$, representing the stationary solution of the proper-time FPE, differs from the standard J\"uttner function $\phi_\J$~\eqref{e:juttner-c} by a prefactor proportional to the inverse energy. As discussed in Ref.~\cite{2007DuTaHa_2}, the modified distribution $\phi_\MJ$ can be derived from a relative entropy principle by using a Lorentz invariant reference measure in momentum space, whereas the J\"uttner function~ $\phi_\J$ is obtained when adopting a constant reference density (Lebesgue measure), cf. Eq.~\eqref{e:entropy_principle-a}.
\par
We may summarize: If a relativistic Langevin-Ito process has been specified in the inertial frame $\Gs$ and is parameterized by the associated $\Gs$-coordinate time $t$, then this process can be reparameterized by its proper-time $\tau$ and the resulting process is again of the Langevin-Ito type but exhibits a modified stationary distribution. With regard to applications~\cite{2005HeRa,2006RaGrHe} the latter fact means that the correct form of the  fluctuation-dissipation relation depends on the choice of the time-parameter in the relativistic Langevin equation. In principle, however, one can -- similar to the case of purely deterministic relativistic equations of motions -- choose freely between different time parameterizations by taking into account that the noise part needs to be transformed differently than the deterministic part.

\subsection{Moving observers}
\label{s:moving_observer}

The preceding discussion has focused on Langevin equations that describe the random motion of a relativistic Brownian particle in the lab frame, defined as the rest frame of the heat bath. In this part we would like to address the following question~\cite{2008ChDe,2008DuHaWe}: Assuming that a Langevin equation of the type~\eqref{e:RLE_post} holds in the lab frame $\Gs$ -- how does the corresponding process appear to a moving observer?
\par
There exist (at least) two different ways to tackle this problem:
Either one uses the Lorentz invariance of the phase space density, or one applies a Lorentz transformation directly to the Langevin equation~\cite{2008ChDe,2008DuHaWe}.

\subsubsection{Lorentz transformation of the phase space density}

An observer at rest in the lab frame would measure the PDF~$f$ governed by the Fokker-Planck equations~\eqref{e:rel-FPE} or~\eqref{e:a-FPE-high-2}. As discussed in Section~\ref{s:probality_densities} the one-particle phase space density transforms as a Lorentz scalar~\cite{1969VK}; i.e.,
\bse\label{a-e:van-Kampen}
\be\label{a-e:van-Kampen-1}
f'(t',\bs x',\bs p')=f(t(t',\bs x'),\bs x(t',\bs x'),\bs p(\bs p')),
\ee
and, conversely,
\be\label{a-e:van-Kampen-2}
f(t,\bs x,\bs p)=f'(t'(t,\bs x),\bs x'(t,\bs x),\bs p'(\bs p)),
\ee
\ese
where $(t',\bs x',\bs p')$ and $(t,\bs x,\bs p)$ are related by the Lorentz transformation
\bse
\be
x'^\gl(t,\bs x)&=&{\Gl^\gl}_0t+{\Gl^\gl}_i x^i,\\
p'^i(\bs p)&=&{\Gl^i}_0(m^2+\bs p^2)^{1/2}+{\Gl^i}_jp^j.
\ee
\ese 
For an observer moving at constant velocity $\bs w$ through the lab frame, the corresponding Lorentz boost matrix elements are given by Eq.~\eqref{e:a-boost}. Therefore, in order to find $f'$, it suffices to solve the Fokker-Planck equations~\eqref{e:rel-FPE} in the lab frame $\Gs$, e.g., for a given $t$-simultaneous initial condition~$f(0,\bs x,\bs p)$, and to insert the solution into~\eqref{a-e:van-Kampen-1}.

\subsubsection{Lorentz transformation of the Langevin equation}
\label{s:lorentz_transformation_of_the_RLE}

Alternatively, in order to obtain explicit SDEs for the Brownian  motion in  $\Gs'$,  one has to apply a Lorentz transformation to the Langevin equations  in the lab frame $\Gs$~\cite{2008ChDe,2008DuHaWe}. This can be done by employing a similar heuristics as in the case of the proper-time reparameterization, cf. Section~\ref{s:propertime}. 
\par
To illustrate this in more detail, we start again from the Langevin equations~\eqref{e:RLE_time-change-ito} and consider a Lorentz  transformation~\cite{SexlUrbantke} from the lab frame $\Gs$ to another inertial frame $\Gs'$, mediated by a constant matrix ${\Gl^\nu}_\mu$ that leaves the metric tensor $\eta_{\ga\gb}$ invariant. For convenience, we exclude time reversal transformations, i.e., we restrict ourselves to proper Lorentz transformations with  ${\Gl^0}_0>0$. It is convenient to  proceed in two steps: First we define 
$$
Y'^\nu(t):={\Gl^\nu}_\mu  X^\mu(t)\csp
G'^\nu(t):={\Gl^\nu}_\mu  P^\mu(t).
$$
Then we replace $t$ by the coordinate time $t'$ of $\Gs'$ to obtain processes
$$
X'^\ga(t')= Y'^\ga(t(t'))
\csp
P'^\ga(t')= G'^\ga(t(t')).
$$
In order to find the SDEs for $X'^\ga(t')$ and $P'^\ga(t')$, we note that
\mbox{$\diff Y'^0(t) ={\Gl^0}_\mu  \dX^\mu(t)$} and, therefore,
\bse\label{e:LT-heuristics}
\be\label{e:LT-heuristics-1}
\dt'(t) 
=
\diff Y'^0(t)
= 
\f{{\Gl^0}_\mu P^\mu}{P^0}\,\dt
=
\f{G'^0}{P^0}\,\dt
= 
\frac{P'^0(t'(t))}{(\Lambda^{-1})^0_{\;\mu} P'^\mu(t'(t))}\, \dt,
\ee
where $\Gl^{-1}$ is the inverse Lorentz transformation. A similar heuristics as in Eq.~\eqref{e:proper-time-heuristics-b} then gives
\be
\dB^j(t)
\simeq
\sqrt{\dt}
=
\biggl(\f{P^0}{P'^0}\biggr)^{1/2}
\sqrt{\dt'}
\simeq
\biggl[\f{(\Gl^{-1})^0_{\;\mu} P'^\mu}{P'^0}\biggr]^{1/2}
*\diff B'^j(t'),
\ee
\ese
where $B'^j(t')$ is a Wiener process with time parameter~$t'$. Defining 
\bse
\be
A^0 &:=&  A_i\frac {p^i}{p^0} + \f{D_{ij} } 2
\left[  \f{\delta^{ij}}{p^0} - \f{p^i p^j}{(p^0)^3} \right],\\
c^0_{~j}  &:=&  \f{p^i}{p^0} c_{ij},
\ee
\ese
we can introduce Lorentz-transformed coefficient functions in $\Sigma'$ by
\bse\label{e:new_coefficients}
\be
 A'^i(x'^0, {\bs x'}, {\bs p'})  
&=& \notag
\left[\f{(\Lambda^{-1})^0_{~\mu} p'^\mu}{p'^0}\right] \;\times
\\&&\quad
\Lambda^i_{~\nu}\, A^\nu \left( (\Lambda^{-1})^0_{~\mu} x'^\mu,  (\Lambda^{-1})^i_{~\mu} x'^\mu, (\Lambda^{-1})^i_{~\mu} p'^\mu  \right),
\\
\notag
c'^i_{~j}(x'^0, {\bs x'}, {\bs p'}) &:=&
\left[\f{(\Lambda^{-1})^0_{~\mu} p'^\mu}{p'^0}\right]^{1/2}\; \times
\\&&\quad
\Lambda^i_{~\nu} \, c^\nu_{~j} \left( (\Lambda^{-1})^0_{~\mu} x'^\mu,  (\Lambda^{-1})^i_{~\mu} x'^\mu, (\Lambda^{-1})^i_{~\mu} p'^\mu  \right).
\ee
\ese
Substituting the formulas~\eqref{e:LT-heuristics}--\eqref{e:new_coefficients} into the Langevin   equations~\eqref{e:RLE_time-change-ito}, one finds that the particle's trajectory $(\bs X'(t'),\bs P'(t'))$ in $\Gs'$ is governed by an SDE of the form~\cite{1997DeMaRi,2005DuHa,2005Zy}
\bse\label{e:RLE'}
\be\label{e:RLE'-a}
\dX'^\ga(t')
&=&(P'^\ga/{P'^0})\; \dt',\\
\dP'^i(t')\label{e:RLE'-b}
&=& {A'^i}\; \dt' + {c'^i}_j * \diff B'^j(t').
\ee
\ese
It should be noted that, due to the $[\,\cdot\,]$-prefactors in Eqs.~\eqref{e:new_coefficients}, the coefficients $A^\mu$ and ${c^\nu}_\mu$ do \emph{not} form Lorentz tensors. This is essentially a consequence of the coordinate-time parameterization in Eqs.~\eqref{e:RLE_time-change-ito} and \eqref{e:RLE'}.\footnote{The  coefficients $\hat{A}^i$ of the corresponding $\tau$-parameterized process are the spatial components of a Lorentz four-vector $\hat{A}^\mu$; similarly, the quantities $\hat{D}^{ij}=\sum_r{\hat{c}^i}_{~r}{\hat{c}^k}_{~r}$ are spatial components of a second rank Lorentz tensor.}
Furthermore, it is straightforward now to obtain the corresponding SDE for the energy $P'^0=(M^2+\bs P'^2)^{1/2}$ in $\Gs'$ as well as the Fokker-Planck equation by replacing unprimed with primed quantities in Eqs.~\eqref{e:RLE_0-b} and~\eqref{e:FPE-time-change}, respectively.\footnote{In particular, if the coefficients of the lab frame Langevin equation satisfy the relativistic Einstein relation~\eqref{e:einstein-time-change}, then the stationary momentum distribution in the moving frame $\Gs'$ is given by a boosted J\"uttner distribution of the form $\phi'(\bs p')\propto \exp(\gb U'_\ga p'^\ga)$, where $U_\ga'$ is the four-velocity of the heat bath in $\Gs'$.}


\subsection{Relativistic binary collision model}
\label{s:relativistic_binary_collision_model}

The preceding subsections have focused on general aspects of relativistic Langevin and Fokker-Planck equations. Similar to the nonrelativistic case, relativistic SDEs present a useful tool for analytical and numerical studies of relaxation processes in relativistic systems. Stochastic models of this type provide a simplified picture of the underlying  microscopic dynamics. In order for the Langevin approach to be successful, one must know in advance which friction coefficient function~$\ga(P)$ and noise amplitude $D(P)$ are appropriate for the system under consideration. Realistic friction models can be obtained, e.g., by deriving Fokker-Planck equations from relativistic Boltzmann equations~\cite{1970Ak,2007ChKr}. In this context, however, it should be noted that the validity of the relativistic Boltzmann equation~\cite{CercignaniKremer} is less understood than that of its nonrelativistic counterpart. In the remainder of this section, we will sketch an alternative procedure for obtaining friction coefficients and noise amplitudes from a simple microscopic interaction model~\cite{2006DuHa}. The latter can be viewed as the direct relativistic generalization of the elastic binary collision model from Section~\ref{s:nonrelativistic_collision_model}. More precisely, we consider a one-dimensional system consisting of a heavy Brownian particle (mass $M$) which is embedded into a heat bath of smaller particles (mass $m\ll M$, total number $N\gg 1$). Our model assumes that the stochastic motion of a Brownian particle arises due to frequent elastic interactions with the surrounding heat bath particles.  Similar to Section~\ref{s:nonrelativistic_collision_model}, we are interested in finding the \lq best\rq\space approximation of the \lq exact\rq\space dynamics within the class of SDEs defined by Eq.~\eqref{e:RLE_post}.

\paragraph*{Relativistic collision kinematics}
To begin with, we consider a single collision of the Brownian particle (momentum $P$, energy $E=P^0$) with a heat bath particle (momentum $p$, energy $\eps$). The relativistic energy, momentum and velocity of the two particles before the collision are given by
\bse\label{e:rel_kinematics}
\begin{align}
P&=MV\,\gc(V),&
E(P)=&\left(M^2+P^2\right)^{1/2},\\
p&=mv\,\gc(v),&
\eps(p)=&\left(m^2+p^2\right)^{1/2}.
\end{align}
\ese
where $\gc(v):=\left(1-v^2\right)^{-1/2}$. Considering elastic interactions, the collision kinematics is governed by the relativistic mass-energy-momentum conservation laws
\be\label{e:kinematics_relativistic}
\hat{M}=M,\qquad
\hat{m}=m,\qquad
E+\eps=\hat{E}+\hat{\eps},\qquad
P+p=\hat{P}+\hat{p},
\ee
where hat-symbols refer to the state after the collision. Inserting Eqs.~\eqref{e:rel_kinematics} into the conservation laws~\eqref{e:kinematics_relativistic}, and solving for the momentum of Brownian particle after the collision, $\hat{P}$, we obtain~\cite{2006DuHa}
\bse
\be
\hat{P}&=&\gc(u)^2\,[2u\, E-(1+u^2)\,P],
\ee
where the collision-invariant center-of-mass velocity $u$ is given by 
\be\label{e:relativistic_relative_velocity}
u(p,P)=\f{P+p}{E+\eps}.
\ee
\ese
Accordingly, the momentum change $\Gd P_r:=\hat{P}-P$ of the Brownian particle in a single collision with the heat bath particle \lq$r$\rq\space is given by
\be\label{e:rel_delta}
\Gd P_r=
-2\gc(u_r)^2\, \f{\eps_r}{E+\eps_r} \;P
+2\gc(u_r)^2\, \f{E}{E+\eps_r}\; p_r,
\ee
where ${u}_r:= u(p_r,P)$ and $\eps_r:= \eps(p_r)$. In the non-relativistic limit case, where $u_r^2\ll 1$, $E\simeq M$ and  $\eps_r\simeq m$, Eq.~\eqref{e:rel_delta} reduces to Eq.~\eqref{e:p-tilde-nonrel}.
\par
Furthermore, by making the same assumptions as in Section~\ref{s:nonrelativistic_collision_model}, we find that the momentum change $\gd P(t):= P(t+\gd t)-P(t) $ of the Brownian particle during a small-but-sufficiently-long time interval $[t,t+\gd t]$ can be approximated by
\bse\label{e:relativistic_almost_langevin}
\be
\gd P(t)
&\approx&\notag
-2\sum_{r=1}^N \gc(u_r)^2\, \f{\eps_r}{E+\eps_r}\; P(t)\;I_r(t,\gd t) +\\
&& \quad
2\sum_{r=1}^N
\gc(u_r)^2\, \f{E}{E+\eps_r} \;p_r\;I_r(t,\gd t).\qquad\quad
\label{e:initial_1_rel}
\ee
Formally, the collision indicator~$I_r(t,\gd t)$ is again given by [cf. Eq.~\eqref{e:indicator_taylor}]
\be\label{e:indicator_taylor_rel}
I_r(t,\gd t)&\approx&
\f{\gd t}{2}\;|v_r-V|\;\gd(x_r-X),
\ee
\ese
but now one has to use the relativistic velocities $V=P/(M^2+P^2)^{1/2}$ and $v_r=p_r/(M^2+p_r^2)^{1/2}$, respectively. Equation~\eqref{e:initial_1_rel} is the relativistic counterpart of Eq.~\eqref{e:binary_almost_langevin_full}. Heuristically, the first term on the rhs. of Eq.~\eqref{e:initial_1_rel} can again be interpreted as \lq friction\rq, while the second contribution may be viewed as \lq noise\rq.

\begin{figure}[t]
\centering
\vspace{0.5cm}
\includegraphics[width=9cm,angle=0]{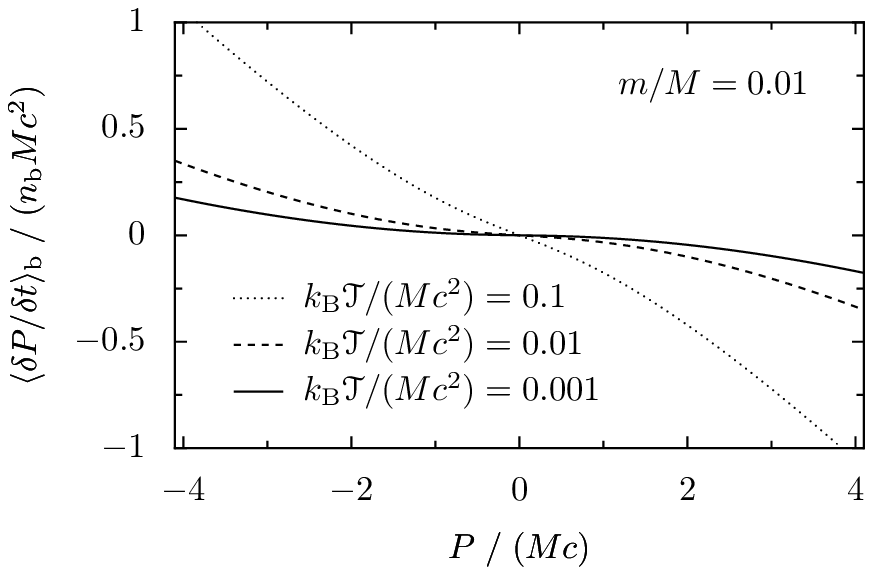}
\caption{\label{fig:relativistic_mean_drift}
Relativistic binary collision model. Mean drift force $\drift(P):=\lan \gd P(t)/\gd t \ran_\bath $ numerically evaluated from Eq.~\eqref{e:relativistic_mean_drift} for different values $\kB\Temp$, with $n_\bath=N/L$ denoting the number density of the heat bath particles.}
\end{figure}

\paragraph*{Bath distribution and drift}
Similar to the nonrelativistic case, Eqs.~\eqref{e:relativistic_almost_langevin} can be used to calculate the statistical moments of the momentum increments -- provided one specifies the phase space distribution of the heat bath particles. We will assume here that the heat bath is in a thermal equilibrium state, so that the one-particle phase space PDF is given by a spatially homogeneous J\"uttner function on $\Volume=[0,L]$, i.e.,
\be\label{e:relativistic_bath} 
f_\bath^1(x_r,p_r) = \label{e:rel_pdf_ansatz} 
(\Z_1 L)^{-1}\exp\bigl[-\gb (m^2+p_r^2)^{1/2}\bigr]\Gt(L-x_r)\Gt(x_r),
\ee 
where $\Temp=(\gb \kB)^{-1}$ is the temperature, and $\Z_1=2m\, K_1(\gb m)$ with $K_1(z)$ denoting the modified Bessel function.  With regard to our subsequent discussion, we are interested in calculating the mean drift force $\drift$, defined by\footnote{In principle, higher moments can be calculated in a similar manner, but then one has to specify the corresponding many-particle heat bath PDFs.}
\be
\drift(p):=\lan\f{\gd P(t)}{\gd t}\;\biggr|\;P(t)=p\ran_\bath
\ee
Inserting $\gd P(t)$ from Eq.~\eqref{e:initial_1_rel}, we find
\be\notag
\drift(p)
&=&
-n_\bath 
\lan \gc(u_r)^2\, \f{\eps_r}{E+\eps_r}\;L\;|v_r-V|\;\gd(x_r-X)\;\biggr|\;P(t)=p\ran_\bath P +\\ 
&&\label{e:relativistic_mean_drift}
n_\bath 
\lan \gc(u_r)^2\, \f{E}{E+\eps_r} \;p_r\;L\;|v_r-V|\;\gd(x_r-X)\;\biggr|\;P(t)=p\ran_\bath
\label{e:rel_drift_1},
\ee
where $n_\bath=N/L$ is the number density of the bath particles. 
In order to determine $\drift(p)$, we note that for some arbitrary function $G(p,P)$, we have
\be
&&\lan G(p_r,P)\,L\;|v_r-V|\;\gd(x_r-X)\;\biggr|\;P(t)=p\ran_\bath
=\notag
\\
&&\qquad
\notag
\Z_1^{-1}
\int \diff p_r\;G(p_r,P)\;
\exp\bigl[-\gb (m^2+p_r^2)^{1/2}\bigr]
\times\\
&&\qquad\qquad\qquad\qquad
\label{e:rel_drift_2}
\biggl|\f{p_r}{(m^2+p_r^2)^{1/2}}-\f{P}{(M^2+P^2)^{1/2}}\biggr|.
\ee
The rhs. of Eq.~\eqref{e:rel_drift_1} involves the functions
\bse
\be
G_1(p_r,P)&:=&\gc(u_r)^2\, \f{\eps(p_r)}{E(P)+\eps(p_r)},\\
G_2(p_r,P)&:=&\gc(u_r)^2\, \f{E(P)}{E(P)+\eps(p_r)} \;p_r.
\ee
\ese
Unfortunately, it is very difficult or perhaps even impossible to analytically evaluate the integral~\eqref{e:rel_drift_2} for the functions $G_{1/2}$. Figure~\ref{fig:relativistic_mean_drift} depicts the mean drift force $\drift$, obtained by numerically integrating the formula~\eqref{e:rel_drift_2} for different values of $P$.

\paragraph*{Langevin approximation} 
We conclude this section by discussing how one could, in principle, approximate Eqs.~\eqref{e:relativistic_almost_langevin} by, e.g., a nonlinear backward  Ito SDE of the form~\eqref{e:RLE_post}, reading
\be\label{e:RLE_post_repeated}
\dP(t)=-\ga(P)\, P\,\dt + [2D(P)]^{1/2}\bullet\dB(t).
\ee
The considerations for the two-component J\"uttner gas from Section~\ref{s:Rel_Thermo} imply that the stationary momentum distribution of the Brownian particle in the binary collision model is given by the J\"uttner function 
\be
\phi_\J(p)=\exp[-\gb(p^2+M^2)^{1/2}]/[2M\;K_1(\gb M)].
\ee
Hence, in order for Eq.~\eqref{e:RLE_post_repeated} to yield the expected J\"uttner  distribution, the functions $\ga$ and $D$ must be coupled by the relativistic Einstein relation~\eqref{e:Einstein_relativistic},
\be\label{e:Einstein_relativistic_repeated}
D(p)=\gb^{-1}\ga(p) \,E(p),
\ee
where $E=(p^2+M^2)^{1/2}$. In order to determine the function $\ga$, we demand that the Langevin equation
yields the same mean drift force $\drift$ as the collision model, i.e.,
\be\label{e:mean_value_criterion_repeated}
\lan \f{\dP(t)}{\diff t}\;\biggl|\; P(t)=p\ran
&\overset{!}{=}&
\lan\f{ \gd P(t)}{\gd t}\;\biggl|\;P(t)=p\ran_\bath.
\ee
For the post-point (backward-Ito) Langevin equation~\eqref{e:RLE_post_repeated}  we know that [cf. Eq.~\eqref{e:backward_ito_expectation}]
\be
\lan \f{\dP(t)}{\diff t}\;\biggl|\; P(t)=p\ran
=-\ga(p)\,p + \f{\diff}{\diff p} D(p).
\ee
Thus, by means of the Einstein relation~\eqref{e:Einstein_relativistic_repeated}, the lhs. of Eq.~\eqref{e:mean_value_criterion_repeated} is given by
\be
\lan \f{\dP(t)}{\diff t}\;\biggl|\; P(t)=p\ran=
-\ga(p)\,p +\gb^{-1} \f{\diff}{\diff p}[\ga(p) \,E(p)],
\ee
and the condition~\eqref{e:mean_value_criterion_repeated} becomes equivalent to the differential equation [cf. Eq.~\eqref{e:nonrelativistic_ODE}]
\be\label{e:mean_value_criterion_repeated_ode}
-\ga(p)\,p +\gb^{-1} \f{\diff}{\diff p}[\ga(p) \,E(p)]=\drift(p).
\ee
In the case of the collision model, where the function $\drift(p)$ is not exactly known, one could, e.g., try to fit $\drift(p)$ by a simple analytic expression and, subsequently, use this approximation in Eq.~\eqref{e:mean_value_criterion_repeated_ode}.


\section{Non-Markovian diffusion processes in Minkowski spacetime}
\label{s:rel_diff}
\label{s:relativistic_diffusion}

The preceding section has focused on relativistic Brownian motions in phase space. In the remainder we will discuss relativistic diffusion models in Minkowski spacetime, i.e., continuous relativistic stochastic processes that do not explicitly depend on the momentum coordinate. On the one hand, such spacetime processes may be constructed, e.g., from a Brownian motion process in phase space by integrating out the momentum coordinates. As a result of this averaging procedure, the reduced process for the position coordinate will be non-Markovian. Alternatively, one can try to derive or postulate a relativistic diffusion equation and/or diffusion propagators in spacetime on the basis of microscopic models~\cite{1950Go,1970VK,1996MaWe,1998BoPoMa} or plausibility considerations~\cite{2007DuTaHa}. Regardless of the approach adopted, in order to comply with the principles of special relativity,  the resulting spacetime process must be non-Markovian, in accordance with the results of Dudley (Theorem~11.3 in~\cite{1965Du}) and Hakim (Proposition~2 in~\cite{1968Ha}). Roughly speaking, this means that any   relativistically acceptable generalization of the classical diffusion equation~\eqref{e:intro_diffusion_equation} should be of at least second order in the time coordinate. 
\par
The construction and analysis of relativistic diffusion models in Minkowski spacetime poses an interesting problem in its own right. Additionally, the investigation of these processes becomes relevant in view of potential analogies with relativistic quantum theory~\cite{2005OrHo,1984GaEtAl}, similar to the analogy between Schr\"odinger's equation and the diffusion equation~\eqref{e:intro_diffusion_equation}  in the nonrelativistic case~\cite{Ryder96,Roepstorff}. The present section intends to provide an overview over classical relativistic diffusion models that have been discussed in the literature~~\cite{1965Ha,2005OrHo,1950Go,1974Ka,1996MaWe,1998BoPoMa,2000KoLi,2001KoLi,2001HePa,2005Ko,2007Ko,2007DuTaHa}. For this purpose, we first recall basic properties of the Wiener (Gaussian) process, which constitutes the standard paradigm for nonrelativistic diffusions in position space (Section~\ref{s:nonrelativistic_diffusion}). Subsequently, relativistic generalizations of the nonrelativistic diffusion equation~\eqref{e:intro_diffusion_equation} and/or the nonrelativistic Gaussian diffusion propagator will be discussed~\cite{2007DuTaHa}.

\subsection{Reminder: nonrelativistic diffusion equation}
\label{s:nonrelativistic_diffusion}

We start by briefly summarizing a few relevant facts about the standard nonrelativistic diffusion equation~\cite{1855Fi,Becker,Gardiner,Roepstorff}
\be\label{e:diffusion}
\f{\p}{\p t} \gr=\D\,\nabla^2 \gr
\csp 
t\ge t_0,
\ee
where $\D>0$ denotes the spatial diffusion constant, and $\gr(t,\bs x)\ge 0$ the one-particle PDF for the particle positions $\bs x\in \R^d$ at time $t$. Within classical diffusion theory, Eq.~\eqref{e:diffusion} is postulated to describe the (overdamped) random motion of a representative particle in a fluctuating environment (heat bath). In particular, Eq.~\eqref{e:diffusion} refers to the rest frame of the bath.
\par
There exist several well-known ways to motivate or derive the phenomenological diffusion  equation~\eqref{e:diffusion} by means of microscopic models (see, e.g., \cite{1855Fi,Becker,Gardiner,Roepstorff}). With regard to our subsequent discussion of relativistic alternatives, it is useful to briefly consider a \lq hydrodynamic\rq\space derivation~\cite{1975Forster}, which starts from the continuity equation
\be\label{e:diffusion_continuity}
\f{\p}{\p t} \gr(t,\bs x)=-\nabla\cdot \bs  j(t,\bs  x),
\ee
where $\bs j(t,\bs x)$ denotes the current density vector. In order obtain a closed equation for the density $\gr$, the current $\bs j$ has to be expressed in terms of $\gr$. One way of doing this is to  postulate the following rather general ansatz \{cf. Eq.(2.81) in~\cite{1975Forster}\}
\be\label{e:current_ansatz}
\bs j(t,\bs x)=-\nabla\int_{t_0}^t\dt'\; K(t-t')\;\gr(t',\bs x),
\ee
where, in general, $K$ may be a memory kernel. However, considering for the moment the memory-less kernel function\footnote{The factor \lq$2$\rq\space in Eq.~\eqref{e:current_ansatz_diffusion} appears because of the convention \mbox{$\int_{t_0}^t\dt'\;\gd(t-t') f(t')= f(t)/2.$}}
\be\label{e:current_ansatz_diffusion} 
K_\mrm{F}(t-t'):=2\D\;\gd(t-t'),
\ee
one finds
\be 
\bs j_\mrm{F}(t,\bs x)=-\D\;\nabla\gr(t,\bs x).
\ee
Upon inserting this expression into the continuity equation~\eqref{e:diffusion_continuity}, we recover the classical diffusion equation~\eqref{e:diffusion}.
\par
\begin{figure}[b]
\centering
\vspace{0.5cm}
\includegraphics[width=7.5cm]{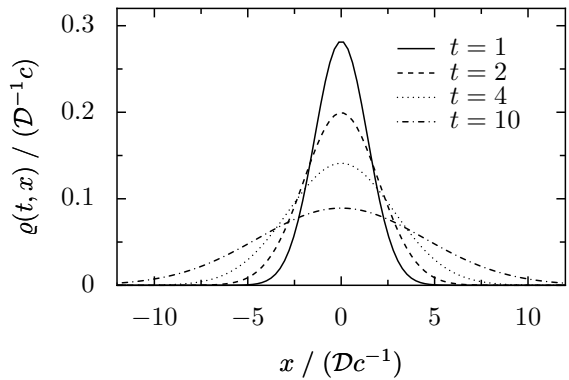}
\caption{Spreading of the Gaussian PDF $\gr(t,x)=p(t,x|0,0)$ from Eq.~\eqref{e:solution_non-rel} at different times~$t$, where~$t$ is measured in units of $\D/c^2$. At initial time $t=t_0=0$, the PDF corresponds to a $\gd$-function centered at the origin.
\label{RelDiff_NRD} }
\end{figure}

Now, it has been well-known for a long time that the diffusion equation~\eqref{e:diffusion} is in conflict with the postulates of special relativity. To briefly illustrate this, we specialize to simplest case of $d=1$ space dimensions, where  $\nabla^2={\p^2}/{\p x^2}$. In this case, the propagator of Eq.~\eqref{e:diffusion} at times $t> t_0$ is given by the Gaussian
\be\label{e:solution_non-rel}
p(t,x|t_0,x_0)=
\left[\f{1}{4\pi\, \D (t-t_0)}\right]^{1/2}\,
\exp\biggl[-\f{(x-x_0)^2}{4\D (t-t_0)}\biggr].
\ee
The propagator~\eqref{e:solution_non-rel} represents the solution of Eq.~\eqref{e:diffusion} for the initial condition 
$$
\gr(t_0,x)=\gd(x-x_0). 
$$
That is, if $X(t)$ denotes the random path of a particle with fixed initial position $X(t_0)=x_0$, then $p(t,x|t_0,x_0)\diff x$ gives the probability that the particle is found in the infinitesimal volume element $[x,x+\diff x]$ at time~$t>t_0$. As evident from Eq.~\eqref{e:solution_non-rel}, for each $t>t_0$ there is a small, but non-vanishing probability that the particle may be observed at distances $|x-x_0|>c(t-t_0)$, where $c=1$ is the speed of light in natural units. The evolution of the nonrelativistic Gaussian PDF from Eq.~(\ref{e:solution_non-rel}) is depicted in~Fig.~\ref{RelDiff_NRD}.
\par
It is worthwhile to summarize a few essential properties of Eqs.~\eqref{e:diffusion} and \eqref{e:solution_non-rel}: Equation~\eqref{e:diffusion} is a linear parabolic partial differential equation. Due to the linearity, more general solutions may be constructed by superpositioning, i.e.,  by integrating the solution~\eqref{e:solution_non-rel} over some given initial PDF $\gr_0(x_0)$. Equation~\eqref{e:diffusion} describes a Markov process which means that  the transition PDF~\eqref{e:solution_non-rel} satisfies the Chapman-Kolmogoroff criterion  
\be\label{e:Markov-condition}
p(t,x|t_0, x_0)=
\int_{\R} \diff x_1\;p(t, x|t_1,x_1)\;
p(t_1, x_1|t_0,x_0)
\quad
\ee
for all $t_1\in (t_0,t)$. The corresponding diffusion process $X(t)$ can be characterized in terms of the following SDE:
\be\label{e:nonrelativistic_diffusion_process}
\diff X(t)=(2\D)^{1/2}*\diff B(t), \qquad X(t_0)=x_0,
\ee
where $B(t)$ is a standard Wiener process as defined in Section~\ref{s:OUP}. Formally, Eq.~\eqref{e:nonrelativistic_diffusion_process} may be obtained from the Langevin equations~\eqref{e:OUP_external} of the classical Ornstein-Uhlenbeck process with  $\Force\equiv 0$ as follows: First we rewrite Eq.~\eqref{e:OUP_external-b} as
\be
\label{e:OUP_langevin_math_rewritten}
\f{\diff V(t)}{M\ga}&=&- V \dt+\left(\f{2D_0}{M^2 \ga^2}\right)^{1/2}*\diff B(t). 
\ee 
Upon letting $(M\ga)\to\infty$ and $D_0\to\infty$ such that $\D=D_0/(\ga M)^2$ remains constant, the lhs. of Eq.~\eqref{e:OUP_langevin_math_rewritten} should become negligible. Then, by making use of $\dX=V\dt$, Eq.~\eqref{e:nonrelativistic_diffusion_process} is recovered.\footnote{Debbasch and Rivet~\cite{1998DeRi} discuss the difficulties that arise when attempting a similar reduction for the relativistic Ornstein-Uhlenbeck process.} This limiting procedure defines the so-called overdamped regime of the Ornstein-Uhlenbeck process. The mean square displacement of the overdamped process~\eqref{e:nonrelativistic_diffusion_process} is given by~\cite{Becker} 
\be
\lan [X(t)-X(t_0)]^2\ran
&:=&\notag
\int \diff x\;(x-x_0)^2\;p(t,x|t_0,x_0)\\
&=&
2\D\,(t-t_0),
\ee
qualitatively similar to the asymptotic behavior of the classical Ornstein-Uhlenbeck process; cf. ~Eq.~\eqref{e:OUP_moments}. Finally, we note that the solution of Eq.~\eqref{e:diffusion} with initial condition
\be
\gr(t_0,x)\equiv\gr_0(x),
\ee
can be expressed in terms of the Feynman-Kac formula~\cite{KaSh91,Grigoriu}
\be
\gr(t,x)=
\lan \gr_0\bigl(x+(2\D)^{1/2} B(t)\bigr) \ran
\label{e:diffusion_representation},
\ee
where $\lan\,\cdot\,\ran$ indicates an average with respect to the Wiener measure of the standard Wiener process $B(t)$ with initial condition $B(t_0)=0$. Equation~\eqref{e:diffusion_representation} yields an efficient Monte-Carlo simulation scheme for computing the solutions of the diffusion equation~\eqref{e:diffusion} for a broad class of initial distributions $\gr_0$.


\subsection{Telegraph equation}
\label{s:telegraph}

The problem of constructing continuous diffusion models which, in contrast to the classical  nonrelativistic equations~\eqref{e:diffusion} and \eqref{e:solution_non-rel}, avoid superluminal velocities, has attracted considerable interest over the past years~\cite{1965Ha,2005OrHo,1950Go,1974Ka,1996MaWe,1998BoPoMa,2000KoLi,2001KoLi,2001HePa,2005Ko,2007Ko,2007DuTaHa,2008AlBeGa}.  Nonetheless, it seems fair to say that a commonly accepted solution is still outstanding. Apart from the theoretical challenge of developing a consistent relativistic diffusion theory, there exist several practical applications including, e.g., the analysis of data from high energy collision experiments~\cite{2004AbGa,2005AbGa,2004Wo} or the diffusion of light through turbid media~\cite{1989Is,2001GrRi,2003SaDaSj} and foams~\cite{2003MiSt,2005MiMaSt,2006MiSaFa}. In this context, a frequently considered alternative to the classical diffusion equation~\eqref{e:diffusion} is given by the telegraph equation~\cite{1922Ta,1950Go,1996MaWe,1989JoPr,1990JoPr,2000KoLi,2001KoLi,2004AbGa,2005AbGa,2005Ko,2007Ko}
\be\label{e:telegraph}
\left(\tau_v \f{\p^2}{\p t^2}+\f{\p}{\p t}\right)\, \gr=
\D\,\nabla^2 \gr.
\ee 
Here, $\D>0$ plays again the role of a diffusion constant, while $\tau_v>0$ is an additional relaxation time parameter. Equation~\eqref{e:telegraph} can also be obtained from the continuity equation~\eqref{e:diffusion_continuity}, provided one uses the exponential memory kernel (Section 2.10 in Ref.~\cite{1975Forster})
\be\label{e:current_ansatz_telegraph} 
K_\mrm{T}(t-t'):=\f{\D}{\tau_v}\;\exp[-(t-t')/\tau_v],
\ee
instead of $K_\mrm{F}$ from Eq.~\eqref{e:current_ansatz_diffusion}.  
Similar to Eq.~\eqref{e:diffusion}, the telegraph equation~\eqref{e:telegraph} refers to a special frame where the background medium, causing the random motion of the diffusing test particle, is at rest (on average). The \lq nonrelativistic limit\rq\space corresponds to letting $\tau_v\to 0$ in Eq.~\eqref{e:telegraph}, which leads back to Eq.~\eqref{e:diffusion}. For $\tau_v>0$,  Eq.~\eqref{e:telegraph} is a linear hyperbolic partial differential equation. Because of the second order time derivative in Eq.~\eqref{e:telegraph}, one now also has to specify the first order time derivative of the initial distribution at time $t_0$. 
\par
Considering particular initial conditions
\be\label{e:telegraph_initial}
\gr(t_0,x)=\gd(x-x_0),\qquad\qquad 
\f{\p}{\p t}\gr(t_0,x)\equiv 0,
\ee
one finds that the corresponding solution of Eq.~\eqref{e:telegraph} is given by~\cite{1950Go,1996MaWe}
\bse\label{e:telegraph_propagator}
\be
p(t,x|t_0,x_0)&=&\notag
\f{e^{-(t-t_0)/(2\tau_v)}}{2}\;
\biggl\{\gd[|x-x_0|-v(t-t_0)]+\\
&&\qquad\qquad\qquad
\f{\Gt(\xi^2)}{2\tau_v v}\left[I_0(\xi)+\f{t}{2\tau_v}\f{I_1(\xi)}{\xi}\right]
\biggr\}
\label{e:telegraph_propagator_1}.
\ee
Here, we have abbreviated
\be
\xi:=\f{1}{2}
\left[\left(\f{t-t_0}{\tau_v}\right)^2-
\left(\f{x-x_0}{\tau_v v}\right)^2\right]^{1/2},
\qquad\qquad
v:=(\D/\tau_v)^{1/2},
\ee
\ese 
and the modified Bessel functions of the first kind, $I_{\nu}(z)$, are defined by
\be\notag
I_\nu(z):=\sum_{k=0}^\infty\f{1}{\Gc(k+\nu+1)\;k!} \left(\f{z}{2}\right)^{2k+\nu}
\ee
with $\Gc(z)$ denoting the Euler gamma function.  According to our knowledge, the solution~\eqref{e:telegraph_propagator} was first obtained by Goldstein in 1938/1939. Actually, Goldstein derived the result~\eqref{e:telegraph_propagator} by considering the continuum limit of a persistent random walk model~\cite{1922Fu}; subsequently, he proved that this function satisfies the telegraph equation~\eqref{e:telegraph}, cf. Section~8 of his paper~\cite{1950Go}. 
\par
The propagator~\eqref{e:telegraph_propagator} is characterized by two salient features:
\begin{enumerate}
\item 
As evident from the $\gd$-function term, the solution exhibits two singular diffusion fronts traveling at absolute velocity $v:=(\D/\tau_v)^{1/2}$ to the left and right, respectively.
\item
 Due to the appearance of the Heaviside $\Gt$-function, the solution is non-zero only within the region~$|x-x_0|\le v(t-t_0)$, i.e.,  upon fixing $\tau_v$ such that $\mbox{$v=c=1$}$ the solution vanishes outside the light cone. 
\end{enumerate}
\par
Thus, in contrast to the nonrelativistic propagator~\eqref{e:solution_non-rel}, Eqs.~\eqref{e:telegraph_propagator} define a relativistically acceptable diffusion model. Because of the second order time derivative, the telegraph equation~\eqref{e:telegraph} describes a \emph{non-Markovian} process, in accordance with the aforementioned theorems of Dudley~\cite{1965Du} and Hakim~\cite{1968Ha}. The non-Markovian character of the propagator~\eqref{e:telegraph_propagator}  can also be proven directly by verifying that this solution does not fulfill the condition~\eqref{e:Markov-condition}.
\par
The linearity of Eq.~\eqref{e:telegraph} implies that more general solutions can be obtained  by integrating the propagator~\eqref{e:telegraph_propagator} over some given initial distribution~$\gr_0(x_0)$. In principle, one may also construct other classes of solutions with \mbox{${\p}\gr(t_0,x)/{\p t}\not\equiv 0$}, e.g., by applying a Laplace-Fourier transformation~\cite{1992MaPoWe,1992MaPoWe_2,1996MaWe} to Eq.~\eqref{e:telegraph}. We note, however, that in order for the solution $\gr(t,x)$ to remain normalized and positive at all times $t>t_0$, additional constraints on the initial conditions must be imposed, cf. Eq.~\eqref{e:telegraph_initial}. Various solutions and extensions of the telegraph equation~\eqref{e:telegraph}, including different types of boundary conditions, additional external sources, etc., have been discussed, e.g., by Goldstein~\cite{1950Go}, Masoliver et al.~\cite{1992MaPoWe,1992MaPoWe_2}, Foong~\cite{1993Fo}, Foong and Kanno~\cite{1992FoKa}, Renardy~\cite{1993Re}, and Dorogovtsev~\cite{1999Do}.
\par
Similar to the nonrelativistic diffusion equation~\eqref{e:diffusion}, the telegraph equation~\eqref{e:telegraph} may be derived and/or motivated in many different ways. A detailed overview  is given by Masoliver and Weiss~\cite{1996MaWe}, who discuss four different possibilities of deducing Eq.~\eqref{e:telegraph} from underlying models; see also Koide~\cite{2005Ko,2007Ko}. During the past decades, the telegraph equation~\eqref{e:telegraph} has been used to describe a number of different phenomena. The applications include:
\begin{itemize}
\item
\emph{Transmission of electrical signals.} 
According to Masoliver and Weiss~\eqref{e:telegraph}, the earliest derivation of the telegraph equation is based on a paper by William Thomson~\cite{1854Kelvin} (who later became Lord Kelvin), published in 1855. He considered the problem of how to transmit electrical signals without distortion, a question closely related to the design of the first transatlantic cable.  
\item
\emph{Continuum models of persistent diffusion.} 
It seems that the concept of persistent diffusion was first proposed in 1917/1922 by F\"urth~\cite{1917Fu,1922Fu} who aimed at describing the random motion of biological objects. Independently, a similar approach was suggested by Taylor~\cite{1922Ta} in an attempt to treat turbulent diffusion~\cite{1974Ka}. F\"urth and Taylor considered discrete models, assuming that a particle moves with constant absolute velocity between neighboring lattice points. At each lattice point, the particle is either back-scattered or transmitted, with the transmission probability being larger than the back-scattering probability (\emph{persistence}). A few decades later, in 1950, Goldstein~\cite{1950Go} demonstrated for the 1D case\footnote{Bogu\~n\'a et al.~\cite{1998BoPoMa} discuss persistent random walks in higher space dimensions.} that, for a suitable choice of the transition probabilities,  the continuum limit of this model leads to the telegraph equation. Hence, in contrast to the ordinary diffusion equation~\eqref{e:diffusion}, the telegraph equation~\eqref{e:telegraph} relies on asymmetric transition probabilities, causing the non-vanishing probability concentration at the diffusion fronts. In more recent years, persistent diffusion models have been employed to describe the propagation of photons in thin slabs and foams~\cite{1989Is,1999BoPoMa,2003MiSt,2005MiMaSt,2006MiSaFa}.
\item
\emph{Heat transport and propagation of heat waves.}  
In this case, the function $\gr(t,x)$ in Eq.~\eqref{e:telegraph} is interpreted as a temperature field and the normalization condition is dropped. For a detailed account of the vast literature on heat waves we refer to the review articles of Joseph and Preziosi~\cite{1989JoPr,1990JoPr}.
\item
\emph{High energy ion collision experiments.}  In recent years, the telegraph equation has been used to estimate the dissipation of net charge fluctuations, which may obliterate signals of QCD phase transitions in nuclear collisions~\cite{2004AbGa,2005AbGa}. In this context, however, the coordinate $x$ in Eq.~\eqref{e:telegraph} is interpreted as a rapidity variable.  
\end{itemize}
Another interesting aspect of the telegraph equation is elucidated in a paper by Kac~\cite{1974Ka}. He observed that the solutions of Eq.~\eqref{e:telegraph} with initial conditions
\be
\gr(t_0,x)\equiv\gr_0(x),\qquad\qquad 
\f{\p}{\p t}\gr(t_0,x)\equiv 0,
\ee
may be expressed in the form\footnote{The result~\eqref{e:poisson_representation} may be generalized to an arbitrary number of space dimensions; cf. pp. 500 in Kac's paper~\cite{1974Ka}.} 
\be
\gr(t,x)&=&\notag
\f{1}{2}\lan \gr_0\biggl(x-v\int_{t_0}^t\ds\;(-1)^{N(s)}\biggr ) \ran+\\
&& \label{e:poisson_representation}
\quad\f{1}{2}\lan \gr_0\biggl(x+v\int_{t_0}^t\ds\;(-1)^{N(s)}\biggr ) \ran,
\ee
where $v=(\D/\tau_v)^{1/2}$, and $\lan\,\cdot\,\ran$ indicates an average with respect to the $\tau_v$-parameterized Poisson process $N(t)$; i.e.,  for any given
time $t>t_0$ we have
\bse\label{e:poisson_process}
\be
\mrm{Prob}\{N(t)=k\}=\f{e^{-(t-t_0)/(2\tau_v)}}{k!}\left(\f{t-t_0}{2\tau_v}\right)^k,\qquad k=0,1,2,\ldots;
\ee
and for any finite sequence $t_0<t_1<\ldots <t_n$ the increments 
\be
N(t_1)-N(t_0),\, N(t_2)-N(t_1),\,\ldots,\, N(t_n)-N(t_{n-1})
\ee
\ese
are independent. Equation~\eqref{e:poisson_representation} is the direct counterpart of Feynman-Kac formula~\eqref{e:diffusion_representation} for the classical diffusion equation. Equations~\eqref{e:poisson_representation} and \eqref{e:poisson_process} together provide an efficient Monte-Carlo simulation scheme for computing solutions of the telegraph equation~\eqref{e:telegraph}. Moreover, the Poisson path integral representation~\eqref{e:poisson_representation} discloses an interesting correspondence between the free-particle Dirac equation~\cite{Peskin} and the telegraph equation~\eqref{e:telegraph}, which was first pointed out by Gaveau et al.~\cite{1984GaEtAl} in 1984: The solutions of both equations may be linked by means of an analytic continuation. This connection is similar to the relation between the diffusion equation~\eqref{e:diffusion} and the free particle Schr\"odinger equation in the nonrelativistic case.\footnote{For further reading about path integral representations of the Dirac propagator we refer to the papers of Ichinose~\cite{1982Ic,1984Ic}, Jacobson and Schulman~\cite{1984JaSc}, Barut and Duru~~\cite{1984BaDu}, and Gaveau and Schulman~\cite{1987GaSc}; see also footnote 7 in Gaveau et al.~\cite{1984GaEtAl} and problem 2-6, pp. 34-36 in Feynman and Hibbs~\cite{FeynmanHibbs}.} The crucial difference is given by the fact that the measures of the functional integration refer to different underlying processes, respectively.
\par
However, the telegraph equation~\eqref{e:telegraph} is not the only possible relativistic generalization of the nonrelativistic diffusion equation~\eqref{e:intro_diffusion_equation} and, recently, there has been some controversy about its applicability and validity~\cite{2000KoLi,2001KoLi,2001HePa}. An early critical discussion of Eq.~\eqref{e:telegraph} in the context of relativistic heat transport was given by van Kampen~\cite{1970VK} in 1970. Starting from a simple microscopic model, consisting of a cloud of material particles that exchange electromagnetic radiation, van Kampen derived an integral equation for the temperature of the particles as function of time and space. He then showed how the telegraph equation~\eqref{e:telegraph} can be recovered as an approximation to the more precise integral equation, but that the validity of this approximation breaks down in the vicinity of the diffusion fronts. 
\par
Similarly, the singular diffusion fronts predicted by Eq.~\eqref{e:telegraph_propagator} represent a source of concern if one wishes to adopt the telegraph equation~\eqref{e:telegraph} as a model for particle transport in a random medium. While these singularities may be acceptable in the case of photon diffusion~\cite{1989Is,1999BoPoMa,2003MiSt,2005MiMaSt,2006MiSaFa}, they seem unrealistic for massive particles, because such fronts would imply that a finite fraction of particles carries a huge amount of kinetic energy (much larger than $mc^2$).  In view of these shortcomings,  it appears reasonable to explore other constructions of relativistic diffusion processes~\cite{1984Ka,2007DuTaHa}. In the next part we will discuss a different approach~\cite{2007DuTaHa} that may provide a viable alternative to the solutions of the telegraph equation.


\subsection{Relativistic diffusion propagator}
\label{s:action}
In principle, one can distinguish two different routes towards constructing relativistic diffusion processes: One can either try to find an acceptable relativistic diffusion equation, or one can focus directly on the structure of the diffusion propagator. In the present part we shall consider the latter approach~\cite{2007DuTaHa}. The basic idea is to rewrite the nonrelativistic diffusion propagator~\eqref{e:solution_non-rel} in such a form that its relativistic generalization follows in a straightforward manner. This can be achieved be reexpressing Eq.~\eqref{e:solution_non-rel} in terms of an integral-over-actions.
\par
For this purpose, we consider a nonrelativistic particle traveling from the event
$\bar x_0=(t_0, x_0)$ to $\bar x=(t, x)$ and assume that the  
particle can experience multiple scatterings on its way, and
that the velocity is approximately constant between two successive 
scattering events. Then the total action (per mass) required along the path 
is given by
\be\label{e:action_nonrelativistic}
a(\bar x|\bar x_0)=\f{1}{2}\int_{t_0}^t\diff t'\; v(t')^2,
\ee
where the velocity $ v(t')$ is a piecewise constant function, satisfying
\be
x=x_0+\int_{t_0}^t\diff t'\; v(t').
\ee
Clearly, the nonrelativistic action~\eqref{e:action_nonrelativistic} becomes minimal for the deterministic (direct) 
path, i.e.,  if the particle does \emph{not} collide at all. In this case, 
it moves with constant velocity 
\mbox{$ v(t')\equiv( x- x_0)/(t-t_0)$} for 
all $t'\in[t_0,t]$, yielding the smallest possible action value
\be\label{e:action_1}
a_-(\bar x|\bar x_0)=\f{(x-x_0)^2}{2(t-t_0)}.
\ee
On the other hand, to match the boundary conditions it is merely required that the mean velocity equals $(x- x_0)/(t-t_0)$.  Consequently, in the nonrelativistic case, the absolute velocity of a particle may become arbitrarily large during some intermediate time interval $[t',t'']\subset [t_0,t]$. Hence, the largest possible action value is \mbox{$a_+(\bar x,\bar x_0)=+\infty$}. These considerations put us in the position to rewrite the Wiener propagator~\eqref{e:solution_non-rel} as an integral-over-actions:
\bse\label{e:action-rep}
\be\label{e:action-rep-a}
p(\bar x|\bar x_0)&\propto&
\int_{a_-(\bar x|\bar x_0)}^{a_+(\bar x|\bar x_0)} 
\diff a\; \exp\biggl(-\f{a}{2\D}\biggr),
\ee 
supplemented by the normalization condition
\be
1=\int\dx\; p(\bar x|\bar x_0).
\ee 
\ese
The representation~\eqref{e:action-rep} may be generalized to the relativistic case in a straightforward manner: One merely needs to insert the corresponding relativistic 
expressions into the boundaries of the integral~\eqref{e:action-rep-a}. 
A commonly considered relativistic generalization of 
Eq.~\eqref{e:action_nonrelativistic}, based on the particle's proper time,  reads~\cite{Weinberg}
\be\label{e:action_relativistic}
a(\bar x|\bar x_0)=-\int_{t_0}^t\diff t'\;\left[1- v(t')^2\right]^{1/2}.
\ee
Analogous to the nonrelativistic case, the relativistic 
action~\eqref{e:action_relativistic} assumes its 
minimum $a_-$ for the deterministic (direct) path 
from $ x_0$ to $ x$, characterized by a constant velocity 
$ v(t')\equiv( x- x_0)/(t-t_0)$. One explicitly obtains
\bse\label{e:rel_PDF_best}
\be\label{e:action_2}
a_-(\bar x|\bar x_0)
=-\left[(t-t_0)^2-( x- x_0)^2\right]^{1/2},
\ee 
i.e.,  $a_-$ is the negative Minkowski distance 
of the two spacetime events $\bar x_0$ and~$\bar x$.
The maximum action value $a_+=0$ is realized for particles 
that move at light speed.\footnote{In general, particles must undergo reflections in order to match the spatial boundary conditions.}
Hence, the transition PDF for the relativistic generalization 
of the Wiener process reads
\be\label{e:rel_PDF_new}
p(\bar x|\bar x_0)=
\mcal{N}^{-1}\;\left\{\exp\biggl[-\f{a_-(\bar x,\bar x_0)}{2\D}\biggr]
-1\right\},
\ee
\ese
if $( x- x_0)^2\le (t-t_0)^2$, and $p(\bar x|\bar x_0)\equiv 0$ 
otherwise, with $a_-$ determined by Eq.~\eqref{e:action_2}. 
\par
The relativistic diffusion process described by Eqs.~\eqref{e:rel_PDF_best} is \emph{non-Markovian}, i.e., it does not fulfill Chapman-Kolmogoroff criterion~\eqref{e:Markov-condition}. The  functional form of the propagator~\eqref{e:rel_PDF_new} remains the same for higher space dimensions $d>1$. The  normalization constants $\mcal{N}_d$ for $d=1,2,3$ read
\be 
\mcal{N}_d&=&\mcal{N}_d'-\f{u^d}{d} O_d,
\ee
where $u:=t-t_0$, $O_d={2\pi^{d/2}}/{\Gc(d/2)}$ is surface area of the $d$-dimensional unit sphere, and $\mcal{N}_d'$ can be expressed in terms of modified Bessel functions of the first kind $I_n$ and modified Struve functions $L_k$ \cite{AbSt72}, as
\bse
\be
\mcal{N}_1'&=& u\;\pi\,
\left[I_1(\chi)+L_{-1}(\chi)\right],\\
\mcal{N}_2'&=& u^2\;\f{2\pi}{\chi^2}\,
\left[1+(\chi-1)\;\exp(\chi)\right],\\
\mcal{N}_3'&=&
u^3\;\f{2\pi^2}{\chi^2}\,
\left\{\chi\left[I_2(\chi)+L_0(\chi)\right]-2L_1(\chi)\right\},
\qquad
\ee
\ese
with $ \chi={u}/{(2\D)}$.
\par
\begin{figure}[t]
\centering
\vspace{0.5cm}
\includegraphics[width=8.5cm]{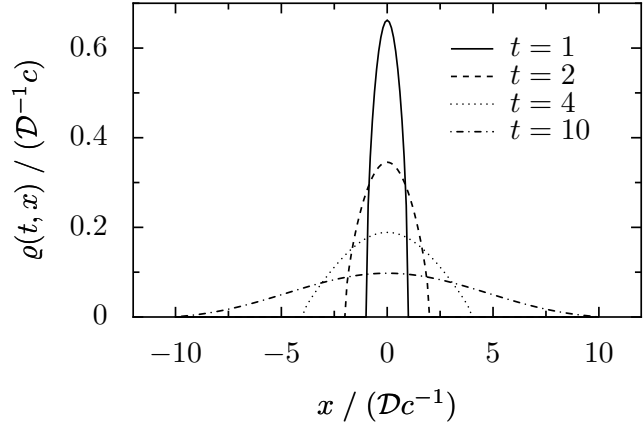}
\caption{Transition PDF $\gr(t,x)= p(t,x|0,0)$ for the one-dimensional ($d=1$) relativistic diffusion process~\eqref{e:rel_PDF_best} at different times $t$ (measured in units of $\D/c^2$). At time $t=t_0=0$, the function $\gr(t,x)$ reduces  to a $\gd$-function centered at $x_0=0$. In contrast to the nonrelativistic diffusion propagator, cf. Fig.~\ref{RelDiff_NRD}, the PDF~\eqref{e:rel_PDF_best} vanishes outside of the light cone.  
\label{RelDiff_PDF}}
\end{figure}
\begin{figure}[t]
\centering
\vspace{0.5cm}
\includegraphics[width=9cm]{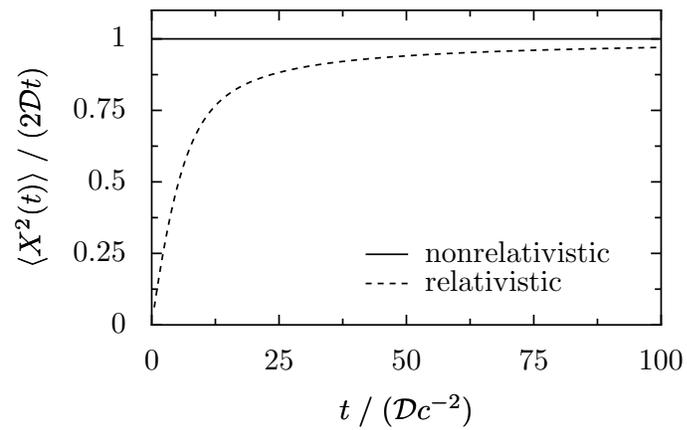}
\caption{Comparison of the mean square displacements $\lan X^2(t)\ran$, divided by $2\D t$, for the one-dimensional (\mbox{$d=1$}) nonrelativistic Wiener process~\eqref{e:solution_non-rel} and its relativistic generalization from Eq.~\eqref{e:rel_PDF_best} with initial condition $(t_0,x_0)=(0,0)$. 
\label{RelDiff_MSD}
}
\end{figure}

In contrast to the solution~\eqref{e:telegraph_propagator} of the telegraph equation, the propagator~\eqref{e:rel_PDF_new} vanishes continuously at the diffusion fronts. Figure~\ref{RelDiff_PDF} depicts the PDF $\gr(t,x)= p(t,x|0,0)$ of the 
diffusion process~\eqref{e:rel_PDF_best} for the one-dimensional 
case \mbox{$d=1$} at different times $t$. The corresponding mean 
square displacement is plotted in Fig.~\ref{RelDiff_MSD} (dashed curve). 
\par
It is also interesting to note that the PDF~\eqref{e:action-rep} 
is a special case of a larger class of diffusion processes, defined by
\be\label{e:embedding}
p_w(\bar x|\bar x_0)&=&
\mcal{N}[w]^{-1}
\int_{a_-(\bar x|\bar x_0)}^{a_+(\bar x|\bar x_0)} \diff a\; w(a),
\qquad
\ee 
where $w(a)\ge 0$ is a weighting function, and $\mcal{N}[w]$ the 
time-dependent normalization constant. In particular, Eq.~\eqref{e:embedding} may be viewed as a path integral definition in 
the following sense: Physically permissible paths from $\bar x_0$ to $\bar x$ 
have action values (per mass) $a$ in the range $[a_-,a_+]$. Grouping the 
different paths together according to their action values, one may assign 
to each such class of paths, denoted by $\mcal{C}(a;\bar x,\bar x_0)$, the 
statistical weight $w(a)$. The integral~\eqref{e:embedding} can then be 
read as an integral over the equivalence classes $\mcal{C}(a;\bar x,\bar x_0)$ 
and their respective weights~$w(a)$. The nonrelativistic Wiener process 
corresponds to the specific choice $w(a)=\exp[-a/(2\D)]$; hence, it appears  
natural to define the relativistic generalization by using the same weighting 
function. It is, however, worth mentioning that a very large class of functions $w(a)$ 
yields an asymptotic growth of the spatial mean square displacement 
that is proportional to $t$, corresponding to \lq ordinary\rq\space 
diffusion. Moreover, Eq.~\eqref{e:embedding} could also be used to 
describe super-diffusion or sub-diffusion processes~\cite{1990BoGe,2000MeKl,1999Ca}, whose asymptotic  mean square displacements grow as $t^\ga,\;\ga\ne 1$.\footnote{This  can be achieved, e.g., by choosing the integral 
boundaries as \mbox{$\tilde a_-= ( x- x_0)^2/(t-t_0)^\ga$}, $\ga\ne 1$ and 
$a_+=\infty$, but then the variable $a$ may not be 
interpreted as a conventional action anymore.}

\section{Outlook}
\label{s:summary}

We would like to conclude our discussion by summarizing problems which, in our opinion, deserve further consideration in the future: 
\begin{itemize}
\item \emph{Microscopic models.} The one-dimensional binary collision model from Section~\ref{s:relativistic_binary_collision_model} appears to be the simplest example for motivating relativistic Langevin-type equations by means of an underlying microscopic model. Future investigations should focus on constructing relativistic Langevin equations from more precise particle-field interaction models; e.g., one could consider the motion of a classical relativistic point particle in quasi-static external random fields. If successful, this approach would yield more precise noise and friction models for relativistic systems.\footnote{In this context, technical and conceptual challenges are likely to arise when considering the energy loss of the particle due to radiation emission~\cite{1938Di,1948El,1999Ro,2000Ro,2000Sp,2006ArMaDo}.} In particular, this would contribute to clarifying under which circumstances stochastic differential equations may provide a feasible approximation to complex relativistic systems.  A useful starting point for future research in this direction could be the work of Blanco et al.~\cite{1983BlPeSa}, Johnson and Hu~\cite{RQBM2000JoHu,RQBM2001JoHu,RQBM2002JoHu} and  Galley et al.~\cite{RQBM2006Ga}, who proposed to model the interaction between quantum test particles and vacuum fluctuations of quantum fields by means of effective Langevin equations.

\item \emph{Relativistic processes driven by non-Gaussian noise.}
The most frequently studied relativistic stochastic differential equations~\cite{2006DuHa,1997DeMaRi,1998DeRi,2001BaDeRi,2001BaDeRi_2,2005OrHo,2005DuHa,2005DuHa_2,2006Fa,2004Zy,2005Zy,2007ChDe_1,2007ChDe_2,2008Ba_1,2008Ba_2} are driven by Brownian motion (Wiener) processes which couple to the momentum coordinates. It would be interesting to also consider other driving processes (e.g., Poisson or L\'evy noise) and to compare with the results of the corresponding nonrelativistic equations~\cite{2005SoKl,1999JeMeFo,2008DeHoHa}. On the one hand, the properties of such generalized relativistic processes deserve to be studied from a purely mathematical point of view, e.g., with regard to potential modifications of normal or anomalous diffusion effects.  On the other hand, it would be desirable to identify classes of physical systems that can be described by these processes. For example, it seems likely that the quasi-random particle acceleration~\cite{1948Fe} in various astrophysical systems may be efficiently modeled by non-Gaussian driving processes. 

\item \emph{Relativistic fluctuation theorems.}
Fluctuation theorems are mathematical relations that may be used to deduce certain thermodynamic properties of equilibrium systems by measuring suitable averages from different realizations of a non-equilibrium process~\cite{1997Ja,2005Se,2006DoEtAl,2007Ja,2007TaLuHa,2008TaHaMo}. During the past two decades, numerous fluctuation theorems have been established for various closed and open nonrelativistic systems. Both from a theoretical and practical point of view, it would be most interesting to generalize these results to the relativistic case. Recently, a first step in this direction was made by Fingerle~\cite{2007Fi}, who derived a relativistic fluctuation theorem for the special relativistic Brownian motion process proposed in~\cite{2005DuHa}. It will be worthwhile to extend these investigations to other, more general relativistic processes (see also Cleuren et al.~\cite{2008ClEtAl}). 

\item \emph{Relativistic diffusion models in Minkowski space.}
The above problems refer to stochastic processes in relativistic phase space. Alternatively, one may focus on constructing and analyzing novel types of relativistic diffusion processes in Minkowski spacetime, similar to those discussed in Section~\ref{s:rel_diff}. One particularly important issue in this context concerns the existence of reasonable path integral representations for the propagators of such processes. The latter question is closely related to the problem of finding path integral formulations of relativistic quantum propagators~\cite{1982Ic,1984Ic,1984BaDu,FeynmanHibbs}.
 
\item \emph{Relativistic quantum Brownian motions.}
Another potential generalization of the above ideas concerns the construction of relativistic quantum Brownian motion processes. Recent efforts in this direction include the aforementioned papers by Johnson and Hu~\cite{RQBM2000JoHu,RQBM2001JoHu,RQBM2002JoHu} and Galley et al.~\cite{RQBM2006Ga}, who considered the motion of quasi-particles in fluctuating quantum fields. Similar concepts have also been studied within the theory of stochastic semiclassical gravity, where the gravitational field plays the role of a \lq Brownian particle\rq\space with the vacuum fluctuations of quantum fields forming a stochastic environment (\lq bath\rq); for details we refer to the recent review by Hu and Verdaguer~\cite{lrr-2004-3}. Another promising starting point may be the work of Pechukas~\cite{1991Pe}, and Tsonchev and Pechukas~\cite{2000TsPe}, who developed a simple elastic collision model of nonrelativistic quantum Brownian motions. In this context, we also mention the work of Breuer and Petruccione~\cite{RQDIFF1998BrPe,RQDIFF1998BrPe_1,RQDIFF1998BrPe_2}, who proposed a  Langevin equation approach to describe quantum state diffusion in the framework of special relativity, see also Diosi~\cite{RQDIFF1998Di}. 

\item \emph{Extensions to general relativity.} 
Last but not least, notwithstanding recent progress~\cite{1998KlSh,2005Ra,2005Ra_1,2007ChDe,2005Ri,2004De,2005RiDe,2007FrLJ,2007Fi,2007ChDe}, the generalization of stochastic concepts and their applications within the framework of general relativity offers many interesting challenges for the future.
\end{itemize}


\section*{Acknowledgments}

It is a great pleasure to thank our colleagues David Cubero, Fabrice Debbasch, Werner Ebeling, Igor Goychuk, Stefan Hilbert, Gert Ingold, Siegmund Kohler, Benjamin Lindner, Jes{\'u}s Casado-Pascual, Lutz Schimansky-Geier, Peter Talkner, Stefan Weber and Martijn Wubs for many helpful discussions and comments.

\addcontentsline{toc}{section}{Symbols}
\section*{Symbols}

\label{s:symbols}

\setlongtables
\begin{longtable}{ll}

$M$        &  rest mass of the Brownian particle\\
$m$        &  rest mass of a heat bath particle\\
$\Gs$        & inertial laboratory frame := rest frame of the heat bath\\
$\Gsm;\; \Gsc$ & moving frame; comoving rest frame of the Brownian particle\\
$\Obs;\; \Obsm$    &  lab observer;  moving observer \\
$t$ & time coordinate\\
$\tau$ & proper time of the Brownian particle\\
$c$ & vacuum speed of light (set to unity throughout, i.e., $c=1$)\\
$d$ & number of space dimensions\\

$\bs X, \bs x $ & position coordinate $\bs X=(X^i)$, $i=1,\ldots, d$\\
$\bs V, \bs v$ & particle velocity\\
$\bs w $ & observer velocity\\
$\bs P, \bs p$ & momentum coordinates\\
$E, P^0, \eps,p^0 $ & particle energy\\

$\eta=(\eta_{\ga\gb}) $ &Minkowski metric tensor\\
$\Gl$ & Lorentz transformation (matrix)\\
$\gc$ & Lorentz factor $\gc(\bs v)=(1-\bs v^2)^{-1/2}$ \\

$\eve$ &spacetime event\\

$X^\ga$ & (contravariant) time-space four-vector  $(X^\ga)=(X^0,\bs X)$, $\ga=0,1,\ldots, d$ \\
$P^\ga$ &  energy-momentum four-vector, $(P^\ga)=(E,\bs P)$\\
$U^\ga$ &  velocity four-vector, $U^\ga=P^\ga/M$ \\

$f$ & one-particle phase space probability density\\
$\gr$ & one-particle position  probability density\\
$\phi$ & one-particle momentum  probability density\\
$\psi$ & one-particle velocity probability density\\

$j^\ga(t,\bs x)$ & four-vector current density\\
$\gt^{\ga\gb}(t,\bs x)$ & energy-momentum tensor\\

$\kB$ & Boltzmann constant\\
$\Temp$  & temperature \\
$\gb$  & inverse thermal energy $\gb:=(\kB\Temp)^{-1}$\\
$\Energy; \Energy^\ga$     & internal energy; thermodynamic energy-momentum four-vector \\
$\Ent$     & thermodynamic entropy \\
$\ent$     & relative entropy \\
$\Press$  & pressure \\
$\Vol$  & volume (number)\\

$\ga $ & friction coefficient \\
$D$  & noise coefficient \\
$\D$  & spatial diffusion constant \\
$\Wie(s)$ & $d$-dimensional standard Wiener process with time parameter $s$\\
$\Prob$ &  probability measure of the Wiener process\\

$\odot$ & placeholder for discretization rule in stochastic integrals, $\odot\in\{*,\circ,\bullet\}$\\
$*$ & Ito (pre-point) interpretation of the stochastic integral\\
$\circ$ & Stratonovich-Fisk (mid-point) interpretation of the stochastic integral\\
$\bullet$ & backward Ito (post-point) interpretation of the stochastic integral\\

$\R$        &  set of real numbers\\
$\Volume\subset\R^d$   & finite $d$-dimensional spatial volume set \\
$\hyp$      &  $d$-dimensional hyperplane in $(1+d)$-dimensional Minkowski spacetime\\
$\cone(\eve)$ & backward lightcone of the event $\eve$\\
$\iso(\xi^0)$ & isochronous hyperplane $\{(t,\bs x)\;|\;t=\xi^0\}$\\

$\gl$   & Lebesgue measure, e.g., $V=\gl(\Volume)$ \\
$\mu,\rho$   & measures \\
$\Ind$    &  indicator function with values in $\{0,1\}$\\
$\EW{X}$    &  expected value of the random variable or process $X$ 

\end{longtable}



 \appendix

\section{Stochastic integrals and calculus}
\label{as:stochastic_calculus}

This appendix summarizes the most commonly considered stochastic integral definitions and the corresponding rules of stochastic calculus. For a more rigorous and comprehensive introduction, we refer to, e.g.,  Refs.~\cite{KaSh91,Grigoriu,Gardiner,1996Risken}.
\par
We consider a Wiener process (standard Brownian motion) $B(t)$ as defined in Section~\ref{s:OUP}; i.e.,  the increments $\diff B(t):=B(t+\dt)-B(t)$ are stochastically independent~\cite{KaSh91,Grigoriu} and characterized by the Gaussian distribution
\be\label{a-e:OUP_langevin_math_density}
\Prob\{\dB(t)\in[y,y+\diff y]\}=\left(2\pi\diff t\right)^{-1/2}
\exp\bigl[-{y^2}/({2\diff t})\bigr]\diff y. 
\ee 
We are interested in defining integrals of the form
\be\label{e:goal}
I=\int_0^t f(Y(s))\odot\diff B(s),
\ee
where $f(y)$ is some real-valued function, $Y(s)$ a real-valued time-dependent process, and $\odot$ signals a discretization rule discussed below. If $B(s)$ were some ordinary differentiable function of $s\in[0,t]$, then the integral in  Eq.~\eqref{e:goal} would simply be given by\footnote{By writing Eq.~\eqref{e:goal_b}, it is implictly assumed that $f(y)$, $Y$ and $\dot{B}$ are sufficiently smooth functions so that this integral exists in the sense of Riemann-Stieltjes; in this case, the value of the integral~\eqref{e:goal_b} is independent of the underlying discretization scheme~\cite{Grigoriu}.}
\be\label{e:goal_b}
I=\int_0^t f(Y(s))\;\dot B(s) \;\diff s,
\ee 
where $\dot B=\dB/\diff s$. Unfortunately, $\dot B(s)$ is not well-defined for the Wiener process~\cite{Gardiner,KaSh91}, but it is possible to generalize the concept of integration to also include the Wiener process and other stochastic  processes~\cite{KaSh91,Grigoriu,Gardiner}. However, in contrast to the standard  Riemann-Stieltjes integral~\eqref{e:goal_b}, the integral with respect to a stochastic process may depend on the choice of the discretization scheme $\odot$  and, in particular, also require modifications of differential calculus.
\par
To illustrate these aspects for the most commonly considered stochastic integral definitions, we will always consider the following equidistant partition $\{t_0, t_1\ldots, t_N\}$ of the time interval $[0,t]$: 
\be\label{e:partition}
\Gd t=t_k-t_{k-1}=t/N, \quad 
k=1,\ldots,N,
\qquad
t_0=0,
\qquad
t_N=t.
\ee

\subsection{Ito integral}
\label{as:Ito_calculus}

We first summarize the properties of Ito's stochastic integral~\cite{1944Ito,1951Ito}. 
Its relationship to other stochastic integrals is discussed in Section~\ref{as:comparison}.

\subsubsection{One-dimensional case} 

The Ito stochastic integral of some real-valued function $f(Y(t))$ with respect to a standard Brownian motion process $B(t)$ over the time-interval $[0,t]$ can be defined by
\be\label{e:ito_definition}
\int_0^t f(Y(s))*\diff B(s)&:=&
\lim_{N\to\infty}  \sum_{k=0}^{N-1} f(Y(t_k)) 
\left[B (t_{k+1})-B(t_k)\right],
\ee
where the partition $\{t_0,\ldots, t_N\}$ is given by~\eqref{e:partition}. The peculiar, defining feature of this integral is that, on the rhs. of Eq.~\eqref{e:ito_definition}, the argument of the function $f$ must be evaluated at the lower boundary points $t_k$ of the discrete intervals~$[t_k,t_{k+1}]$; i.e., the definition of the Ito integral is \emph{non-anticipating}. Accordingly, the Ito discretization scheme is also known as the \emph{pre-point} rule. 
\par
Now consider a stochastic process $Y(t)$ which, for two given functions $A(y)$ and $C(y)$, is defined by
\be\label{e:stochastic_integral_ito}
Y(t)=
Y(0)+\int_0^t A(Y(s))\,\diff s +
\int_0^t C(Y(s))*\diff B(s),
\ee
and where the last term is interpreted as an Ito integral~\eqref{e:ito_definition}. Stochastic integral equations like Eq.~\eqref{e:stochastic_integral_ito} are usually abbreviated by rewriting them as an Ito \emph{stochastic differential equation} (I-SDE)
\be
\diff Y(t)&=&\label{e:I-SDE}
A(Y)\;\diff t+ C(Y)*\diff B(t),
\ee
complemented by the initial condition $Y(0)$. From the non-anticipating definition~\eqref{e:ito_definition} of the Ito integral and the properties of the Wiener process it follows that~\cite{Gardiner}\footnote{$\EW{\;\cdot\;|\; Y(t)=y} $ denotes the conditional expectation with respect to the Gaussian measure of the Wiener process $B(t)$.}
\be\label{e:conditional_ito}
\EW{C(Y)*\diff B(t)\;|\; Y(t)=y}=0.
\ee
The Fokker-Planck equation for the PDF $f(t,y)$ of the stochastic process defined by Eq.~\eqref{e:I-SDE} reads
\be
\parder{f}{t}=\parder{}{y}\left[-Af+\f{1}{2}\, \parder{}{y} ( C^2 f )\right],
\ee
where $A=A(y)$ and $C=C(y)$. A deterministic initial condition $Y(0)=y_0$ translates into $f(0,y)=\gd(y-y_0)$. 
\par
Finally, an important peculiarity arises when one considers nonlinear transformations $G$ of the stochastic process $Y(t)$. More precisley, assuming that $Y$ is defined by the I-SDE~\eqref{e:I-SDE}, then the differential change of the process $Z(t):=G(Y(t))$ is given by (see, e.g., Section 4.3.2 in \cite{Gardiner}) 
\be
\diff Z(t)\notag
&=&G'(Y)* \diff Y +\f{1}{2}\;C(Y)^2\; G''(Y)\;\dt\\
&=&\notag
\left[A(Y)\;G'(Y) +\f{1}{2}\;C(Y)^2\; G''(Y)\right]\dt
+C(Y)\;G'(Y)*\diff B(t),\\
\label{e:ito-formula}
\ee
where $G'(y)=\diff G(y)/\diff y$ and  $G''(y)=\diff^2 G(y)/\diff y^2$. Within ordinary differential calculus, the term containing $G''$ is absent. Equation~\eqref{e:ito-formula} is usually referred to as ~\emph{Ito formula}.

\subsubsection{The $n$-dimensional case} 
Consider the $n$-dimensional stochastic process $\bs Y(t)=(Y^1(t),\ldots, Y^n(t))$, defined by the following $n$-dimensional generalization of Eq.~\eqref{e:I-SDE}:
\be
\diff Y^i(t)&=&\label{e:I-SDE_d}
A^i(\bs Y)\;\diff t+ {C^i}_r(\bs Y)*\diff B^r(t),
\ee
where $i=1,\ldots,n$ and $r=1,\ldots, K$.  In Eq.~\eqref{e:I-SDE_d}, the Wiener processes $B^r(t)$ represent $K$ independent noise sources, and each term ${C^i}_r(\bs Y)*\diff B^r(t)$ symbolizes an Ito integral. The Fokker-Planck equation for the PDF $f(t,y^1,\ldots,y^n)$ reads 
\be
\parder{f}{t}=\label{e:I-FPE_d}
\parder{}{y^i}\left[- A^i f+
\f{1}{2}\,\parder{}{y^j}( {C^i}_r{C^j}_{r} f) \right].
\ee
The generalized Ito-formula reads (see, e.g., Section 4.3.2 in \cite{Gardiner})
\be
\diff G[\bs Y(t)]=
\left[A^i\;\p_i G +
\f{1}{2}\;{C^i}_r{C^j}_{r}\; \p_i\p_j G \right]\dt +
{C^i}_r\;\p_i G*\diff B^r(t),
\ee
where $\p_i:=\p/\p y^i$.


\subsection{Stratonovich-Fisk  integral}
\label{as:SF_calculus}

Next we summarize the properties of an alternative stochastic integral definition which was proposed by Stratonovich~\cite{1964St,1966St,1968St} and Fisk~\cite{1963Fisk,1965Fisk}. In contrast to the non-anticipating Ito integral, the Stratonovich-Fisk (SF) integral is \emph{semi-anticipating}, but satisfies the rules of ordinary stochastic calculus.

\subsubsection{One-dimensional case} 

The SF stochastic integral of some real-valued function $f(Y(t))$ with respect to a standard Brownian (Wiener)  motion process $B(t)$ over the time-interval $[0,t]$ can be defined by
\be
\int_0^t f(Y(s))\circ\diff B(s)&:=&\notag
\lim_{N\to\infty}  \sum_{k=0}^{N-1} 
\f{1}{2}\, [f(Y(t_{k+1}))+f(Y(t_{k}))]\times\\
&&\qquad\qquad\quad
\left[B (t_{k+1})-B(t_k)\right]
 \label{e:sf_definition}
\ee
where the partition $\{t_0,\ldots, t_N\}$ is given by~\eqref{e:partition}. In contrast to Ito's integral~\eqref{e:ito_definition}, the SF definition~\eqref{e:sf_definition} uses the mean of the boundary values of $f$ on the intervals~$[t_k,t_{k+1}]$; i.e., the definition of the SF integral is \emph{semi-anticipating}. This discretization scheme is also known as the \emph{mid-point} rule. 
\par
Similar to Eq.~\eqref{e:stochastic_integral_ito}, we may consider a stochastic process $Y(t)$ defined by
\be\label{e:stochastic_integral_sf}
Y(t)=
Y(0)+\int_0^t A(Y(s))\,\diff s +
\int_0^t C(Y(s))\circ\diff B(s),
\ee
where now the last term is interpreted as an SF integral~\eqref{e:sf_definition}. The integral equation~\eqref{e:stochastic_integral_sf} can be abbreviated in terms of the equivalent SF stochastic differential equation (SF-SDE)
\be
\diff Y(t)&=&\label{e:SF-SDE}
A(Y)\;\diff t+ C(Y)\circ\diff B(t),
\ee
with initial condition $Y(0)$. From the semi-anticipating definition~\eqref{e:sf_definition} of the SF integral and the properties of the Wiener process it follows that~\cite{Gardiner}
\be
\EW{C(Y)\circ\diff B(t)\;|\; Y(t)=y}=\f{1}{2}\, C(y)\,C'(y)\,\dt,
\ee
where $C'(y)=\diff C(y)/\diff y$. The Fokker-Planck equation for the PDF $f(t,y)$ of the stochastic process~\eqref{e:SF-SDE} reads 
\be
\parder{f}{t}
&=&\parder{}{y}\left[-Af+\f{1}{2}\, C \parder{}{y} ( C f )\right]
\ee
where $A=A(y)$, and $C=C(y)$. The deterministic initial condition $Y(0)=y_0$ translates into $f(0,y)=\gd(y-y_0)$. 
\par
It can be shown~\cite{Gardiner,Grigoriu} that the SF integral definition preserves the rules of ordinary stochastic calculus; i.e., if $Y(t)$ is defined by the SF-SDE~\eqref{e:SF-SDE}, then the differential change of the process $Z(t):=G(Y(t))$ is given by (see, e.g., Section 4.3.2 in \cite{Gardiner}) 
\be
\diff Z(t)\notag
&=&G'(Y)\circ \diff Y \\
&=&\notag
A(Y)\;G'(Y)\;\dt
+C(Y)\;G'(Y)\circ\diff B(t),\\
\label{e:sf-formula}
\ee
where $G'(y)=\diff G(y)/\diff y$. 
\par
However, as will be discussed in Section~\ref{as:comparison}, for a given SF-SDE with sufficently smooth coefficient functions $A$ and $C$ one can always find an I-SDE, which yields the same Fokker-Planck equation. Hence, in order to describe a certain physical process, one may choose that integral definition which is most convenient for the problem under consideration.

\subsubsection{The $n$-dimensional case} 
Consider the $n$-dimensional stochastic process $\bs Y(t)=(Y^1(t),\ldots, Y^n(t))$, defined by the following $n$-dimensional generalization of Eq.~\eqref{e:SF-SDE}:
\be
\diff Y^i(t)&=&\label{e:SF-SDE_d}
A^i(\bs Y)\;\diff t+ {C^i}_r(\bs Y)\circ \diff B^r(t),
\ee
where $i=1,\ldots,n$ and $r=1,\ldots, K$.  In Eq.~\eqref{e:SF-SDE_d}, the Wiener processes $B^r(t)$ represent $K$ independent noise sources, and each term ${C^i}_r(\bs Y)\circ \diff B^r(t)$ symbolizes an SF integral. The Fokker-Planck equation for the PDF $f(t,y^1,\ldots,y^n)$ reads 
\be
\parder{f}{t}=\label{e:SF-FPE_d}
\parder{}{y^i}\left[
- \left(A^i +\f{1}{2} {C^j}_r\parder{}{y^j} {C^i}_{r}\right) f+
\f{1}{2}\parder{}{y^j}({C^i}_r  {C^j}_{r}f) \right],
\ee
and the transformation rules of ordinary differential calculus apply. 


\pagebreak
\subsection{Backward Ito integral}
\label{as:hk_calculus}

We still consider a third stochastic integral definition which is also known as the backward Ito (BI) integral~\cite{KaSh91,MacKean}. 
Its relationship to the other stochastic integrals is discussed in Section~\ref{as:comparison}.

\subsubsection{One-dimensional case} 

The BI stochastic integral of some real-valued function $f(Y(t))$ with respect to $B(t)$ over the time-interval $[0,t]$ can be defined by
\be\label{e:hk_definition}
\int_0^t f(Y(s))\bullet\diff B(s)&:=&
\lim_{N\to\infty}  \sum_{k=0}^{N-1} f(Y(t_{k+1})) 
\left[B (t_{k+1})-B(t_k)\right],
\ee
where the partition $\{t_0,\ldots, t_N\}$ is given by~\eqref{e:partition}.  On the rhs. of Eq.~\eqref{e:hk_definition}, in contrast to the Ito and SF integrals, the argument of the function $f$ must be evaluated at the upper boundary points $t_{k+1}$ of the discrete intervals~$[t_k,t_{k+1}]$; i.e., the definition of this integral is \emph{anticipating}. This discretization scheme is also known as the \emph{post-point} rule. 
\par
Similar to above, we may consider a stochastic process $Y(t)$ which, for two given functions $A(y)$ and $C(y)$, is defined by
\be\label{e:stochastic_integral_hk}
Y(t)=
Y(0)+\int_0^t A(Y(s))\,\diff s +
\int_0^t C(Y(s))\bullet\diff B(s),
\ee
and where the last term is now interpreted as a BI integral~\eqref{e:hk_definition}. Equation~\eqref{e:stochastic_integral_ito} can be abbreviated by rewriting it as a backward Ito stochastic differential equation (BI-SDE)
\be
\diff Y(t)&=&\label{e:HK-SDE}
A(Y)\;\diff t+ C(Y)\bullet\diff B(t),
\ee
complemented by the deterministic initial condition $Y(0)$. From the anticipating definition~\eqref{e:hk_definition} of the BI integral and the properties of the Wiener process it follows that~\cite{Gardiner}
\be\label{e:backward_ito_expectation}
\EW{C(Y)\bullet\diff B(t)\;|\; Y(t)=y}= C(y)\,C'(y)\,\dt.
\ee
The Fokker-Planck equation for the PDF $f(t,y)$ of the stochastic process defined by Eq.~\eqref{e:HK-SDE} reads
\be
\parder{f}{t}=\parder{}{y}\left[-Af+\f{1}{2}\,C^2 \parder{}{y}f \right],
\ee
where $A=A(y)$ and $C=C(y)$. The deterministic initial condition $Y(0)=y_0$ translates into $f(0,y)=\gd(y-y_0)$. 
\par
It can be shown that, similar to the Ito integral, also the BI integral requires a modification of differential calculus. More precisley, assuming that $Y$ is defined by the BI-SDE~\eqref{e:HK-SDE}, the differential change of the process $Z(t):=G(Y(t))$ is given by
\be
\diff Z(t)\notag
&=&G'(Y)\bullet \diff Y -\f{1}{2}\;C(Y)^2\; G''(Y)\;\dt\\
&=&\notag
\left[A(Y)\;G'(Y) -\f{1}{2}\;C(Y)^2\; G''(Y)\right]\dt
+C(Y)\;G'(Y)\bullet\diff B(t),\\
\label{e:hk-ito-formula}
\ee
where $G'(y)=\diff G(y)/\diff y$ and  $G''(y)=\diff^2 G(y)/\diff y^2$. 

\subsubsection{The $n$-dimensional case} 
Consider the $n$-dimensional stochastic process $\bs Y(t)=(Y^1(t),\ldots, Y^n(t))$, defined by the following $n$-dimensional generalization of Eq.~\eqref{e:HK-SDE}:
\be
\diff Y^i(t)&=&\label{e:HK-SDE_d}
A^i(\bs Y)\;\diff t+ {C^i}_r(\bs Y)\bullet\diff B^r(t),
\ee
where $i=1,\ldots,n$ and $r=1,\ldots, K$.  In Eq.~\eqref{e:HK-SDE_d}, the Wiener processes $B^r(t)$ represent $K$ independent noise sources, and each term ${C^i}_r(\bs Y)\bullet\diff B^r(t)$ symbolizes a BI integral. The Fokker-Planck equation for the associated PDF $f(t,y^1,\ldots,y^n)$ reads 
\be\label{e:HK-FPE_d}
\parder{f}{t}=
\parder{}{y^i}\left[
- \left(A^i + {C^j}_r\parder{}{y^j} {C^i}_{r}\right) f+
\f{1}{2}\parder{}{y^j}({C^i}_r  {C^j}_{r}f) \right].
\ee
The generalized backward Ito-formula reads 
\be
\diff G[\bs Y(t)]=
\left[A^i\;\p_i G -
\f{1}{2}\;{C^i}_r{C^j}_{r}\; \p_i\p_j G \right]\dt +
{C^i}_r\;\p_i G\bullet\diff B^r(t),
\ee
where $\p_i:=\p/\p y^i$.


\subsection{Comparison of stochastic integrals} 
\label{as:comparison}

As anticipated in the preceding sections, the three different stochastic integrals/SDEs may be transformed into each other. In particular, a given Fokker-Planck equation can usually be realized by any of three SDE types, upon choosing the coefficient functions appropriately. To illustrate this by example, we reconsider the $n$-dimensional SDEs from above, assuming identical noise coefficients ${C^i}_r$ but different drift coefficients $A^i_{*|\circ|\bullet}(\bs Y)$, respectively, i.e.
\bse\label{e:SDEs}
\be
\diff Y^i(t)&=&\label{e:SDE-1}
A_*^i(\bs Y)\;\diff t+ {C^i}_r(\bs Y)*\diff B^r(t),\\
\diff Y^i(t)&=&\label{e:SDE-2}
A_\circ^i(\bs Y)\;\diff t+ {C^i}_r(\bs Y)\circ\diff B^r(t),\\
\diff Y^i(t)&=&\label{e:SDE-3}
A_\bullet^i(\bs Y)\;\diff t+ {C^i}_r(\bs Y)\bullet\diff B^r(t),
\ee
\ese
where $i=1,\ldots,n$ and $r=1,\ldots,K$. We would like to determine the drift coefficients such that these three different types of SDEs describe the same $n$-dimensional stochastic process $\bs Y(t)=(Y^1(t),\ldots, Y^n(t))$ on the level of the Fokker-Planck equations\footnote{For most practical purposes, two Markovian stochastic processes can be considered as physically equivalent if their PDFs are governed by the same Fokker-Planck equation.}, which can be compactly summarized as follows  
\be\label{e:fpe-combined}
\parder{f_\odot}{t}=
\p_i\left[
- \left(A_\odot ^i +\gl_{\odot}\, {C^j}_r\p_j {C^i}_{r}\right) f_\odot+
\f{1}{2}\p_j({C^i}_r  {C^j}_{r}f_\odot) \right],
\ee
where $\p_i:=\p/\p y^i$, and $\gl_*=0$, $\gl_\circ=1/2$, and $\gl_\bullet=1$. We distinguish three cases.

\paragraph*{Eq. \eqref{e:SDE-1} is given:}
Eq.~\eqref{e:fpe-combined} implies that Eqs.~\eqref{e:SDE-2} and \eqref{e:SDE-3} describe the same process like Eq.~\eqref{e:SDE-1}, if we fix
\be\label{e:comparison_results_1}
A_\circ^i= A_*^i-\f{1}{2}\,{C^j}_r \p_j {C^i}_{r}
\csp
A_\bullet^i= A_*^i-{C^j}_r \p_j {C^i}_{r}.
\ee

\paragraph*{Eq. \eqref{e:SDE-2} is given:}
Eqs.~\eqref{e:SDE-1} and \eqref{e:SDE-3} describe the same process like Eq.~\eqref{e:SDE-2}, if we fix
\be\label{e:comparison_results_2}
A_*^i= A_\circ^i+\f{1}{2}\,{C^j}_r \p_j {C^i}_{r}
\csp
A_\bullet^i= A_\circ^i-\f{1}{2}\,{C^j}_r\p_j {C^i}_{r} .
\ee

\paragraph*{Eq. \eqref{e:SDE-3} is given:}
Eqs.~\eqref{e:SDE-1} and \eqref{e:SDE-2} describe the same process like Eq.~\eqref{e:SDE-3}, if we fix
\be\label{e:comparison_results_3}
A_*^i= A_\bullet^i+ {C^j}_r\p_j {C^i}_{r} 
\csp
A_\circ^i= A_\bullet^i+\f{1}{2}\,{C^j}_r\p_j {C^i}_{r} .
\ee  
\par
To summarize, by means of Eqs.~\eqref{e:comparison_results_1}, \eqref{e:comparison_results_2} and~\eqref{e:comparison_results_3} one can change between the different forms of  stochastic integration and stochastic differential calculus, respectively. Each SDE type has advantages and disadvantages:  The Ito formalism is well suited for numerical simulations~\cite{1996Risken,1999KloedenPlaten,2004Glasserman} and yields a vanishing noise contribution to conditional expectations of the form~\eqref {e:conditional_ito}. The Stratonovich-Fisk approach is more difficult to implement numerically, but preserves the rules of ordinary differential calculus (in contrast to Ito/backward Ito integration). Finally, within the backward Ito scheme, fluctuation dissipation relations may take a particularly elegant form~(cf. Section 6.2 in Ref.~\cite{1982HaTh}, and Ref.~\cite{1994Kl}).

\subsection{Numerical integration}
\label{appendix:numerics}

A detailed introduction to the numerical simulation of SDEs can be found in~\cite{1996Risken,1999KloedenPlaten,2004Glasserman}. A simple Monte-Carlo algorithm for numerically integrating  Eqs. \eqref{e:SDEs} follows directly from the definition of the stochastic integrals. The corresponding discretization scheme, which works sufficiently well for many purposes, reads
\bse\label{e:SDE-num}
\be
Y^i(t+\Gd t)-Y^i(t)&=&\label{e:SDE-1-num}
A_*^i(\bs Y(t))\;\Gd t+ {C^i}_r(\bs Y(t))\;\Gd B^r(t),\\
Y^i(t+\Gd t)-Y^i(t)
&=&\notag
A_\circ^i(\bs Y(t))\;\Gd t+\\\label{e:SDE-2-num}
&&\quad\f{1}{2}[{C^i}_r(\bs Y(t+\Gd t))+
{C^i}_r(\bs Y(t))]\,\Gd B^r(t),\\
Y^i(t+\Gd t)-Y^i(t)
&=&\label{e:SDE-3-num}
A_\bullet^i(\bs Y(t))\;\Gd t+ {C^i}_r(\bs Y(t+\Gd t))\;\Gd B^r(t).
\ee
\ese
Here, the $\Gd B^r(t)$ are random numbers, sampled from a Gaussian normal distribution with density
\be\label{a-e:OUP_langevin_math_density-discrete}
\Prob[\Gd B^r(t)]=\left(\f{1}{2\pi\Gd t}\right)^{1/2}
\exp\biggl\{-\f{[\Gd B^r (t)]^2}{2\Gd t}\biggr\}. 
\ee 
As evident from Eqs.~\eqref{e:SDE-num}, for given functions $A_*^i$ and ${C^i}_r$, the discretized I-SDE~\eqref{e:SDE-1-num} allows for calculating the values $Y^i(t+\Gd t)$ directly from the preceding values $Y^i(t)$. By contrast, the discretized SF-SDEs~\eqref{e:SDE-2-num} and BI-SDEs~\eqref{e:SDE-3-num} are implicit equations, which must be solved for $Y^i(t+\Gd t)$. The latter difficulty can be avoided by transforming a given SF/BI-SDE to the corresponding I-SDE by means of Eqs.~\eqref{e:comparison_results_1}, \eqref{e:comparison_results_2} and~\eqref{e:comparison_results_3}.


\section{Surface integrals in Minkowski spacetime}
\label{appendix:surface_integrals}

We would like to integrate a tensorial quantity $\gt^{\mu\ga\gb\ldots}(t,\bs x)$ over a fixed $d$-dimensional hyperplane $\hyp$ in $(1+d)$-dimensional Minkowski space, e.g.,\footnote{If $\gt^{\mu\ga\gb\ldots}$ is a tensor of rank $n$ then the quantity $\Mom^{\ga\gb\ldots}[\hyp]$ from Eq.~\eqref{e:surface_integral} is tensor of rank $(n-1)$.}
\be\label{e:surface_integral}
\Mom^{\ga\gb\ldots}[\hyp]&:=&\int_\hyp \diff \gs_\mu\; \gt^{\mu\ga\gb\ldots}(t,\bs x).
\ee
Relevant examples are integrals over the energy-momentum tensor $\gt^{\ga\gb}$ or the current density $j^\ga$, as discussed in Appendix~\ref{s:relativistic_thermodynamics}. We outline the general procedure for the case of $d=3$ space dimensions, considering Cartesian coordinates with metric tensor  $(\eta_{\ga\gb})=\diag(-1,1,1,1)$. In this case, the surface element $\diff \gs_\mu$ may be expressed in terms of the alternating differential form~\cite{1967Ha_2}
\be\label{e:surface_integral_element}
\diff \gs_\mu=
\f{1}{3!}\veps_{\mu\ga\gb\gc}\dx^\ga\wedge \dx^\gb\wedge \dx^\gc,
\ee
where $\veps_{\mu\ga\gb\gc}$ is the Levi-Cevita tensor\footnote{The total antisymmetric covariant Levi-Cevita tensor $\veps_{\ga\gb\gc\gd}$ is 
$0$ if two or more indicies are equal, $+1$ for even permutations of the indices $(0123)$ and $-1$ for odd permutations. Similar to the metric tensor, the Levi-Cevita tensor is numerically invariant under Lorentz transformations with determinant $+1$. For a general discussion of the properties of  Levi-Cevita tensors we refer to Section 5.5 of Sexl and Urbantke~\cite{SexlUrbantke}.} and \lq$\wedge$\rq\space denotes the antisymmetric product
\bse\label{e:wedge_product}
\be\label{e:wedge_product_a}
\dx^\ga \wedge \dx^\gb=-\dx^\gb \wedge \dx^\ga,
\ee
implying that
\be\label{e:wedge_product_b}
\dx^0 \wedge \dx^0=\dx^1 \wedge \dx^1=\ldots=0. 
\ee
\ese
Inserting \eqref{e:surface_integral_element} and ordering differentials with the help of Eq.~\eqref{e:wedge_product_a}, the integral~\eqref{e:surface_integral} can be rewritten as
\be\label{e:surface_integral_2}
\Mom^{\ga\ldots}[\hyp]
&:=&\notag
\int_\hyp \dx^1\wedge \dx^2\wedge \dx^3\;\gt^{0\ga\ldots} - 
\int_\hyp \dx^0\wedge \dx^2\wedge \dx^3\;\gt^{1\ga\ldots} +\\
&&
\int_\hyp \dx^0\wedge \dx^1\wedge \dx^3\,\gt^{2\ga\ldots} -
\int_\hyp \dx^0\wedge \dx^1\wedge \dx^2\,\gt^{3\ga\ldots}.
\ee
With regard to thermodynamics, we are particularly interested in integrating over space-like or time-like surfaces $\hyp$ given in the form
\be
x^0=t=g(\bs x)=g(x^1,\ldots,x^d).
\ee
Typical examples are the isochronous hyperplane~$\iso(\xi^0)$ of an  inertial frame $\Gs$, defined by 
\bse
\be\label{e:isochronous}
\iso(\xi^0):=\{\;(t,\bs x)\;|\; t = \xi^0=:g(\bs x)\; \},
\ee
or the backward lightcone $\cone[\eve]$ of some spacetime event $\eve$  with coordinates $(\xi^0,\bs \xi)$ in $\Gs$, which is given by
\be\label{e:lightcone}
\cone(\eve)
:=
\{\;(t,\bs x)\;|\; t = \xi^0-|\bs x-\bs\xi|=:g(\bs x)\;\}.
\ee
\ese
Given such an explicit representation of the hyperplane, we may express the differential $\dx^0$ in Eq.~\eqref{e:surface_integral_2} in terms of $\diff x^i$ by using
\be
\dx^0=\p_i g\;\dx^i.
\ee
Inserting this expression into Eq.~\eqref{e:surface_integral_2} and taking into account Eqs.~\eqref{e:wedge_product}, we obtain
\be
\Mom^{\ga\ldots}[\hyp]
&=&\notag
\int_\hyp \dx^1\wedge \dx^2\wedge \dx^3\;\left[
\gt^{0\ga\ldots} - 
(\p_i g)\; \gt^{i\ga\ldots} \right]\\
&:=&
\int\diff^d x
\left[
\gt^{0\ga\ldots} (g(\bs x),\bs x) - (\p_i g)\; \gt^{i\ga\ldots}(g(\bs x),\bs x) \right].
\ee
Hence, in the case, we may write for the covariant surface element four-vector
\be
(\diff \gs_\ga)=n_\ga \diff^d x
\csp
n_\ga =(1,-\p_i g).
\ee
In particular, for the isochronous hyperplane  $\iso(\xi^0)$  from Eq.~\eqref{e:isochronous}, we have $g(\bs x)=\xi^0$ and $\p_i g=0$ in $\Gs$, yielding
\be
\Mom^{\ga\ldots}[\xi^0]
:=
\Mom^{\ga\ldots}[\iso(\xi^0)]
&=&
\int\diff^d x\;\gt^{0\ga\ldots}(\xi^0,\bs x).
\ee
For comparison, when integrating over the lightcone $\cone(\eve)$ from Eq.~\eqref{e:lightcone}, one has to use
\be
\p_i g=-\f{x^i-\xi^i}{|\bs x-\bs\xi|},
\ee
such that $n_\ga n^\ga=0$, yielding explicitly
\be\label{a-e:lc-average}
\Mom^{\ga\ldots}[\eve]
&=&\notag
\int \d^dx\;\biggl\{
\gt^{0\ga\ldots}(\xi^0-|\bs x-\bs\xi|,\bs x) \;+
\\ &&\qquad\qquad
\f{x^i-\xi^i}{|\bs x-\bs\xi|}\;
\gt^{i\ga\ldots}(\xi^0-|\bs x-\bs\xi|,\bs x)\biggr\}.
\ee

\paragraph*{Example 1: Gas in a cubic container.}
We calculate $\Mom^\ga[\eve]$ for the energy-momentum tensor $\gt^{\ga\gb}$ from Eq.~\eqref{e:stress-tensor-covariant-gas}, describing an isotropic, stationary,  spatially homogeneous gas in the lab frame $\Gs$. Considering a cubic vessel $\Volume=[-L/2,L/2]^d$ in $\Gs$, the components of $\gt^{\ga\gb}$ in $\Gs$ read\footnote{$\gt^{\ga\gb}$ from Eq.~\eqref{a-e:stress-tensor-covariant-gas} is the energy-momentum tensor per particle and has to be multiplied by the particle number $N$ to obtain the total energy and momentum of an $N$-particle system.}
\be\label{a-e:stress-tensor-covariant-gas}
\gt^{\ga\gb}(t,\bs x)
=
\Vol^{-1}\,\Ind(\bs x;\Volume)
\begin{cases} 
\energy,    &\qquad\ga=\gb=0,\\
\vir /d ,    &\qquad\ga=\gb=1,\ldots, d,\\
0,& \qquad\ga\ne\gb,
\end{cases}
\ee
where $V=L^d$. Since $\gt^{\ga\gb}$ is diagonal and time-independent in~$\Gs$, we find 
\be
\Mom^{0}[\eve]
=
\int\diff^d x\; \gt^{00} (g(\bs x),\bs x)
=
\Vol^{-1}\energy\, \int\diff^d x\; \Ind(\bs x;\Volume)
=
\energy.
\ee
The spatial components $\Mom^{i}[\eve]$ can be calculated by means of a partial integration
\be
\Mom^{i}[\eve]
&=&\notag
-\int\diff^d x \; (\p_i g)\;\gt^{ii}(g,\bs x) \\
&=&\notag
-(Vd)^{-1} \vir
\int\diff^d x \; (\p_i g)\;\Ind(\bs x;\Volume)\\
&=&\notag
(Vd)^{-1} \vir
\int\diff^d x \; g(\bs x)\;\p_i\Ind(\bs x;\Volume)\\
&=&
(Vd)^{-1} \vir
\int\diff^d x \; (\xi^0-|\bs x-\bs\xi|)\;
\p_i\Ind(\bs x;\Volume),
\ee
where, cf. Eq.~\eqref{e:em-divergence-b},
\be\notag
\p_i\Ind(\bs x;\Volume)
=
\left[\gd(L/2+x^i)-\gd(L/2-x^i)\right]
\prod_{j\ne i}\Gt(L/2+x^j)\,\Gt(L/2-x^j)
\ee
is antisymmtric with respect to $x^i$; hence
\be
\Mom^{i}[\eve]
&=&\notag
- (Vd)^{-1} \vir
\int\diff^d x \; |\bs x-\bs\xi|\;
\p_i\Ind(\bs x;\Volume).
\ee
In the one-dimensional case, where $V=L$ and  $d=1$, this integral can be solved exactly, and one finds
\be
\Mom^{1}[\eve]
&=&
- L^{-1}  \EW{p^1v^1} (|L/2+\xi^1|-|L/2-\xi^1|).
\ee
In particular, for a lab observer located in the origin $\xi^i=0$ we have $\Mom^{1}[\eve]=0$. In this example, the origin may be interpreted as the photographic center-of-mass, which can be defined by the condition  $\Mom^{1}[\eve]=0$ in the lab frame.
\par
If the picture is taken from outside the volume $\Volume=[-L/2,L/2]$, corresponding to the condition $L/|\xi^1|\le 1/2$, then
\be\label{a-e:photographic-drift-one-dimensional}
\Mom^{1}[\eve]
=
- \EW{p v}\;( {\xi^1}/{|\xi^1|});
\ee 
i.e., if the mean momentum is naively sampled from a photo taken outside of the photographic center-of-mass, then even the lab observer (who is not moving with respect to the system)  will find a non-vanishing momentum average value always pointing away from herself. In the case of a J\"uttner gas with $\vir=d/\gb$, the magnitude of this photographic drift effect is proportional to the temperature (and particle number).
\par
In the $d$-dimensional case, performing the $\dx^i$-integration first, we find
\be
\Mom^{i}[\eve]
&=&\notag
\f{\vir}{Vd}
\int\diff^{d/i} x \; 
[(x^1-\xi^1)^2+\dots(L/2-\xi^i)^2+\ldots]^{1/2}-\\
&&
\f{\vir}{Vd}
\int\diff^{d/i} x \; 
[(x^1-\xi^1)^2+\dots(L/2+\xi^i)^2+\ldots]^{1/2},
\ee
with the remaining $(d-1)$-dimensional integration (denoted by $\diff^{d/i} x $) ranging over $[-L/2,L/2]^{d-1}$.
Again, $\Mom^{i}[\eve]$ vanishes if $\xi^{i}=0$. 
In the limit $L/|\bs \xi|\ll 1/2$, corresponding to an observer position  far outside the volume, one may expand the integrands, yielding
\be
\Mom^{i}[\eve]
&\simeq&\notag
-(Vd)^{-1}\;\vir
\int\diff^{d/i} x \; L\;\xi^i\;
[(x^1-\xi^1)^2+\dots(\xi^i)^2+\ldots]^{-1/2}\\
&\simeq&\notag
-(Vd)^{-1}\;\vir\;({\xi^i}/{|\bs \xi|})\,
\int\diff^{d/i} x \; L\;\\
&=&\label{a-e:photographic-drift}
- d^{-1}\;{\vir}\;({\xi^i}/{|\bs \xi|}).
\ee
In the one-dimensional case this reduces to Eq.~\eqref{a-e:photographic-drift-one-dimensional}.

\paragraph*{Example 2: Arbitrary confinement.}
In the preceding example, the spatial density was taken to be piecewise constant with a singular drop-off at the boundaries (walls of the vessel). As a slightly more general example, consider an equilibrated gas in a smooth confinement, e.g., due to some external potential. Moreover, assume that there exists a preferred inertial frame $\Gs$, where the gas is stationary with an isotropic momentum density, so that it can be described by a Lorentz-scalar one-particle density of the form
\be\label{a-e:PDF_stationary}
\vphi(\bs x,\bs p)=\gr(\bs x)\;\phi(\bs p),
\ee
with normalized marginal distributions $\gr$ and $\phi$, and $\phi$ being spherically symmetric. In this case, the (kinetic) energy-momentum tensor 
\be
\gt^{\ga\gb}(t,\bs x):
=
\int \f{\diff^dp}{p^0}\;p^\ga p^\gb\;\vphi(\bs x,\bs p)
\ee
has components
\be\label{a-e:stress-tensor-covariant-gas-2}
\gt^{\ga\gb}(t,\bs x)
=
\gr(\bs x)
\begin{cases} 
\energy,    &\qquad\ga=\gb=0,\\
\EW{p^i v^i},    &\qquad\ga=\gb=i=1,\ldots, d,\\
0,& \qquad\ga\ne\gb,
\end{cases}
\ee
Then, from Eq.~\eqref{a-e:lc-average}, one obtains
\be\label{a-e:lc-0}
\Mom^{0}[\eve]
=
\energy \int \d^dx\; \gr(\bs x)=\energy,
\ee
and
\be\label{a-e:lc-i}
\Mom^{i}[\eve]
&=&
\EW{p^i v^i}\int \d^dx\;
\f{x^i-\xi^i}{|\bs x-\bs\xi|}\;
\gr(\bs x)=:\Mom^{i}(\bs \xi)
\ee
In this case, the photographic center-of-mass position $\bs \xi_*$ in $\Gs$ is defined by
\be
\Mom^{i}(\bs \xi_*)=0
\csp
i=1,\ldots, d;
\ee 
e.g., if $\gr$ is rotationally invariant with respect to some space point $\bs z$ in $\Gs$, then $\bs \xi_*=\bs z$. On the other hand, assuming  that the density $\gr$ vanishes outside a finite region $\Volume\in\R^d$ in $\Gs$ and that a stationary observer is located at a position $\bs \xi$ far away from $\Volume$, then we may approximate
\be\label{a-e:lc-i-outside}
\Mom^{i}[\eve]
&\simeq&
\EW{p^i v^i}\int \d^dx\;
\f{x^i-\xi^i}{|\bs\xi|}\;\gr(\bs x)
=
\EW{p^i v^i}\f{\EW{x^i}-\xi^i}{|\bs\xi|}.
\ee



\section{Relativistic thermodynamics}
\label{s:relativistic_thermodynamics}

In this appendix, we shall refer to the  J\"uttner gas~\cite{1957Sy} from Section~\ref{s:juettner_gas} in order to compare different formulations of relativistic thermodynamics~\cite{1907Ei,1908Pl,1963Ott,1968VK,1969VK_2}. This serves to illustrate how different definitions and conventions imply different Lorentz transformation formulas for, e.g., the temperature.

\subsection{Reminder: nonrelativistic thermodynamics}
\label{s:nonrelativistic_thermodynamics}

Traditionally, nonrelativistic thermodynamics intends to describe a many-particle systems by means of a few macroscopic control parameters. According to Callen~\cite{1974Ca,Callen}, the most \lq natural\rq~\space candidates for thermodynamic variables are either conserved quantities (such as total internal energy $\Energy$ in the rest frame, particle number $N$, etc.) or parameters that characterize the breaking of symmetries  (e.g., the volume parameter $\Vol$ characterizes the breaking of translation invariance, external magnetic fields may break isotropy, etc.). Within the axiomatic formulation of thermodynamics~\cite{1974Ca,Callen,2001JoOt}, one further postulates the existence of an extensive thermodynamic potential $\Ent(\Energy,\Vol,\ldots)$ called entropy, required to be a monotonous function of $\Energy$. Given the entropy $\Ent$,  intensive thermodynamic quantities (temperature $\Temp$, pressure $\Press$, etc.) are defined by a differentiation of $\Ent$ with respect to the extensive control parameters $(\Energy,\Vol,\ldots)$, yielding the \emph{first law} of thermodynamics
\bse\label{e:first_law_non}
\be\label{e:first_law_1}
\diff \Ent=\Temp^{-1}(\diff \Energy + \Press\, \diff \Vol+\ldots).
\ee
Upon introducing a heat differential form $\dbar \Heat$ and a work contribution  $\dbar \Work$ by~\cite{1968VK,SommerfeldTD}
\be\label{e:heat_definition}
\dbar \Heat:=\Temp\;\diff \Ent
\csp
\dbar \Work&:=&- \Press\, \diff \Vol-\ldots,
\ee
the first law~\eqref{e:first_law_1} can be rewritten in the form
\be\label{e:first_law_2}
\dbar \Heat= \diff\Energy -\dbar \Work.
\ee
\ese
In the nonrelativistic framework, it is rather straightforward to generalize  Eqs.~\eqref{e:first_law_non} to also describe moving systems~\cite{1968VK}. To illustrate this, consider a nonrelativistic thermodynamic system of constant total mass $M$, moving at constant mean velocity $\bs w$. In this case, one can interpret the mean velocity $\bs w'$ as an additional intensive variable, and \emph{define} the internal energy $\Energy$ by~\cite{1968VK}
\be
\HamEn'=\Energy+{M\bs w'^2}/{2},
\ee 
where $\HamEn'$ is the total energy of the moving system; hence,
\bse\label{e:first_law}
\be
\diff\Energy
= \diff \HamEn'- \bs w' \cdot \diff (M \bs w')
= \diff \HamEn'- \bs w' \cdot \diff \bs G',
\ee
where $\bs G':=M\bs w'$ is the total (mean) momentum of the system.
Combining this with Eq.~\eqref{e:first_law_non}, the first law can be expressed as
\be
\dbar \Heat= \diff\HamEn' - \bs w'\cdot \diff \bs G'-\dbar \Work,
\ee
\ese
where the second term on the rhs. may be interpreted as acceleration work.

\subsection{Relativistic case}

The relativistic generalization of Eqs.~\eqref{e:first_law_non}--\eqref{e:first_law} is less trivial due to the following reasons~\cite{1968VK} 
\begin{itemize}
 \item One cannot decompose the relativistic energy $\HamEn$ of a thermodynamic system into the sum of a kinetic part and a term $\Energy$ depending only on the internal state.
 \item In general, the rest mass (energy) of a thermodynamic system is not constant in thermodynamic processes, since, e.g., any heat transfer $\dbar\Heat$ represents a change of energy and, therefore, of mass.
\item  The transfer of energy and momentum between moving systems implies the transfer of mass and, hence, of  momentum.
\item There exist several different, reasonable ways to define heat and work in relativistic systems.
\end{itemize}
When adopting a statistical approach\footnote{Equations~\eqref{e:first_law_non} and \eqref{e:first_law} represent relations between macroscopic observables. The microscopic, statistical  justification of these equations rests on the assumption that thermodynamic variables can be related to well-defined expectation values with respect to a phase space probability distribution~\cite{Becker,2007Ca}. A statistical distribution (ensemble) with density $f$ provides a \lq\lq good\rq\rq\space model of thermodynamics if the expectation values satisfy the differential relation~\eqref{e:first_law} for a suitable entropy functional $\Ent[f]$. Recently, it has been  shown~\cite{2007Ca} that thermodynamic relations of the type~\eqref{e:first_law_non} hold for a rather wide class of distributions (not only microcanonical and canonical ones), provided the entropy functional is chosen appropriately.} towards relativistic thermodynamics, i.e., when identifying thermodynamic variables with statistical averages, one can add one more source of difficulty to this list~\cite{1970Yu}:
\begin{itemize}
 \item Expectation values such as the mean energy or the mean momentum   of a many-particle system are nonlocal quantities whose definition requires choosing (i.e., fixing) a specific hyperplane in spacetime. The choice of the hyperplane may single out a preferred frame of reference.
\end{itemize}
In particular, the last two aspects have led to considerable debate and confusion during the past century, e.g., concerning the transformation of temperature under Lorentz transformations~\cite{1963Ott,1965Ar,1965Ar_1,1965Bo,1965Ga,1965Su,1966La,1966La_1,1966Ki,1966Ar,1966Fr,1966Pa,1966Pa_2,1966Ro,1967Ga,1967La,1967Mo,1967No,1967Re,1967Wi,1967Wi_2,1968Ba,1968Ge,1968LaJo,1968Li,1968Na,1969Ba,1969Ha,1970Ge,1970LaJo,1970Yu,1971CaHo,1977Ag,1978Kr,1981La,1995Ko,1996LaMa,2005ArLoAn} 
\par
\begin{figure}[t]
\centering
\includegraphics[width=10cm]{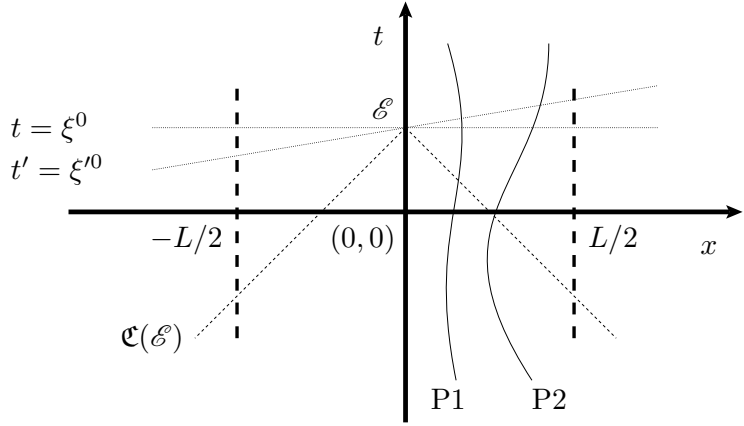}
\caption{
\label{fig:HyperPlanes}
The different hyperplanes as defined in Eq.~\eqref{e:hyperplanes}.
The worldlines of two particles are labeled by ``P1'' and ``P2''; the worldlines of the container walls correspond to vertical lines at $x=-L/2$  and $x=L/2$ (thick dashed lines), respectively.
Assume a lab observer, resting at position $x=0$ in $\Gs$, takes a photograph of the system at the spacetime event $\eve$ with coordinates $(t,x)=(\xi^0,0)$ in~$\Gs$. This photograph will reflect the state of the system along the lightcone $\cone(\eve)$.
}
\end{figure}

In the remainder of this section we briefly summarize three different formalisms of relativistic thermodynamics, as proposed by a Planck~\cite{1908Pl}, Ott~\cite{1963Ott} and van Kampen~\cite{1968VK,1969VK_2}, respectively.
This serves to illustrate how different definitions and conventions yield, e.g., different transformation formulas for the temperature. For this purpose, we consider a confined system with fixed particle number $N$ that, in the lab frame~$\Gs$, can be described by a spatially homogeneous, stationary J\"uttner distribution~\eqref{e:juttner}. Aiming to identify  thermodynamic quantities with statistical averages, we distinguish three different types of hyperplanes in Minkowski spacetime, as illustrated in Fig.~\ref{fig:HyperPlanes}.:
\par
We define the $\Gs$-isochronous hyperplane~$\iso(\xi^0)$, corresponding to events $(t, \bs x)$ with $t=\xi^0$=constant in the lab frame $\Gs$, by 
\bse\label{e:hyperplanes}
\be\label{e:isochronous-sigma}
\iso(\xi^0):=\{\;(t,\bs x)\;|\; t = \xi^0,\;\bs x\in\R^d\; \}.
\ee
Similarly, we define for a frame $\Gs'$ moving at velocity $\bs w=(w,0,\ldots,0)$ relative to the lab frame $\Gs$, the  $\Gs'$-isochronous hyperplane~$\iso'(\xi'^0)$  by 
\be\label{e:isochronous-moving}
\iso'(\xi'^0):=\{\;(t',\bs x')\;|\; t' = \xi'^0,\;\bs x'\in\R^d\; \}.
\ee
We denote the backward lightcone $\cone(\eve)$ of some spacetime event $\eve$, having coordinates  $\bar\xi(\eve)=(\xi^0,\bs \xi)$ in $\Gs$ and coordinates $(\xi'^\ga)=({\Gl^\ga}_\gb\xi^\gb)$  in $\Gs'$, by
\be\label{e:cone}
\cone(\eve):=
\{\;(t,\bs x)\;|\; t = \xi^0-|\bs x-\bs\xi|,\;\bs x\in\R^d\;\}.
\ee
\ese
We next introduce relativistic thermodynamic variables as integrals 
over these hyperplanes.
 
\paragraph*{Energy and momentum}
We distinguish energy-momentum mean values defined along the different hyperplanes:\footnote{Although the energy-momentum tensor was introduced above with the help of the one-particle phase space PDF, Eqs.~\eqref{e:td_energy} can also be used to develop more general  thermodynamic theories based on the energy-momentum tensor (and other conserved currents) constructed, e.g., from field theories~\cite{Weinberg,WeinbergQFT1,1976Is,2002Ga}.}
\bse\label{e:td_energy}
\be
\Energy^\gb[\xi^0]& :=&\label{e:td_energy_a}
N\int_{\iso(\xi^0)}  \diff \gs_\ga\; \gt^{\ga \gb}(t,\bs x),
\\
\Energy^\gb[\xi'^0] &:=&\label{e:td_energy_b}
N\int_{\iso'(\xi'^0)} \diff \gs_\ga\; \gt^{\ga \gb}(t,\bs x),\\
\Energy^\gb[\eve] &:=&\label{e:td_energy_c}
N\int_{\cone(\eve)} \diff \gs_\ga\; \gt^{\ga\gb}(t,\bs x).
\ee
\ese
In the presence of an arbitrary external confinement, the energy-momentum tensor is \emph{not}  divergence-free everywhere,\footnote{Similar problems arise if one wants to construct a relativistic continuum model for the  electron~\cite{1966Ro,1967Ga,1986CaJi}.} cf. Eq.~\eqref{e:em-divergence}. In this case, each of the three quantities from Eq.~\eqref{e:td_energy} represents a different\footnote{The three four-vectors from Eq.~\eqref{e:td_energy} would coincide if the energy-momentum tensor were divergence-free everywhere~\cite{1970Yu}.} nonlocal four-vector, since they refer to different hyperplanes, respectively. In principle, either of the three four-vectors could be used as thermodynamic variable, giving rise to different thermodynamic formalisms~\cite{1970Yu}.
\par
The lab-isochronous four-vector $\Energy^\gb[\xi^0]$ from Eq.~\eqref{e:td_energy_a} is most easily calculated in the lab frame~$\Gs$; cf. Appendix~\ref{appendix:surface_integrals}. Using Eq.~\eqref{e:em-vector-t-c}, one finds
\bse\label{e:em-lab}
\be\label{e:em-lab-a}
\Energy^\gb[\xi^0]
=
N\lan p^\gb\ran_{\xi^0}=
N\begin{cases}
\energy,&\quad  \gb=0,\\
0,        &\quad  \gb\ne 0.
\end{cases}
\ee
In the case of a J\"uttner distribution, the one-particle expectation value $\energy=\energy_d$ is given by Eqs.~\eqref{a-e:Juttner-energy-b}--\eqref{a-e:Juttner-energy-d}. In another frame $\Gs'$, moving with velocity $w$ along the $x^1$-axis of $\Gs$, one finds the transformed energy-momentum vector  \mbox{$\Energy'^\gb[\xi^0]={\Gl(w)^\gb}_\nu \Energy^\nu[\xi^0]$} as
\be\label{e:em-lab-b}
\Energy'^\gb[\xi^0]
=
N \begin{cases}
\gc\energy, &\qquad  \gb=0,\\
-\gc w \energy, &\qquad  \gb=1,\\
0, &\qquad  \gb>1,
\end{cases}
\ee
\ese
where $\gc(w)=(1-w^2)^{-1/2}$ and ${\Gl(w)^\gb}_\nu$  denotes the Lorentz transformation matrix. As discussed below, Ott~\cite{1963Ott} and van  Kampen~\cite{1968VK,1969VK_2} use  Eqs.~\eqref{e:em-lab} as starting point for their formulations of relativistic thermodynamics, but 
consider different definitions of heat and temperature, respectively.

\par
For comparison, the $\Gs'$-isochronous energy-momentum four-vector, defined in  Eq.~\eqref{e:td_energy_b} by integration along the hyperplane $\iso'(\xi'^0)$, is most conveniently calculated in $\Gs'$, 
yielding, by virtue of Eq.~\eqref{e:em-vector-t-prime-explicit},
\bse\label{e:em-moving}
\be\label{e:em-moving-a}
\Energy'^\gb[\xi'^0]=
N\lan p'^\gb\ran_{\xi'^0}
=
N
\begin{cases}
\gc (\energy+w^2 \vir /d),&\qquad \gb=0,\\
-\gc w(\energy+ \vir /d),&\qquad \gb= 1,\\
0,                   &\qquad \gb>1,
\end{cases}
\ee
where  $\vir =d/\gb$ for the J\"uttner distribution, cf. Eq.~\eqref{e:single_virial}. Equation~\eqref{e:em-moving-a} presents the basis of the Planck-Einstein formulation~\cite{1907Ei,1908Pl} of relativistic thermodynamics, cf. discussion below. Applying an inverse Lorentz 
transformation $\Gl(-w)$ to Eq.~\eqref{e:em-moving-a} gives
\be\label{e:em-moving-b}
\Energy^\gb[\xi'^0]
=N
\begin{cases}
\energy,&\qquad \gb=0,\\
-  w \vir /d,&\qquad \gb= 1,\\
0,         &\qquad \gb>1.
\end{cases}
\ee
\ese
The non-vanishing component with $\gb=1$ reflects the fact that the integration was performed along the hyperplane \lq\lq$t'$=constant\rq\rq, which introduces an apparent asymmetry in the lab frame, cf.~Fig.~\ref{fig:HyperPlanes},  and results in a spurious mean momentum.\footnote{Some authors~\cite{1963Ott,1969VK,1969VK_2} have interpreted the $w^2$-term in Eq.~\eqref{e:em-moving} as an energy contribution due to presence of the walls. This view, although objected to by others~\cite{1970Yu}, seems to be at least partially correct. The above derivation shows that the difference between $\Energy'^\gb[\xi^0]$ and $\Energy'^\gb[\xi'^0]$ can be attributed to the different underlying hyperplanes $\iso(\xi^0)$ and $\iso'(\xi'^0)$, respectively. However, Eqs.~\eqref{e:td_energy_a} and \eqref{e:td_energy_a} would give the same result if the energy-momentum tensor were divergence-free everywhere~\cite{1970Yu}. In the above example of a homogeneous gas, the divergence is non-zero only on the boundary due to the appearance of the $\Gt$-function in the phase space density, cf. Eq.~\eqref{e:em-divergence}; i.e., the difference between $\Energy'^\gb[\xi^0]$ and $\Energy'^\gb[\xi'^0]$ is indeed related to the presence of the boundary. It seems that Einstein was aware of this problem, cf. his remarks in~\cite{1907Ei}, but at that time did not follow up this issue.} 
\par
In fact, the most frequently discussed versions of relativistic thermodynamics (see, e.g.,~\cite{1907Ei,1908Pl,1963Ott,1965Ar,1965Ar_1,1965Bo,1965Ga,1965Su,1966La,1966La_1,1966Ki,1966Ar,1966Fr,1966Pa,1966Pa_2,1966Ro,1967Ga,1967La,1967Mo,1967No,1967Re,1967Wi,1967Wi_2,1968Ba,1968Ge,1968LaJo,1968Li,1968Na,1969Ba,1969Ha,1970Ge,1970LaJo,1970Yu,1971CaHo,1977Ag,1978Kr,1981La,1995Ko,1996LaMa,2005ArLoAn}) adopt -- implicitly or explicitly -- either  Eqs.~\eqref{e:em-lab} or  Eqs.~\eqref{e:em-moving} as the starting point. From an experimental point of view, these equations refer to different, nonequivalent measurements procedures: 
\par
In order to determine $\Energy'^\gb[\xi^0]$ a moving observer (at rest $\Gs'$) had to reconstruct the velocities and/or momentum values along the hyperplane $t=\xi^0$, whereas to obtain $\Energy'^\gb[\xi'^0]$ velocities must be sampled $\Gs'$-simultaneously along $t'=\xi'^0$. 
On the other hand, as pointed out by Gamba~\cite{1967Ga}, even for a lab observer it will be very difficult (if not impossible in practice) to reconstruct the velocity data along the hyperplane $t=\xi^0$ due to the finiteness of the speed of light. 

\paragraph*{Isochronous \textit{vs.} lightcone averages}
A more natural way of performing measurements (e.g., in astronomy) is to  \lq take a photograph\rq\space of a given system. A photograph recorded by an observer at the event $\eve$ reflects the state of the system along the lightcone $\cone[\eve]$, see Fig.~\ref{fig:HyperPlanes}. Let us assume that an idealized photograph encodes both position and velocity\footnote{In practice, velocities can be reconstructed from color variations due the Doppler shift of spectral lines.} of the particles in the system. Then, the empirical energy and momentum averages which can be sampled from this data correspond to the lightcone average $\Energy^\gb[\eve]$ as defined in Eq.~\eqref{e:td_energy_c}. 
\par
Calculating $\Energy^\gb[\eve]$ for a stationary lab distribution of the form $\vphi(\bs x,\bs p)=\gr(\bs x)\phi(\bs p)$, with isotropic momentum PDF $\phi$, gives in the lab frame (cf. Appendix~\ref{appendix:surface_integrals})
\bse\label{e:lightcone-average}
\be
\Energy^{0}[\eve]
&=&\label{e:lightcone-average-a}
N\energy,\\
\Energy^{i}[\eve]
&=&\label{e:lightcone-average-b}
N\,d^{-1}\vir \int \d^dx\;
\f{x^i-\xi^i}{|\bs x-\bs\xi|}\;
\gr(\bs x)
\csp
i=1,\ldots,d,
\ee
\ese
where $(\xi^0,\bs \xi)$ are the spacetime coordinates $\eve$ in $\Gs$.
By comparing with Eqs.~\eqref{e:em-lab-a} and~\eqref{e:em-moving-b},
we see that the 0-components are identical,
\be
\Energy^{0}[\eve]=\Energy^{0}[\xi^0]=\Energy^{0}[\xi'^0];
\ee
i.e. when sampling energy values it does not matter whether this is done from a photograph or from simultaneously collected (i.e., reconstructed) data.
\par
The situation is different, when estimating the mean momentum. As evident from Eq.~\eqref{e:lightcone-average-b}, even for a lab observer at rest in $\Gs$, the lightcone average depends on the position $\bs \xi$ of the observer.\footnote{Clearly, averages in the lab frame do not depend on the specific value $\xi^0$ of the time coordinate if the PDF is stationary in this frame.}  A distinguished  \lq\lq photographic center-of-mass\rq\rq\space  position $\bs \xi_*$ in $\Gs$ can defined by
\be
\Energy^{i}[\eve]\bigr|_{\bs \xi=\bs \xi_*}=0
\csp
i=1,\ldots, d.
\ee 
For example, if $\gr$ is symmetric with respect to the origin of $\Gs$, then $\bs \xi_*=\bs 0$; this would correspond to a lightcone as drawn in  Fig.~\ref{fig:HyperPlanes}.
\par
To illustrate how $\Energy^{i}[\eve]$ depends on the observer position, assume that the density $\gr$ vanishes outside a finite region $\Volume\in\R^d$ in $\Gs$ and that a stationary observer is located at a position $\bs \xi$ far away from $\Volume$. In this case, one can approximate $|\bs x-\bs\xi|\simeq|\bs\xi|$ in the integrand of   Eq.~\eqref{e:lightcone-average-b}, yielding
\be
\Energy^{i}[\eve]
&=&\label{e:lc-i-outside}
Nd^{-1}\vir
\f{\EW{x^i}-\xi^i}{|\bs\xi|}.
\ee
In particular, when considering a homogeneous J\"uttner gas with position mean value $\EW{\bs x}=\bs 0$ and $\vir=d/\gb$, then\footnote{Note that Eqs.~\eqref{e:lightcone-average-b} and~\eqref{e:apparent_drift} are  consistent with Eq.~\eqref{e:em-moving-b}, as can be seen by letting $\xi^1\gg 0$, $\xi^2=\ldots=\xi^d=0$ and $w\to 1$ in  Eq.~\eqref{e:em-moving-b}.} 
\be\label{e:apparent_drift}
\Energy^{i}[\eve]=-\f{\xi^i}{|\bs\xi|}\f{N}{\gb};
\ee
i.e., an outside observer gas, who naively estimates $\Energy^{i}[\eve]$ from his photographic data, could erroneously conclude that the gas is moving away from him (at a velocity proportional to the temperature). This effect\footnote{The effect becomes neglible if $\gb mc^2\gg 1$.} should be taken into consideration when estimating the velocities  of astrophysical objects from photographs.
\par
From the conceptual point of view, it is worthwhile to note that the hyperplane \lq\lq backward lightcone $\cone(\eve)$\rq\rq\space  is a relativistically invariant object which is equally accessible for any inertial observer. Put differently, if a second observer, moving relative to the first one, takes a snapshot at the same event $\eve$ then her picture will reflect the same state of the system -- even though the colors will be different due to the Doppler effect caused by the observer's relative motion~\cite{Weinberg}.
\par
Another advantage of lightcone averages as defined in Eq.~\eqref{e:td_energy_c} lies in the fact that they can be generalized to general relativity~\cite{MiThWe00,Weinberg} in a straightforward manner, whereas it becomes very difficult to single out a globally  acceptable \lq\lq simultaneous\rq\rq\space hyperplane in curved spacetime.\footnote{In the nonrelativistic limit case $c\to\infty$, the lightcone \lq\lq opens up\rq\rq\space so that photographic measurements become isochronous in any frame in this limit.} In view of these benefits, lightcone averages appear to be the better -- if not the best -- suited candidates if one wishes to characterize a many-particle system by means of nonlocally defined, macroscopic variables within a relativistic  framework.
\par 
However, since historically most authors considered either lab-simultaneously or observer-simultaneously defined quantities, we will restrict ourselves  in the remainder to discussing the  implications of these two choices.

\paragraph*{Entropy}
Having identified the potential candidates for the thermodynamic state variables \lq energy\rq\space and \lq momentum\rq, one still needs to specify \lq entropy\rq. In the case of a J\"uttner gas, one can define an entropy density four-current per particle by~\cite{2008De,CercignaniKremer} 
\be\label{e:entropy-4-current}
s^\ga(t,\bs x)=
-\int \f{\diff^dp}{p^0}\;p^\ga\; 
f(t,\bs x,\bs p) \ln[ h^df(t,\bs x,\bs p)].
\ee
The \lq logarithmic\rq~form of this entropy current is specifically  adapted~\cite{2007Ca}  to the exponential form of the J\"uttner distribution and/or \textit{vice versa}, cf.  Eq.~\eqref{e:entropy_principle-a}. Inserting the J\"uttner function~\eqref{e:juttner} into Eq.~\eqref{e:entropy-4-current}, one finds that the entropy current is stationary in the lab frame $\Gs$ and given by 
\be\label{e:entropy-4-current-juttner}
s^\ga(t,\bs x)=
\Vol^{-1}\;\Ind(\bs x;\Volume)
\begin{cases}
\ln(\Vol \Z_d /h^{d})+\gb\lan\eps\ran_d, &\qquad\ga=0,\\
0,  &\qquad\ga = 1,\ldots,d.
\end{cases}
\ee
Hence, the current~\eqref{e:entropy-4-current-juttner} satisfies the conservation law 
\be\label{e:entropy_conservation}
\p_\ga s^\ga=0.
\ee
The thermodynamic entropy $\Ent$ is obtained by integrating Eq.~\eqref{e:entropy-4-current-juttner} over some space-like or light-like hyperplane $\hyp$, yielding the Lorentz invariant quantity
\be\label{e:entropy_definition}
\Ent[\hyp] 
:= 
N\int_{\hyp} \diff \gs_\ga\; s^\ga(t,\bs x)
=
N\int_{\hyp} \diff \gs'_\ga\; s'^\ga(t',\bs x') 
=:\Ent'[\hyp].
\ee
The conservation law~\eqref{e:entropy_conservation} implies that the integral~\eqref{e:entropy_definition} is the same for the hyperplanes $\iso(\xi^0)$, $\iso'(\xi'^0)$ and $\cone(\eve)$, i.e., 
\be
\Ent[\eve]=\Ent'[\eve]=
\Ent[\xi^0]=\Ent'[\xi^0]=
\Ent[\xi'^0]=\Ent'[\xi'^0].
\ee
The integral~\eqref{e:entropy_definition} is most conveniently calculated along $\hyp=\iso[\xi^0]$ in $\Gs$, yielding
\be
\Ent
=
N\int\diff^dx\; s^0(\xi^0,\bs x)
=
N\ln(\Vol \Z_d/ h^{d} )+\gb\; N\lan\eps\ran.
\ee

\paragraph*{Heat and temperature}
By means of the preceding considerations, we can now summarize and compare the most 
commonly discussed versions of relativistic thermodynamics~\cite{1907Ei,1908Pl,1963Ott,1965Ar,1965Ar_1,1965Bo,1965Ga,1965Su,1966La,1966La_1,1966Ki,1966Ar,1966Fr,1966Pa,1966Pa_2,1966Ro,1967Ga,1967La,1967Mo,1967No,1967Re,1967Wi,1967Wi_2,1968Ba,1968Ge,1968LaJo,1968Li,1968Na,1969Ba,1969Ha,1970Ge,1970LaJo,1970Yu,1971CaHo,1977Ag,1978Kr,1981La,1995Ko,1996LaMa,2005ArLoAn}. As mentioned above, these versions are based on simultaneously defined averages, respectively. In spite of this common feature, differences exist regarding 
\begin{itemize}
\item 
the choice of the underlying hyperplane and/or
\item
the adopted definition of heat, 
\end{itemize}
leading, e.g., to  different temperature transformation laws. Taking into account the historical order, we begin by recalling the Planck-Einstein formulation~\cite{1908Pl,1907Ei} of relativistic thermodynamics.
\par
Guided by Eqs.~\eqref{e:first_law}, Planck and Einstein opted in 1907/08  for the following definition of heat~\{cf. Eq.~(23) in \cite{1907Ei}\}
\bse
\be\label{e:planck-a}
\dbar \Heat'(w'):=
\diff\HamEn'  -w'\diff\Mom'+ \Press'\diff\Vol'
=:\Temp'\diff\Ent',
\ee
where $w'=-w$ is the constant velocity\footnote{For simplicity, we again  consider a thermodynamic system that moves with velocity $\bs w'$ along the $x'^1$-axis of $\Gs'$, so that $\bs w'=(w',0,\ldots,0)=(-w,0,\ldots,0)=-\bs w$ with $\bs w$ denoting the velocity of $\Gs'$ relative to the restframe $\Gs$ of the system.} of the thermodynamic system in the frame $\Sigma'$, and\footnote{The scalar transformation~\eqref{e:planck-d} of the pressure $\Press$ is implied by the transformation laws of force and area~\cite{1908Pl,1968VK}.}\footnote{After reinstating constants $c$ in Eq.~\eqref{e:planck-f}, it becomes evident that $\Mom'(w')\to w'M$ in the nonrelativistic limit case $c\to\infty$.}
\be\label{e:planck-b}
\Vol'(w')&=&\Vol/\gc,\label{e:planck-c}\\
\Press'(w')&=&\Press,\label{e:planck-d}\\
\Ent'(w')&=&\Ent,\\
\HamEn'(w')&=&\gc\left(\HamEn+w'^2\,\Press \Vol\right),\\
\Mom'(w')&=&\gc w'\left(\HamEn+\Press \Vol\right),
\label{e:planck-f}
\ee
with $\gc=\gc(w')=\gc(w)$ denoting the Lorentz-factor. The  choice~\eqref{e:planck-b}--\eqref{e:planck-f} corresponds to defining thermodynamic energy and momentum observer-simul\-taneously as in Eqs.~\eqref{e:em-moving-a} and identifying the pressure as
\be
\Press\Vol= N \vir/d,
\ee
so that for a J\"uttner gas we have
\be\notag
\Press\Vol=N\gb^{-1}.
\ee
Furthermore, substituting Eqs.~\eqref{e:planck-b}--\eqref{e:planck-f} into Eq.~\eqref{e:planck-a} gives
\be
\dbar \Heat'(w')
= \gc^{-1} \left(\diff\HamEn +\Press\,\diff \Vol \right)
= \gc^{-1}\;\dbar \Heat
= \gc^{-1}\;\Temp\,\diff\Ent,
\label{e:planck-g}
\ee
and comparing Eq.~\eqref{e:planck-g} with \eqref{e:planck-a} leads to the  temperature transformation formula of Planck~\cite{1908Pl}\footnote{It seems that, in the later stages of his life, Einstein changed his opinion about the correct transformation laws of thermodynamic quantities, favoring formulas which were later independently derived by Ott~\cite{1963Ott} and Arzelies~\cite{1965Ar,1965Ar_1};  cf. the corresponding discussion by Liu~\cite{1992Liu,1994Liu}, Schr\"oder and Treder~\cite{1992ScTr}, and Requardt~\cite{2008Re}.}
\be
\Temp'=\gc^{-1}\,\Temp=(1-w'^2)^{1/2}\,\Temp,
\ee
\ese
stating that \emph{a moving body appears cooler} (for the J\"uttner gas $\Temp=\gb^{-1}$). This formalism was criticized by Ott~\cite{1963Ott} in 1963 and later also by van Kampen~\cite{1968VK,1969VK_2} and Landsberg~\cite{1966La,1967La} -- mostly because the quantities $(\HamEn',\Mom')$ and  $(\HamEn,\Mom)=(\HamEn'(0),\Mom'(0))$ are \emph{not} related by a Lorentz transformation. As discussed above, this drawback can be traced back to the fact that $\HamEn'$ and $\Mom'$, if taken at different values of $w'$, correspond to different hyperplanes, respectively~\cite{1970Yu}.

\par
To overcome this deficiency, van Kampen~\cite{1968VK,1969VK_2} proposed to define, instead of Eq.~\eqref{e:planck-a}, a thermal energy-momentum transfer four-vector by
\bse\label{e:vk}
\be\label{e:vk-a}
\dbar\Heat^\ga:=\diff \HamEn^\ga-\dbar\Work^\ga,
\ee
where $(\dbar\Work^\ga):=(\Press \diff \Vol,\bs 0)$ in the lab frame $\Gs$, and $\HamEn^\ga$ is chosen to be the lab-simultaneous energy-momentum vector from Eqs.~\eqref{e:em-lab}, i.e.,
\be\label{e:vk-b}
\HamEn^\ga&:=&\Energy^\ga[\xi^0].
\ee
The particular choice~\eqref{e:vk-b} singles out the lab-isochronous hyperplane $\iso[\xi^0]$,  and in a moving frame one then has
\be
\dbar\Heat'^\ga
&=&\notag
\diff \HamEn'^\ga-\dbar\Work'^\ga,\\
(\diff \HamEn'^\ga)
&=&\label{e:vk-e}
(\gc \HamEn^0,\gc\bs w'\HamEn^0),\\
(\dbar\Work'^\ga)
&=&\notag
(\gc\Press \diff \Vol,\gc\bs w'\Press \diff \Vol),
\ee
\ese
with unprimed quantities referring to the lab frame~$\Gs$, defined by $\bs w'=\bs 0$. While essentially agreeing on Eqs.~\eqref{e:vk-a}--\eqref{e:vk-e} and also on the scalar nature of entropy $\Ent'=\Ent$, van Kampen and Ott differ in how to define the temperature of a moving system, i.e., how to  formulate the (first part\footnote{The \lq\lq second part\rq\rq\space~\cite{SommerfeldTD} of the second law states that $\diff\Ent\ge 0$ in a closed system.} of the) \emph{second law}. More precisely, Ott opted for the definition 
\bse\label{e:ott-temperature}
\be\label{e:ott-a}
\Temp'\,\diff\Ent'
:=\dbar \Heat'^0
= \gc \left(\diff\HamEn^0 +\Press\,\diff \Vol \right)
=\gc\, \dbar \Heat^0
= \gc\;\Temp\,\diff\Ent,
\ee
yielding the modified temperature transformation law\footnote{See also Eddington~\cite{1923Ed} and Arzelies~\cite{1965Ar}.}
\be\label{e:ott-b}
\Temp'=\gc\,\Temp=(1-w'^2)^{-1/2}\,\Temp;
\ee
\ese
i.e., according to Ott's definition of heat and temperature \emph{a moving body appears hotter}.
\par 
However, van Kampen was able to demonstrate by example~\cite{1968VK} that the Eqs.~\eqref{e:ott-temperature} are not well-suited if one wants to describe  heat and energy-momentum exchange between systems that move at different velocities (hetero-tachic processes). To obtain a more convenient description, he proposed to characterize the heat transfer  by means of the scalar quantity~\cite{1968VK,1969VK_2} 
\bse\label{e:vk-temperature}
\be
\dbar \Heat'
:=-w'_\ga \dbar\Heat'^\ga
=-w_\ga \dbar\Heat^\ga
=\dbar \Heat=\dbar\Heat^0,
\ee
with  $(w'_\ga)=(-\gc,\gc \bs w')$ denoting the velocity four-vector of the system in~$\Gs'$, and reducing to $(w_\ga)=(-1,\bs 0)$ in the lab frame~$\Gs$. He then defined the temperature by
\be
\Temp' \diff \Ent':=\dbar \Heat'=\dbar \Heat=\Temp \diff \Ent,
\ee
so that, in view of $\Ent'=\Ent$, van Kampen's temperature is a scalar
\be
\Temp'=\Temp;
\ee
\ese
i.e., according to this definition \emph{a moving body neither hotter nor colder}. Adopting this temperature definition, one can define an inverse temperature four-vector by
\bse
\be
\gb'_\ga:=\Temp'^{-1}\,w'_\ga=\Temp^{-1}\,w'_\ga,
\ee
which allows us to rewrite the second law in the compact covariant form
\be
\diff\Ent' 
=
-\gb'_\ga \dbar\Heat'^\ga.
\ee
\ese
Thus, the thermodynamic formalisms proposed by Ott~\cite{1963Ott} and van Kampen~\cite{1968VK,1969VK_2} are based on the same lab-isochronous
hyperplane~\cite{1970Yu} but both formulations differ only with regard to their respective temperature definition. By contrast, the Einstein-Planck formulation~\cite{1907Ei,1908Pl} is based on an  observer-dependent hyperplane.
\par
To summarize, the above discussion shows that whether a moving body appears hotter or not  depends on how one defines thermodynamic quantities. Of course, before comparing the results of experimental observations with theoretical predictions one should make sure that the applied measurement procedures are consistent with the definitions employed in the theory (choice of the hyperplane,  definitions of heat and work, etc.).




\bibliographystyle{unsrt}
\bibliography{BrownianMotion,BrownianMotors,CosmicRays,QuantumBM,Debbasch,DiverseBooks,Einstein,Entropy,FinanceBooks,FT,Hakim,Hanggi,LorentzDirac,Numerics,PathIntegrals,PhotonDiffusion,RelKin,RelStatMech,RelMany,RelConstraint,RelFPE,RelStochQuant,RelTD,RBM,RBMapplied,RBMappliedHE,RBMappliedAstro,RBMradiation,SpecRelativity,StochProc,Telegraph,VanKampen,TD,Journals,Proceedings,MathBooks,ActiveBM,Anomalous,Schimansky,Ebeling,Theses,Selfenergy,StochRes,Added}






\end{document}